%% file: background.tex
\documentclass[11pt]{JHEP3}
\usepackage{graphicx}
\usepackage{amssymb}

\input{newcomm}
\newcommand{\RGB}{R_{\rm GB}}
\newcommand{\zg}{Z_0}

\newcommand{\Rs}{R_{\rm s}}
\newcommand{\Rd}{R_{\rm d}}
\newcommand{\sss}{s_{\rm s}}
\newcommand{\sd}{s_{\rm d}}
\newcommand{\kpu}{\kp_{\rm R}}
\newcommand{\kps}{\kp_{\rm R}^*}
\newcommand{\Lmu}{\Lm_0}
\newcommand{\Gmu}{\Gm}
\newcommand{\Sgu}{\Sg}
\newcommand{\Ru}{R}
\newcommand{\Su}{S}
\newcommand{\rnotc}{r_{0{\rm c}}}
\newcommand{\rplusc}{r_{+\rm c}}
\newcommand{\rhta}{\rh_{\rm m}^{(1,16000)}}

\newcommand{\Rav}{R_{\rm av}}
\newcommand{\chb}{\ch_{\rm b}}
\newcommand{\chp}{\ch_{\rm p}}

\newcommand{\Rc}{R_{\rm c}}

\newcommand{\mO}{\mathcal{O}}

\newcommand{\numc}{n_{\rm c}}

\newcommand{\Reff}{R_{\rm eff}}
\newcommand{\RR}{R_{\rm R}}
\newcommand{\Rosc}{R_{\rm osc}}

\newcommand{\kpc}{\kappa_2^{\rm c}}

\newcommand{\dlat}{d_{\ell}}
\newcommand{\rlat}{r_{\rm \ell}}
\newcommand{\dcon}{d_{\rm c}}
\newcommand{\rcon}{r_{\rm c}}
\newcommand{\rstat}{r_{\rm stat}}
\newcommand{\Vcon}{V_{\rm c}}

\newcommand{\rmm}{r_{\rm m}}
\newcommand{\rav}{r_{\rm av}}
\newcommand{\veff}{v_{\rm eff}}
\newcommand{\rfitmin}{r_{\rm min}}
\newcommand{\rfitmax}{r_{\rm max}}
\newcommand{\ellt}{\tilde\ell}
\newcommand{\Df}{d_{\rm H}}
\newcommand{\Ds}{d_{\rm s}}

\preprint{}

\title{Continuum interpretation of the dynamical-triangulation formulation of\\ quantum Einstein gravity}

\author{Jan Smit\\
Institute for Theoretical Physics, University of Amsterdam, \\
Science Park 904, P.O.\ Box 94485, 1090 GL, Amsterdam, the Netherlands.\\
}

\keywords{quantum gravity, lattice field theory}

\abstract
{
In the time-space symmetric version of dynamical triangulation, a non-perturbative formulation of quantum Einstein gravity, numerical simulations without matter have shown two phases, with spacetimes that are either crumpled or elongated like branched-polymers, with strong evidence of a first-order transition between them. These properties have generally been considered unphysical. Using previously unpublished numerical results, we give an interpretation in terms of continuum spacetimes that have constant positive and negative curvature, respectively in the `elongated' and `crumpled' phase. The magnitude of the positive curvature leads naturally to average spacetimes consisting solely of baby-universes in a branched-polymer structure, whereas the negative curvature accommodates easily a large mother universe, albeit with a crumpling singularity. Nevertheless, there is evidence for scaling in the crumpled phase, which we compare with the well-known scaling in the elongated phase. Using constraint effective-action models we analyze existing numerical susceptibility-data of the phase transition and determine the behavior of the average Regge-curvature. We propose a renormalization of the Regge curvature and compare it to the curvature of the above continuum spacetimes, and also to the curvature implied by the Gauss-Bonnet theorem in the continuum. The latter involves a more benign multiplicative renormalization and suggests that simulations at larger volumes are needed to settle to order of the phase transition.
}

\begin{document}
\section{Introduction}
\label{secintro}

Causal Dynamical Triangulation (CDT) \cite{Ambjorn:1998xu,Ambjorn:2001cv} appears to be a viable direction in the quest of a non-perturbative formulation of quantum Einstein gravity (QEG). Numerical simulations have shown
the existence of a phase region in parameter space
bounded by first and second order critical lines
with non-trivial physical properties: an average geometry with semiclassical as well as fractal features and distant-dependent spectral dimensions \cite{Ambjorn:2005qt,Ambjorn:2008wc,Ambjorn:2010fv,Ambjorn:2012ij}. A basic aspect of the formulation is a foliation of spacetime by space-like hypersurfaces -- hence the name `causal' -- which facilitates a clear analytic continuation between real and imaginary time; for a review see \cite{Ambjorn:2012jv}. Prior to CDT a formulation of dynamical triangulation that is symmetric between space and (imaginary) time -- Symmetric Dynamical Triangulation\footnote{Also known as Euclidean Dynamical Triangulation (EDT).} (SDT)
\cite{Agishtein:1991cv,Ambjorn:1991pq} -- was investigated vigorously. This direction to QEG was largely abandoned because the average background geometries unveiled by numerical simulation were deemed unphysical; see e.g.\ \cite{Thorleifsson:1998jr,Krzywicki:1999ai} for reviews.
A better understanding of SDT results from a continuum viewpoint is desirable, and as a step in this direction we propose here a continuum interpretation of its average geometries.

Without matter the SDT model has two phases,
depending on the value of the bare Newton constant, a {\em crumpled} phase and an {\em elongated} phase.
The spacetimes in these phases do not appear to be physical. In the elongated phase they have tree-like (`branched polymer') characteristics \cite{Ambjorn:1995dj,Ambjorn:1996ny,Gabrielli:1997zy,Gionti:1998jy,AmbjornDJ1997} whereas those in the crumpled phase contain `singular structures', vertices and links connected to macroscopically large volumes in a single lattice step \cite{Hotta:1995ud,Hotta:1995ca,Catterall:1995ig,Catterall:1997xj}.
Strong evidence was found that the transition between the phases is of first order \cite{Bialas:1996wu,deBakker:1996zx}.

Arguments were put forward that a sufficient amount of matter might ameliorate the results \cite{Jurkiewicz:1996yd,Antoniadis:1996pb} (see also \cite{Antoniadis:1992xu}).
With gauge fields the SDT model acquires a third phase, called {\em crinkled} \cite{Bilke:1997sc,Bilke:1998vj} or {\em smooth} \cite{Horata:2000eg,Horata:2003hm}. This phase was also found without gauge fields upon adding a so-called `measure term' \cite{Bruegmann:1992jk}
to the action, with a new parameter that represents the number of gauge fields \cite{Bilke:1997sc,Bilke:1998vj,Ambjorn:1999ix}. Study of this new phase led to a confusing state of affairs: whereas some authors did not convincingly find physically attractive behavior \cite{Bilke:1998vj,Ambjorn:1999ix}, others found evidence for a higher-than-first order transition \cite{Horata:2000eg} with a susceptibility exponent suggestive of emerging graviton degrees of freedom \cite{Horata:2003hm,Antoniadis:1992xu}. Recent studies of the third
phase found that its spectral dimension also
depends on the distance scale \cite{Laiho:2011ya,Laiho:2011zz}.

The continuum interpretation developed here is at scales fairly close to the UV regulator.
Such a situation is known in phenomenological particle theory, in the continuum but with a cutoff on the momenta, for example quark models of nuclear matter \cite{Nambu:1961tp,Buballa:2003qv} or meson models. At fixed cutoff, such continuum models can have first-order phase transitions as a function of the bare parameters.
With approximate $SU(n)\times SU(n)$ chiral symmetry a limit can be taken in which the masses of pseudo Nambu-Goldstone bosons (e.g.\ $\pi$, $K$, $\et$ for $n=3$) become arbitrarily small. However, other physical quantities such as the pion and kaon decay constants then still remain of order of the cutoff.
In CDT and SDT, which a have remnant of diffeomorphism invariance consisting of permutations of lattice coordinates, a similar phenomenon may occur: massless gravitons emerging in the large lattice limit\footnote{In this respect an interesting field-theory example is $Z(n)$ lattice gauge theory. The model is completely discrete but for  $n\geq 6$ it has a `Coulomb phase' with massless photons
\cite{Elitzur:1979uv,Horn:1979fy,Ukawa:1979yv,Yankielowicz:1981ug,Alessandrini:1982ju}.},
with a Plank length of order of the lattice spacing. We do not consider a first-order transition to be an obstruction to a continuum interpretation of SDT.

Our interpretation is based on fitting continuum observables to lattice ones, similar to what is sometimes done in lattice field theory\footnote{For example, the values of a scalar field propagator $G_\ell(x,y)$ on a hypercubic lattice
($x$ and $y$ are integer 4-tuples)
may be arrayed as a decreasing function of $r=|x-y|$ and fitted by a continuum propagator
$G_{\rm c}(r) = z K_1(m r)/r$ in a certain fitting domain; $G_{\rm c}(r)$ is then a continuum approximation to $G_\ell(x,y)$.}.
In earlier work together with B.V.\ de Bakker a continuum curvature was derived from a volume-distance correlator \cite{deBakker:1994zf}. This curvature was found negative in the crumpled phase, positive in the elongated phase and changed sign near the transition. Examples of scaling were also given. The present work elaborates on this, using numerical results obtained shortly after \cite{deBakker:1994zf} that were left unpublished and
incorporating also other work on scaling \cite{Ambjorn:1995dj} and the structure of SDT spacetimes \cite{Bialas:1996eh,Egawa:1996fu}.

In section \ref{secintroSDT} we summarize some basic formulas of SDT.
Section \ref{secobservables} recalls the Regge curvature $\RR$ that appears in the action, the definition of the lattice geodesic-distance and the volume-distance correlator, the previous curvatures derived from these \cite{deBakker:1994zf}, and shows some numerical results for larger volumes. In section \ref{secsmooth} we mention some properties of smooth spacetimes of constant curvature which form the basis of a more elaborate observable (essentially a local proper-time metric) to be fitted to the numerical data. The result of such fits, shown in section \ref{secLocMet}, is a continuum curvature $\Rc$ which changes sign at the phase transition. Section \ref{secmotherbaby} discusses the relevance of $\Rc$ to the baby- and mother-universes found in simulations of SDT.
Evidence of scaling in the crumpled phase is presented in section \ref{secscaling} and compared with that in the elongated phase.

To compare $\Rc$ with $\langle \RR\rangle$ we study in section \ref{seceffact} the {\em constrained effective-action} of the volume-averaged $\RR$. This gives us a tool to obtain the average Regge-curvature -- up to a constant -- in the region of the phase transition from  the data in \cite{deBakker:1996zx}. This is done in section \ref{secPT}, where also the constant is determined by comparison with data in \cite{deBakker:1995yb}. We propose additive and multiplicative renormalizations of the Regge curvature to enable comparison with continuum curvatures, such as $\Rc$ and the curvature implied by the Gauss-Bonnet theorem.
The multiplicative renormalization-constant following from the Gauss-Bonnet curvature comes out much closer to 1 than that following from $\Rc$.
The constraint effective-action contains a renormalized gravitational constant $1/G$ which passes through zero at the transition.
Conclusions are in section \ref{secdisc}.

The continuum interpretation is guided by examples from regular lattices. While conceptually relevant, this material is delegated to appendix \ref{appmodels} since it is rather `lattice technical'. In practical terms it leads to a simple conversion factor linking lattice and continuum geodesic distances. Appendix \ref{appAB} contains alternative ways of fitting the volume-distance correlator with results that differ quantitatively (but not qualitatively) from those in the main text and in appendix \ref{appquartic} we give some details on the quartic-potential model used in the effective action.

\section{Symmetric dynamical triangulation}
\label{secintroSDT}

In the (imaginary-)time-space symmetric version of dynamical triangulation, spacetime consists of flat equilateral four-simplices `glued' together to form a simplicial manifold. On a manifold without boundary, the Einstein-Hilbert action without matter,
\be
S = \intx \sqrt{g}\left(\frac{-R + 2\Lm_0}{16 \pi G_0}\right)
\ee
becomes translated into the form \cite{Agishtein:1991cv,Ambjorn:1991pq}
\bea
S&=& -\kp_2 N_2 + \kp_4 N_4,\\
\kp_2 &=& \frac{2\pi v_2}{8\pi G_0},\quad \kp_4 = \frac{\Lm_0 v_4 + 10 \,\theta v_2}{8\pi G_0}.
\label{Gbare}
\eea
Here $G_0$ and $\Lm_0$ are the bare Newton and cosmological constants,
$N_2$ is the number of triangles, $N_4$ the number of $4$-simplices, $\theta=\arccos(1/4)$ and $v_i$ is the volume of an $i$-simplex,
\be
v_i=\frac{\ell^i\sqrt{i+1}}{i!\sqrt{2^i}},
\qquad
\ell=\sqrt{10}\, \ellt;
\label{V4}
\ee
$\ell$ is the edge length of the simplices and $\ellt$ is that of the dual lattice.
The formal {\em grand-canonical} partition function in the continuum is represented as a sum over all\footnote{Every simplicial configuration is meant to be counted only once. Usually a weight factor $1/C({\cal T})$ -- with $C({\cal T})$ the order of the automorphism group of a configuration -- is inserted to avoid over-counting in discrete coordinates.} triangulations $\mathcal{T}$ obeying manifold conditions, at fixed topology, usually taken to be that of the four-sphere $S^4$ \cite{Agishtein:1991cv,Ambjorn:1991pq},
\bea
Z&=&\int \frac{Dg}{\mathcal{V}(\mbox{diff})}\, e^{-S} \\
&\to&\nonumber\\
Z(\kp_2,\kp_4) &=& \sum_{\cal T} e^{\kp_2 N_2 -\kp_4 N_4}
= \sum_{N_4} e^{-\kp_4 N_4} Z(\kp_2,N_4),
\label{partition1}\\
Z(\kp_2,N_4)&=& \sum_{{\cal T}(N_4)} e^{\kp_2 N_2}.
\label{partition2}
\eea
For large $N_4$, the {\em canonical} partition function (\ref{partition2}) behaves  exponentially \cite{Bruegmann:1995qz,Ambjorn:1996ny}, e.g.\ in the elongated phase
\be
Z(\kp_2,N_4) \sim (N_4)^{\gm(\kp_2)-3}\, e^{\kp_4^{\rm c}(\kp_2) N_4}, \qquad N_4 \to\infty.
\label{defgamma}
\ee
Hence, the grand-canonical partition function (\ref{partition1}) converges at large $N_4$ for $\kp_4 > \kp_4^{\rm c}(\kp_2)$.
For algorithmic reasons, simulations of the canonical partition function (\ref{partition2}) were done with a term $\propto (N_4-\bar N_4)^2$ added to the action to allow for fluctuations around the desired number of simplices $\bar N_4$. It is then possible to select configurations at $N_4=\bar N_4$ for the computation of averages. The phase transition occurs at a pseudo-critical point $\kpc(N_4)$, the crumpled phase is in $\kp_2< \kpc$ and the elongated phase in $\kp_2>\kpc$.

The {\em susceptibility exponent} $\gm$ that characterizes the leading correction to the exponential behavior in (\ref{defgamma})
is found to have the branched-polymer value $1/2$ deep in the elongated phase
($\kp_2 \gg \kpc$) \cite{Ambjorn:1995dj,Ambjorn:1996ny,Gabrielli:1997zy,Gionti:1998jy},
and to decrease to values near zero as $\kp_2 \searrow \kpc$ \cite{Ambjorn:1995dj}.
In the crumpled phase the nature of the subleading behavior less clear \cite{Ambjorn:1995dj,Ambjorn:1998ec}. With additional matter $\gm$ can be negative in the crinkled/smooth phase \cite{Bilke:1997sc,Bilke:1998vj,Ambjorn:1999ix,Horata:2000eg}, and its values at the crinkled/smooth-crumpled transition have been found to correspond to a coefficient in the conformal anomaly of continuum QEG \cite{Horata:2003hm,Antoniadis:1992xu}.

The average of observables $O$ is defined in the usual way
\be
\langle O\rangle = \frac{1}{Z(\kp_2,\kp_4)}\sum_{\cal T} e^{\kp_2 N_2 -\kp_4 N_4}\,
O({\cal T}),
\ee
for the grand canonical average, and
\be
\langle O\rangle = \frac{1}{Z(\kp_2,N_4)}\sum_{{\cal T}(N_4)} e^{\kp_2 N_2}\,
O({\cal T}),
\ee
for the canonical average.

\section{Volume, distance and curvature}
\label{secobservables}

In this section we recall some observables for volume, distance and curvature that are invariant under discrete transformations of lattice coordinates.
The basic definition of four-volume of a set of $n_4$ of four-simplices is $V= n_4 v_4$, with $v_4$ the volume (\ref{V4}) of a single simplex. In the canonical averages, the total volume is fixed to $N_4 v_4$.

A natural notion of geodesic distance is the length of a geodesic path through the interior of the simplices in the piecewise flat manifold.
However, the computation of a such geodesic distance is complicated.
In practice a convenient definition of geodesic distance between two four-simplices $x$ and $y$ has been used: $\dlat(x,y)=$ the minimal number of steps, going from $x$ to $y$ from one simplex to the next, times $\ellt$. It is the minimal length of a path on the links of the dual lattice.
A problem with $\dlat$ is that lattice artifacts do not diminish at distances $\gg \ellt$.
This is illustrated in appendix \ref{appmodels} by simple lattice models in flat spacetime.

The Regge curvature is concentrated on triangles. Its integral around a single triangle and its average over the volume are given by
\bea
\int_{\triangle}d^4 x\,\sqrt{g}\,R
&=& 2 v_2[2\pi -n_4(\triangle)\theta],
\label{RRegge}
\\
\frac{\intx\sqrt{g}\, R}{\intx\sqrt{g}}
&=&
\frac{4\pi v_2}{v_4}\left(\frac{N_2}{N_4} - \frac{10\theta}{2\pi}\right)
\equiv \bar \RR,
\label{RRegge2}
\eea
where $n_4(\triangle)$ is the number of simplices containing the triangle $\triangle$. In (\ref{RRegge2}) it is used that $\sum_\triangle n_4(\triangle) = 10 N_4$, every four-simplex contains 10 triangles.
At the phase transition $\langle \bar \RR\rangle$ does not vanish, but
it `jumps' with a derivative $\langle\partial \RR\rangle/\partial \kp_2$ that has a sharp peak.
The form (\ref{RRegge2}) suggests that $R$ mixes with the cosmological constant (which' contribution to volume-averaged action is just a constant like the $\theta$ term) under renormalization and that it needs an additive renormalization; it may also need a multiplicative renormalization. We return to these ideas in section \ref{secPT} .

In \cite{deBakker:1994zf} an observable for curvature was proposed, based on the behavior of the volume $V(r)$ of a ball of geodesic radius $r$ from an arbitrary origin, in the continuum,
\bea
V(r) &=& \frac{\pi^2}{2}\, r^4 \left[1-\frac{R r^2}{36} + \mathcal{O}(r^4)\right],
\\
V'(r) &=& 2\pi^2 r^3 \left[1-\frac{R r^2}{24} + \mathcal{O}(r^4)\right].
\label{Vpr1}
\eea
This was mimicked in SDT by assuming $V(r)$ to be proportional to $N(r)$, the average number of simplices within lattice geodesic-distance $r$. Its discrete derivative, using units\footnote{These units $\ellt=1$ will be used throughout this work, unless otherwise indicated; $r$ in (\ref{Vpr1}) and (\ref{defn}) have a different meaning and from section \ref{secLocMet} onwards the first (continuum) one will be denotes by $\rcon$.} in which the distance between the centers of two neighboring simplices is 1,
\be
n(r) \equiv N(r)-N(r-1) = \left\langle \sum_y \dl_{r,\dlat(x,y)}\right\rangle,
\label{defn}
\ee
is then `the average volume at distance $r$'. The average is independent of $x$, which can be made explicit, e.g.\ in the probability to find the geodesic distance $r$,
\be
p(r) = \frac{1}{N_4}\, n(r) = \left\langle\frac{1}{N_4^2}\sum_{xy} \dl_{r,\dlat(x,y)}\right\rangle,
\qquad \sum_r p(r) = 1.
\label{pr}
\ee
Introducing an effective volume $v_{\rm eff}$, curvature `observables' were based on the tentative correspondence
\be
v_{\rm eff} n(r) \leftrightarrow V'(r).
\label{Vpr2}
\ee
For this to work out well the quantum fluctuations should not be too large.
The distribution (\ref{pr}) has long-distance tails caused by fluctuations like `baby universes'. Examples of $n(r)$ are shown in figure \ref{fignp260290}.

\FIGURE{\includegraphics[width=8cm]{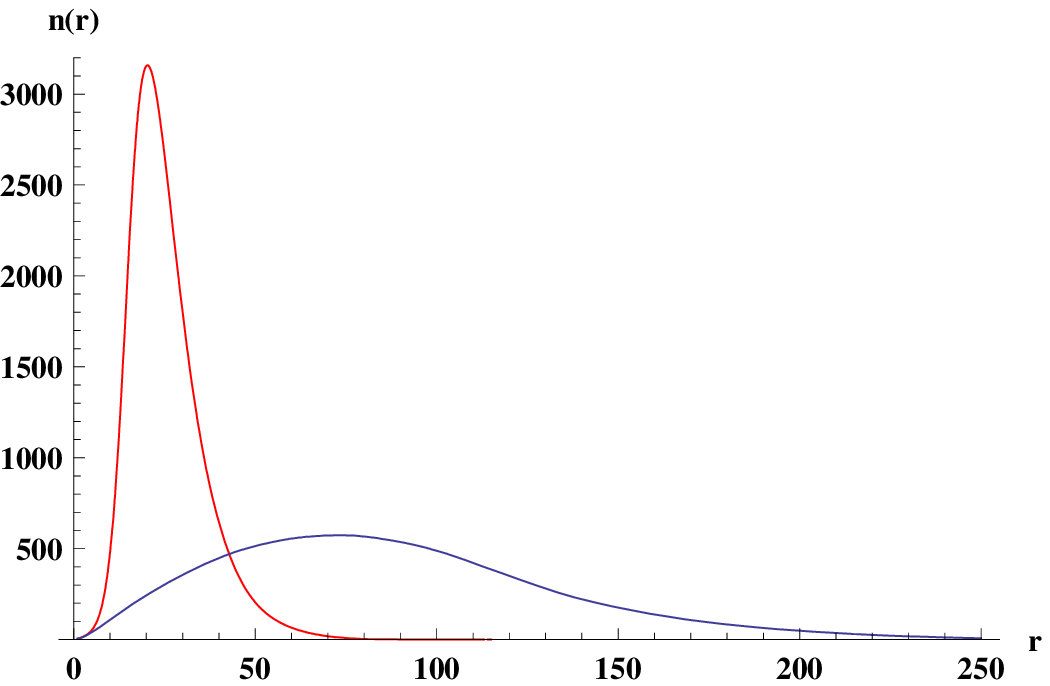}
\caption{$n(r)$ for $N_4 = 64000$; red: crumpled phase, $\kp_2=1.26$; blue: elongated phase, $\kp_2=1.29$.}
\label{fignp260290}
}

A fit $n(r)\approx \al r^3 + \bt r^5$ representing the first two terms in the short-distance expansion (\ref{Vpr1}) gave a curvature $R_V \equiv -24 \bt/\al$, which was
found positive in the elongated phase and negative in the crumpled phase, passing as a function of $\kp_2$ through zero near $\kpc$  \cite{deBakker:1994zf}.
The left plot in figure (\ref{figRV}) shows the result of such fits for $N_4=32000$ and 64000.

The curvature $R_V$ depends sensitively on the $r$ interval used in the fit.
On the one hand it is desirable to use only small values of $r$ in the fit to suppress the higher order terms in (\ref{Vpr1},\ref{Vpr2}), but on the other hand one feels uneasy about the small $r$ region because of lattice artifacts.
In \cite{deBakker:1994zf} this sensitivity of $R_V$ was exploited by restricting the fitting interval to just two subsequent values of $r$, writing $n(r) = \al(r) r^3 + \bt(r) r^5$, $n(r+1)=\al(r) (r+1)^3 + \bt(r)(r+1))^5$, and
$R_{\rm eff}(r+1/2) \equiv -24\bt(r)/\al(r)$.
Plots of this `running' curvature $\Reff(r)$ showed a minimum or approximate $r$-independence in the region $6\lesssim r\lesssim 10$.
The right plot of figure \ref{figRV} shows the values of $\Reff$ at these minima, $R_{\rm eff,\,min}$, for the current larger volumes. This data is somewhat shifted compared to that of $R_V$ but the general trend is similar.
The pseudo-critical transition points as defined by the position of the peak in the node susceptibility are at $\kpc = 1.257(1)$ and $1.280(1)$, respectively for $N_4=32$ k and 64 k \cite{deBakker:1996zx}.

\FIGURE{\includegraphics[width=5cm,angle=-90]{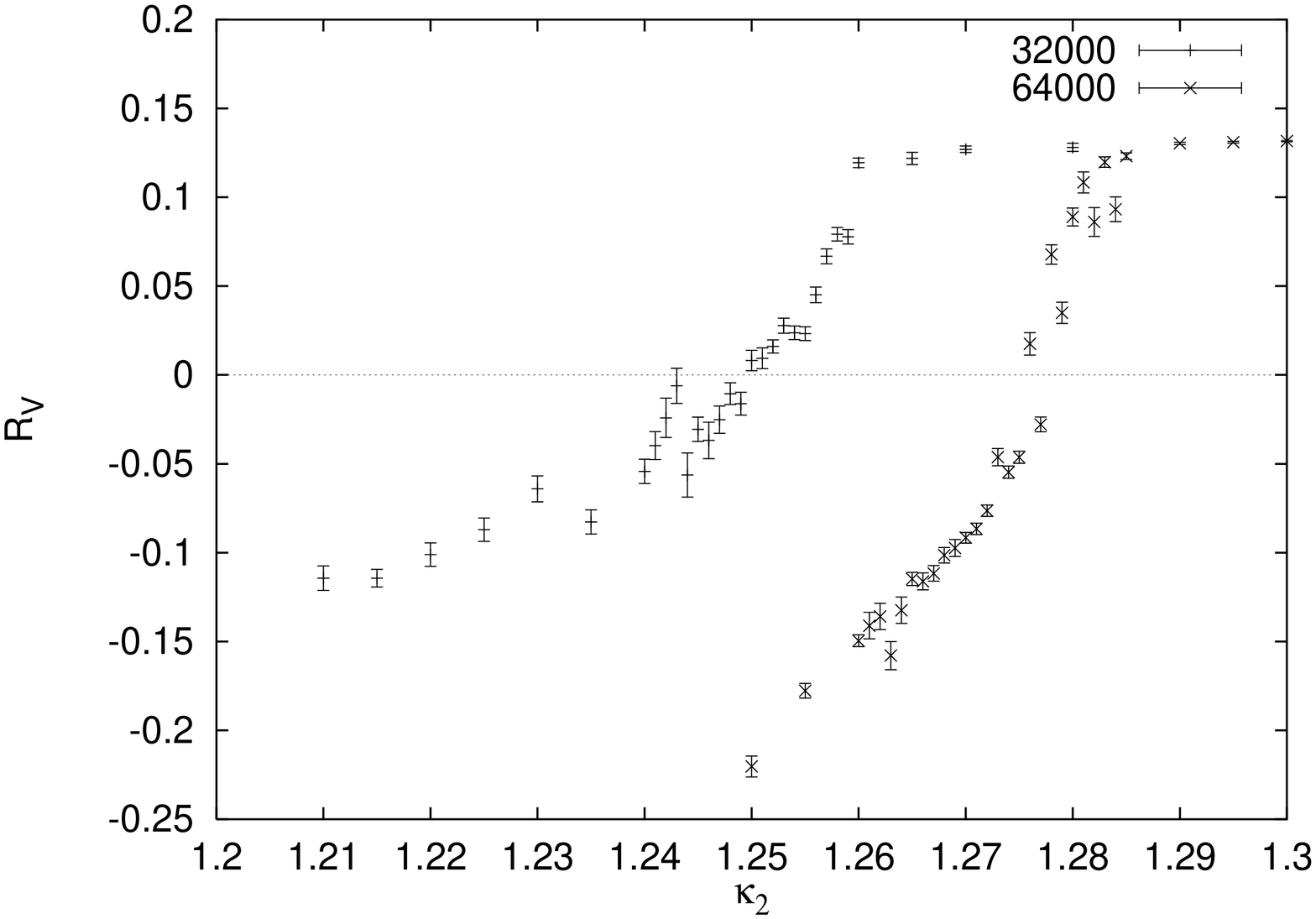}
\includegraphics[width=5cm,angle=-90]{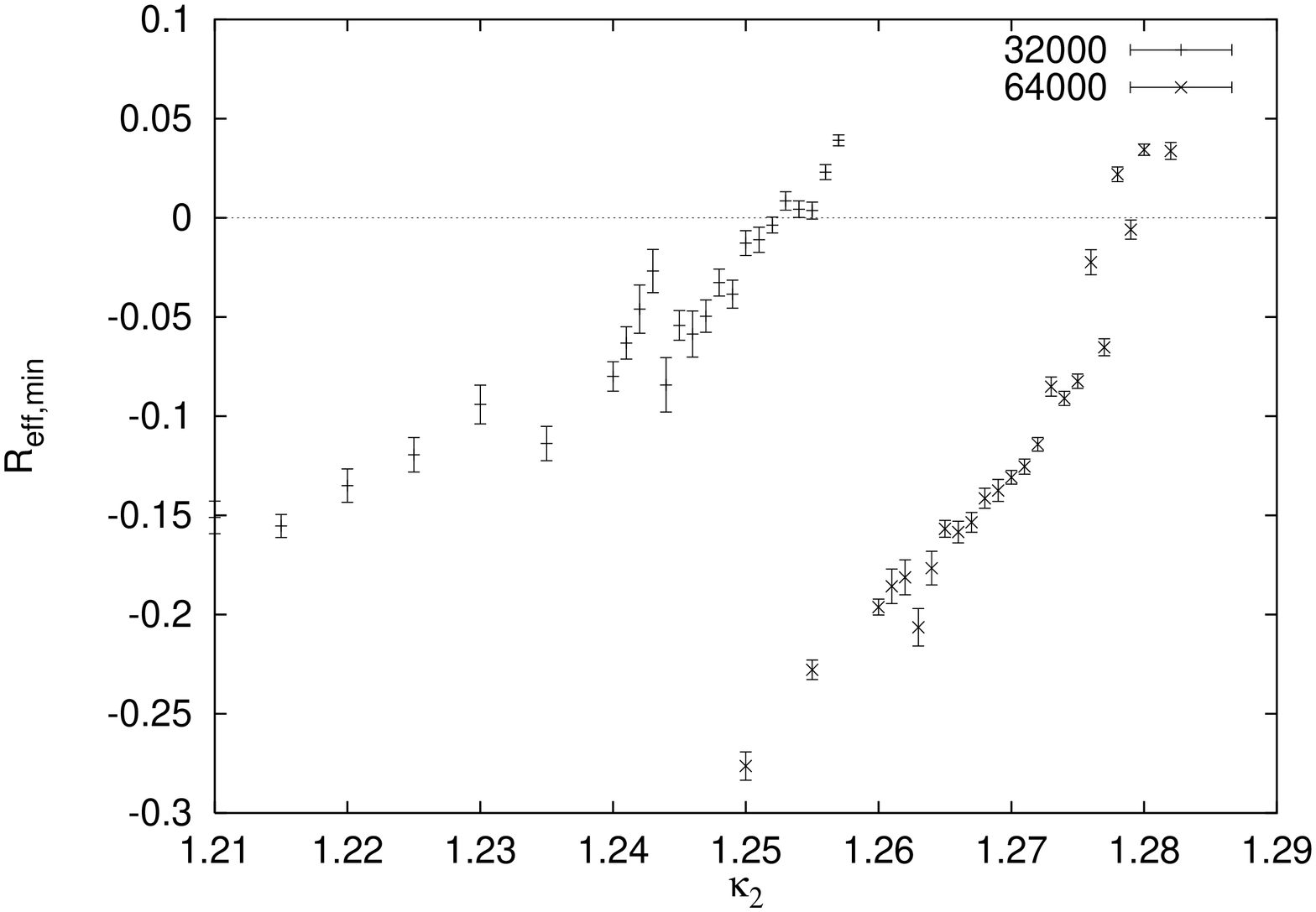}
\caption{Left: curvature $R_V$ as a function of $\kp_2$ from a least-squares fit to the $n(r)$ data in the region $1\leq r\leq 11$, for $N_4 = 32$ k and 64 k.
Right: values of $R_{\rm eff,\, min}$.
Errorbars are obtained with the jackknife method.}
\label{figRV}
}

In the next section the correspondence (\ref{Vpr2}) is explored further, guided by a study of regular lattices in flat space that allow for analytic calculations in appendix \ref{appmodels}.

\section{Geometry from the volume-distance relation}

\subsection{Smooth spacetimes}
\label{secsmooth}

To analyze the average SDT spacetimes from a continuum point of view we compare the quantum expectation value of the volume at distance $r$
with the form it takes in smooth classical spacetimes. Because of the averaging in
$n(r) = \left\langle \sum_{y} \dl_{r,\dlat(x,y)}\right\rangle$
we assume these spacetimes to be homogeneous and isotropic
with constant scalar curvature $R$.
Given the constant curvature, their Euler index $\ch_{\rm E}$ is related to the volume $V$ by the Gauss-Bonnet relation
\be
\ch_{\rm E}
= \frac{R^2 V}{192\pi^2}.
\label{chiGB}
\ee
This can be checked by integrating the Gauss-Bonnet invariant,
\bea
E &=& R^2 -4 R_{\mu\nu}R^{\mu\nu} + R_{\kp\lm\mu\nu}R^{\kp\lm\mu\nu},
\label{GB}
\\
\ch_{\rm E} &=&\frac{1}{32\pi^2}\intx \sqrt{g}\; E,
\eea
over spacetimes in which locally
\be
R_{\kp\lm\mu\nu} = \frac{R}{12}\, \left(g_{\kp\mu} g_{\lm\nu} - g_{\kp\nu} g_{\lm\mu}\right),
\label{Riemcc}
\ee
from which follows that
\be
E = \frac{R^2}{6}.
\ee
Since the SDT simulations used here have $S^4$ topology we are especially interested in the case $\ch_{\rm E} = 2$.
Locally, the metric line element can be expressed in terms of the geodesic distance $r$ from an arbitrary origin,
\bea
ds^2 &=& dr^2 + a(r)^2 d\Om_3^2,
\label{aRW}
\\
R(r) &=& 6\left[-\frac{a^{\prime\prime}(r)}{a(r)} - \frac{a^\prime(r)^2}{a(r)^2} + \frac{1}{a(r)^2}\right],
\label{RRW}
\eea
with $d\Om_3^2$ the line element on the unit three-sphere. Here $a(r)$ plays the role of a (dimension-{\em full}) Robertson-Walker-like scale factor in which $r$ is the imaginary propertime.
With $R(r)=R=$ constant (\ref{RRW}) becomes a differential equation for $a(r)$, with the boundary conditions $a(0)=0$ and $a'(0)=1$.
For positive $R$ the solution is
\be
a(r) = r_0 \sin\frac{r}{r_0},
\qquad
R(r)= \frac{12}{r_0^2},
\label{asphere}
\ee
where $r_0$ is the curvature radius.
The three-volume at distance $r$ is
\be
V'(r) = a(r)^3 \int d\Om_3 = 2\pi^2 a(r)^3.
\label{Va}
\ee
The domain of $r$ can be extended to $0<r<\pi r_0$, resulting in
the $S^4$ geometry with $\ch_{\rm E}=2$, the imaginary-time version of De Sitter space.
Its four-volume is
\be
 V=\int_0^{\pi r_0} dr\ V'(r) = \frac{8}{3}\, \pi^2\, r_0^4.
 \label{Vposcurv}
\ee

\FIGURE{
\includegraphics[width=7cm]{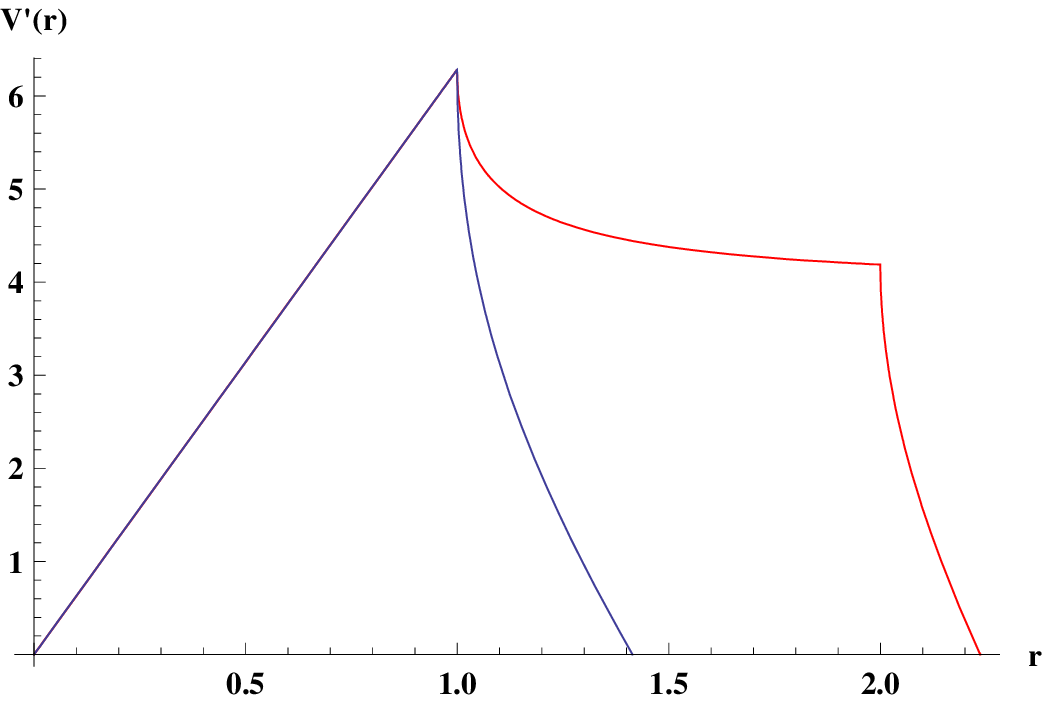}
\caption{$V'(r)$ for flat 2D tori: symmetric $2\times 2$ (blue) and $2\times 4$ (red).}
\label{figvptorus}
}

\noindent For negative $R$ we have
\be
a(r) = r_0 \sinh\frac{r}{r_0},\qquad R=-\frac{12}{r_0^2}.
\label{ahyper1}
\ee
In this case the domain of $r$ can be extended all the way to $0< r< \infty$, resulting in the hyperbolic space $H^4$ which has infinite volume. Finite volume is also possible, by division by a discrete transitive subgroup of the symmetry group $O(4,1)$ of $H^4$ \cite{Ratcliffe2006}.
This leads to an infinite number of nonisometric spaces, which are obtained by gluing the sides of 4D polytopes. These may have {\em cusps}, regions of finite volume extending to infinite distance.
There are 1171 nonisometric spaces with Euler index 1, 22 of them orientable, which have five cusps \cite{RatcliffeVolumeSpectrum}. From double coverings, orientable `gravitational instantons' can be obtained with Euler index 2 \cite{Ratcliffe:1998rb}. The cusps can be `Dehn-filled', resulting in many different topologies \cite{Anderson:2003js}.
Such finite-volume spacetimes are relevant to a semiclassical evaluation of the formal path integral in the continuum \cite{Carlip:2001wq}.

Globally, the volume $V'(r)$ can be a complicated function. Figure \ref{figvptorus} shows examples for flat tori in two dimensions, $R=0$, $\ch_{\rm E}=0$.
In finite volume (\ref{ahyper1}) can only be valid in a finite interval, say $0<r<r_+$, and the maximal value $r_+$ will depend on the particular spacetime. For $\ch_{\rm E}=2$ it has to be smaller than the value for which the volume of the coordinate patch,
\be
2\pi^2\int_0^{r_+} a(r)^3= \frac{8\pi^2}{3}\, r_0^4\, \left(2+\cosh\frac{r_+}{r_0}\right)\sinh^4\frac{r_+}{2r_0},
\label{vpatch}
\ee
equals the total volume $(8\pi^2/3)r_0^4$ set by the Gauss-Bonnet formula (\ref{chiGB}):
\be
r_+|_{\ch_{\rm E}=2} < 2r_0\, \mbox{arccosh}\sqrt{3/2} \simeq 1.317\, r_0.
\label{rpatchmax}
\ee
If $r_+$ would be equal to this maximal value,
then the boundary region at $r_+$ -- being no longer homogeneous and isotropic -- would have to be singular, perhaps crumpled to a point.

\subsection{Local average geometry in SDT}
\label{secLocMet}

The simple relation (\ref{Va}) between the volume at distance $r$ and the scale factor in the metric of smooth homogeneous spacetimes, and the earlier made correspondence (\ref{Vpr2}) between $n(r)$ of SDT and $V'(r)$, invites one using $n(r)$ to derive an average metric from the SDT simulation results,
\be
a(r) \approx [n(r) \veff/2\pi^2]^{1/3}.
\label{aNappr}
\ee
A scalar curvature $R(r)$ can then be computed by the standard formula (\ref{RRW}).
For smooth spacetimes the relation between $a(r)$ and $V'(r)$ can be valid only in a local patch (as figure \ref{figvptorus} illustrates) and the region to the left of the maximum of the $n(r)$ curves is the place to attempt the deduction of a metric scale factor from $n(r)$.

In making (\ref{aNappr}) precise it is helpful to turn to simple lattice models which can be studied analytically. We do this in appendix \ref{appmodels} for triangular and hypercubic lattices in flat spacetime. Applying the insight obtained there to SDT, we write
\bea
\rcon &=& \lm (r-s),  \qquad a_{\rm c} (\rcon) = \lm \, a(r-s),
\label{rconr}
\\
a(r-s) &=& \left[\frac{\veff}{2\pi^2}\, n(r) \right]^{1/3}.
\label{adefSDT}
\eea
The subscript c denotes a continuum distance and scale factor, and $\lm$ is a is a conversion factor that compensates for effects of using the lattice geodesic distance $r$. The shift $s$ is of order of $\ellt$, the lattice spacing on the dual lattice (we continue to set $\ellt=1$).
Discretization effects at small $r$ are expected to be reduced somewhat by interpolating $N(r)$ and replacing $n(r)$ by $N'(r-1/2)$ (cf.\ (\ref{Npvsn}), (\ref{ncubic}) and figure \ref{fcubicnpthrd}).
We tried this with piece-wise polynomial interpolations of $N(r)$ at $r=1$, 2, \dots. Denoting these for simplicity also by $N(r)$, its derivative is $N'(r)$. Using $N'(r-1/2)$ turned out to make no significant difference in the subsequent analysis, even when using high orders of interpolation (for linear interpolation $n(r)=N'(r-1/2)$).
The breaking of rotation invariance in SDT is probably less than, for example, a hypercubic lattice, similar to the notion that a hexagon is a better approximation to a circle than a square. The average over the random configurations will reduce the breaking further.

For the models in appendix \ref{appmodels}, $\veff$ is defined by the continuum limit. Such a limit is not possible of the SDT results in this paper and $\veff$ was determined by fitting the constant-curvature forms (\ref{asphere}) and (\ref{ahyper1}) to the $n(r)$ data at $r=\rfitmin$, \ldots, $\rfitmax$, as in (\ref{veffdet1}).
Then (\ref{veffgm})) gives a value of $\lm$,
\be
\lm = (v_4/\ellt\veff)^{1/4},
\label{deflm}
\ee
where $v_4$ is defined in (\ref{V4}).
The effective volume $\ellt \veff$ is equal to the volume of a four-simplex scaled by the conversion factor $\lm^{-4}$.
In the following we shall work mostly with the lattice $r$ and $a(r)$ and not with the continuum $\rcon$ and
$a_{\rm c}(\rcon)$, keeping in mind that continuum lengths involve the conversion factor $\lm$.

\FIGURE{\includegraphics[width=7cm]{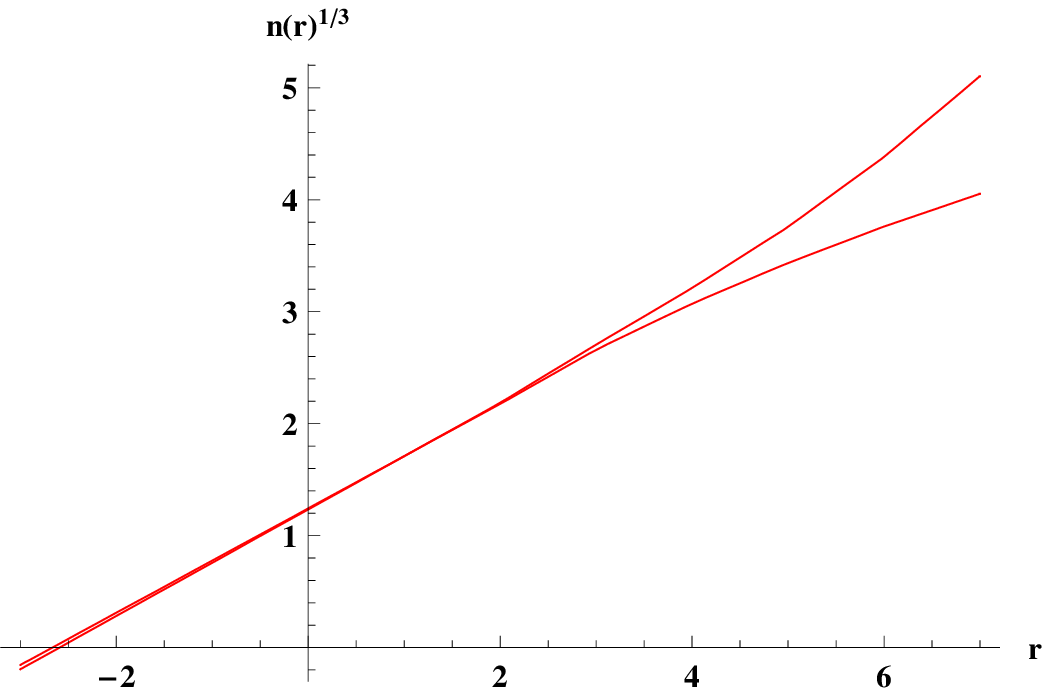}
\caption{Plot of $n(r)^{1/3}$, linearly interpolated and extrapolated into the region $r<1$, for $\kp_2 =1.26$ in the crumpled phase (upper) and $1.29$ (lower) in the elongated phase, $N_4=64$ k. The zeros are at $s_0(1.26)=-2.59$, $s_0(1.29)=-2.67$.}
\label{figpnthrd260290}
}

We have done fits without shift, $s=0$, called A-fits, and with $s\neq 0$, called B-fits. The idea behind the A-fit is to avoid lattice artifacts as much as possible and use only a minimum fitting domain $\rfitmin \gg 1$. Instead, with the B-type, the shift $s$ is chosen such that the data can be approximated reasonably well with least-squares fits at distances all the way down to $r=1$, choosing $\rfitmin=1$. To see that this makes sense, consider figure \ref{figpnthrd260290}, in which linearly interpolated $n(r)^{1/3}$ data is extrapolated to the region where it has a zero point.
The zeros points $s_0$ are nearly independent of $\kp_2$ and the plot indicates that using a negative shift $s=s_0$ can improve the fits in the small distance region. Note that already from $r=3$ onwards the crumpled- and elongated-phase curves deviate. Using $s$ as an additional fitting parameter gives unstable results, the three parameters turn out to be too correlated.
Therefore we fixed $s=s_0$ in B-fits.

The results of a A- and B-fits are described in appendix \ref{appAB}. The choice of fitting domain and type of fit influences the resulting curvatures quantitatively rather strongly, but not their qualitative dependence on $\kp_2$. Curvature is a sensitive observable; the form (\ref{RRW}) leads it to depend on {\em second} derivatives of $n(r)$ through (\ref{adefSDT}).

In the following we describe a fitting method that may be called `discrete osculation' of type B (DOB), its fitting domain consists of just two points\footnote{Interpolating $n(r)$ and letting $\rfitmax\to\rfitmin=r$ would result in matching value and first derivative at $r$.}, $\rfitmax=\rfitmin+1$. It is an improvement on $\Reff$ (section \ref{secobservables}) that uses constant-curvature forms for the fitting function with the shift $s=s_0$. Specifically, in the elongated phase $c$ and $r_0$ are determined by the two equations
\be
c^{1/3} \sin[(r-s)/r_0] = n(r)^{1/3},\quad
c^{1/3} \sin[(r+1-s/r_0] = n(r+1)^{1/3},\;\;
r\equiv\rfitmin,
\label{osceq}
\ee
where $c= 2\pi^2/\veff$,
and similarly with $\sin\to\sinh$ in the crumpled phase.
The resulting curvature $\pm 12/r_0^2$ is denoted\footnote{The $r$ dependence here is not to be confused with that in (\ref{RRW}).} by $\Rosc(r+1/2)$ and these values are interpolated. It is gratifying that $\Rosc(r)$ has stationary points: a minimum in the crumpled phase and a maximum in the elongated phase (the latter is absent in $\Reff(r)$). By the `principle of minimum sensitivity' the value of $r$ is chosen to be the stationary point $r_{\rm stat}$ of $\Rosc(r)$: $(d/dr)\Rosc(r)|_{r=r_{\rm stat}}=0$, $\Rosc\equiv\Rosc(\rstat)$,
$r_{0{\rm osc}}\equiv\sqrt{12/|\Rosc|}$.

\FIGURE{\includegraphics[width=7cm]{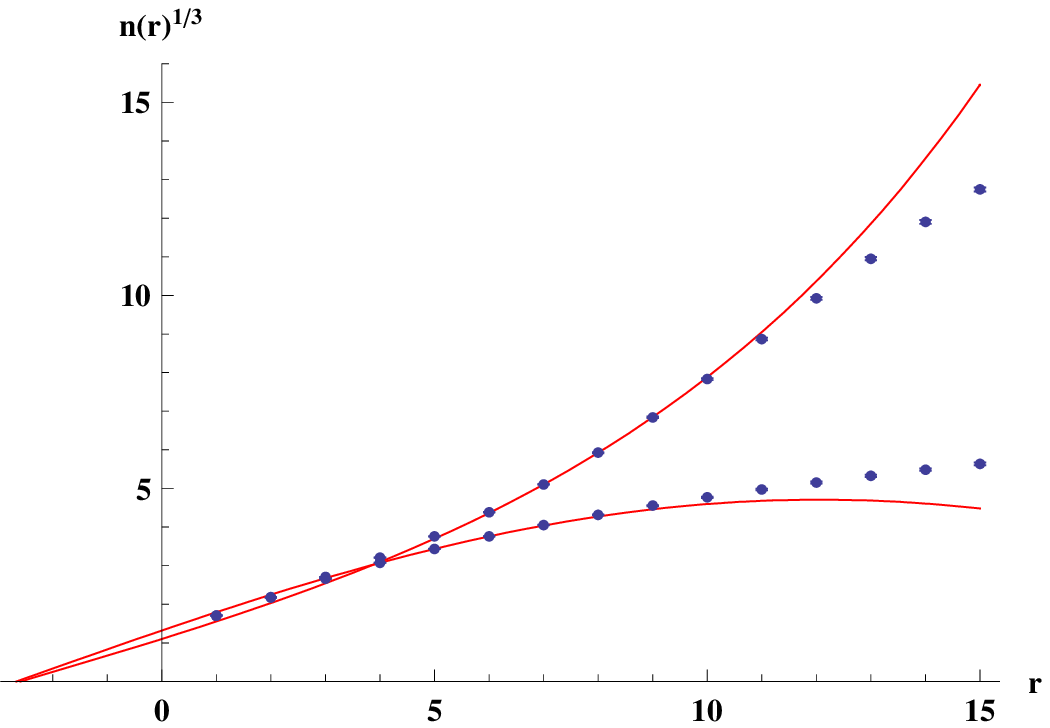}
\includegraphics[width=7cm]{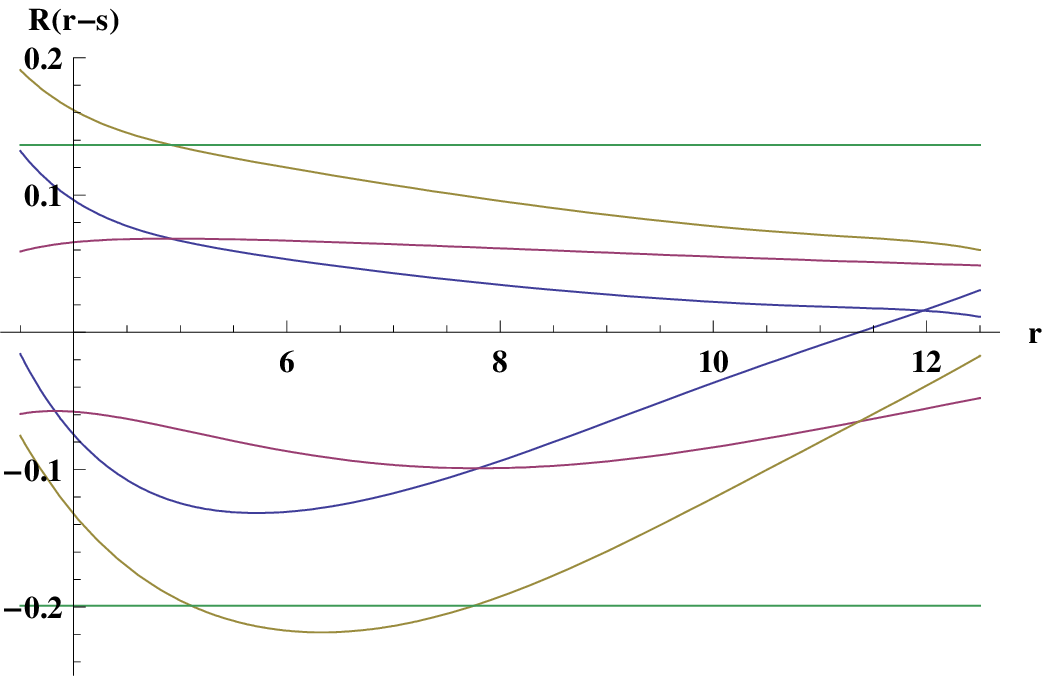}
\caption{Left: results of DOB fits at $\kp_2=1.26$ (upper data in blue and $c\sinh[(r-s)/r_0]$ in red) and at $1.29$ (lower data and $c\sin[(r-s)/r_0]$), for $N_4=64$ k. Right: curvature $R(r-s)$ (green-brown) and its components $R_1=6/a(r-s)^2 - 6 a^{\prime}(r-s)^2/a(r-s)^2]$ (red) and $R_2=- 6 a^{\prime\prime}(r-s)/a(r-s)$ (blue). The green horizontal lines represent $\Rosc$.
}
\label{figpthrd260290osc}
}

Figure \ref{figpthrd260290osc} shows results of such a fit. In the left plot, the upper/lower data show $n(r)^{1/3}$ in the crumpled/elongated phase and the fitting curves. Plotting the third root of $n(r)$ enhances the short distance region. The hardly visible error bars are inherited from jackknife errors on $n(r)$. With $c=2\pi^2/\veff$ determined, the metric $a(r)$ follows from (\ref{adefSDT}). Since $a(r-s)=[n(r)/c]^{1/3}$, it is just a scaled and shifted version of the data in the left plot, interpolated.
The curvature $R(r)$ of (\ref{RRW}) and its components
$6/a^2-6 a^{\prime 2}/a^2\equiv R_1$ and $-6 a^{\prime\prime}/a\equiv R_2$, $R=R_1 + R_2$,
are shown in the right plot. The lower half applies to $\kp_2=1.26$ were curves cross at $\rstat\simeq 7.83$, the upper half is for $\kp_2=1.29$ where curves cross at $\rstat\simeq 4.89$. Evidently $R(r)$ is not constant. Its component $R_1(r)$ has a region of slow variation around $\rstat$, where it touches $\Rosc/2$. The double derivative component $R_2(r)$ varies much more.
For constant curvature, the $6/a^2$ cancels part of $-6a^{\prime 2}/a^2$ in $R_1$ and the sum of the two is equal to $R_2$.

Results at $N_4=64$ k for DOB-fit parameters including other values of $\kp_2$ (chosen from `non-outlying' data in figure \ref{figRV}) are listed below,
\be
\begin{array}{cccccccccc}
\mbox{crumpled}&\mbox{phase}&&&
\mbox{elongated}&\mbox{phase}&&&&\\
\kp_2&s&r_0&c&\kp_2&s&r_0&c\\
1.255&-2.58&7.50&0.0734&1.282&-2.65&11.0&0.121\\
1.260&-2.59&7.76&0.0727&1.283&-2.66&10.0&0.124\\
1.266&-2.60&8.16&0.0723&1.285&-2.66&9.78&0.125\\
1.270&-2.61&8.54&0.0727&1.290&-2.67&9.38&0.127\\
1.277&-2.63&9.88&0.0727&1.300&-2.68&9.21&0.126
\end{array}
\label{tDOBfit}
\ee
The constant $c=2\pi^2/\veff$
differs somewhat between both phases.
It is the ratio of the surface of the unit 4D ball to the effective volume $\veff$ that characterizes the average hyper-surface at fixed geodesic lattice distance $r$.
The lattice-continuum conversion factor (cf.\ (\ref{deflm}))
$\lm=(v_4 c/2\pi^2)^{1/4}$ comes out as $\simeq 0.405$ (0.465) in the crumpled (elongated) phase, and the resulting continuum curvature
$\Rc=\Rosc/\lm^2=\pm 12/(r_{0{\rm osc}}\lm)^2$ is plotted in figure \ref{figpRcosc}. Remarkably, its linear interpolation goes through zero at the pseudo critical point $\kpc=1.280(1)$ as determined by the peak in the node susceptibility in \cite{deBakker:1996zx}.

\FIGURE{\includegraphics[width=12cm]{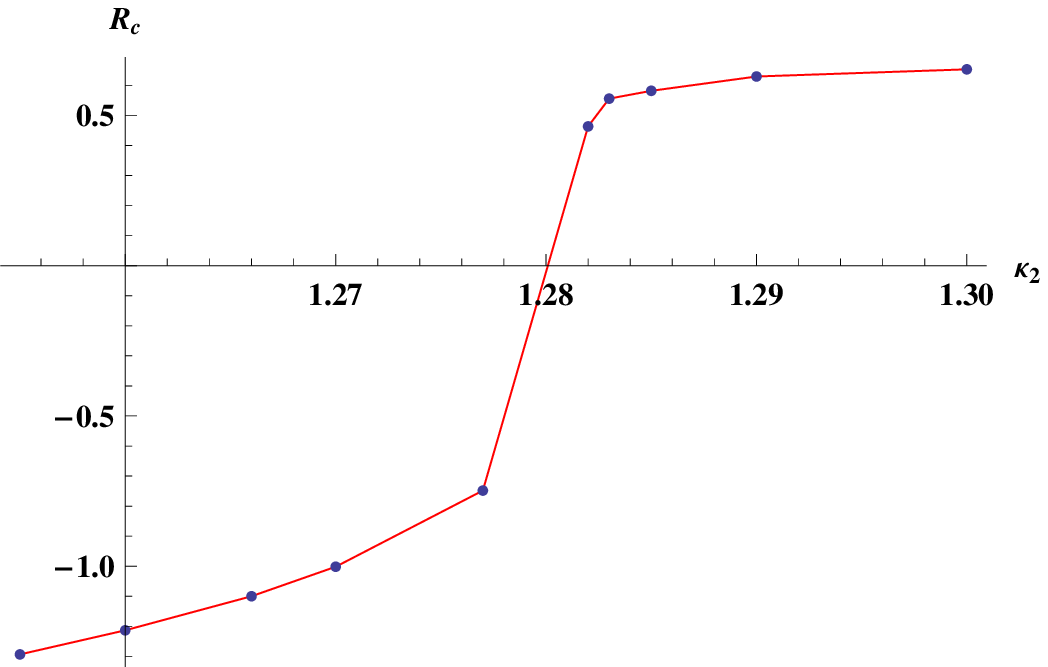}
\caption{Curvature $R_{\rm c}$ from the DOB-fits as a function of $\kp_2$ with linear interpolation.}
\label{figpRcosc}
}

\section{Baby- and mother-universes}
\label{secmotherbaby}

In the previous section we found the average SDT spacetimes to be be locally close to constant-curvature spaces of negative and positive curvature, respectively
in the crumpled and elongated phase. As noted in section \ref{secsmooth}, in case of positive curvature the form of the metric scale factor
$r_{0{\rm c}}\sin(\rcon/r_{0{\rm c}})$ can be naturally extended to $\rcon = \pi r_{0{\rm c}}$ and the resulting spacetime is then globally similar to $S^4$. However, its volume
$V_{\rm c}= (8\pi^2/3)r_{0{\rm c}}^4$ is much smaller than the total volume
$N_4 v_4$. Their ratio is
\be
\numc=\frac{v_4 N_4}{(8\pi^2/3)r_{0{\rm c}}^4}=\frac{\veff N_4}{(8\pi^2/3)r_0^4}
=\frac{3 N_4}{4 c r_0^4},
\label{numcomp}
\ee
where we used (\ref{deflm}). This ratio is about 49 for $(\kp_2,\, N_4) =(1.29,\, 64000)$ in the elongated phase away from the transition. The discrepancy can be explained in the continuum interpretation by assuming that the total spacetime consists of small constant-curvature components `glued' together. The gluing regions contribute negatively to the Euler number such that for the total spacetime $\chi_{\rm E}=2$. If the volume of these `necks' is small enough, their relative contribution to the Einstein-Hilbert action can still be small. The components are average-size baby universes,
$\numc$ in total, and each of them consists on the average of $N_4/\numc$ four-simplices (for the example at $\kp_2=1.29$, $N_4/\numc\approx 1300$).

In the negative-curvature case the scale factor $r_{0{\rm c}}\sinh(\rcon/r_{0{\rm c}})$ can be extended to the infinite domain $0<r<\infty$ corresponding to the hyperbolic space $H^4$. The finite total volume $N_4 v_4$ puts of course a limit on $\rcon$, by (\ref{vpatch}):
\be
\left(2+\cosh\frac{\rplusc}{\rnotc}\right)\sinh^4\frac{\rplusc}{2\rnotc} < \numc,
\ee
where now $\numc$ just stands for the ratio (\ref{numcomp}) and does {\em not} have the interpretation of a number of components. Using the asymptotic form of the hyperbolic functions it follows that $\rcon$ cannot be larger than $\rplusc \approx (r_{0{\rm c}}/3)\ln(32 \numc)$,  For $\kp_2=1.26$ in the crumpled phase, $\numc\simeq 182$ and $\rplusc\simeq 2.9\,\rnotc$. However, according to the derivations in section \ref{secsmooth}, the contribution to the Euler number of a hyperbolic ball with radius $\rplusc$ follows from the Gauss-Bonnet formula (\ref{GB}), which gives
$(-12/\rnotc^2)^2 N_4 v_4 /192\pi^2 = 2\numc$. To get down to $\ch_{\rm E}=2$ of the total space this has to be compensated in some way. Apparently this happens by gluing to baby universes and by crumpling to `singular structures'. Numerical studies of `boundary volume distributions'
\cite{Egawa:1996fu},
and of branching-order distributions in `minbu trees' \cite{Bialas:1996eh}, have led to a picture of SDT spacetimes in the crumpled phase that consists of one large component, the `mother universe', which contains the singular vertices and links, connected to many small baby universes. In the elongated phase there is no mother component. This qualitative difference between the two phases appears to be a natural consequence of the sign of the average curvature: positive inducing small spheres, negative inducing large chunks of hyperbolic space.

The transition between the two phases can be modeled by a `balls in boxes' model \cite{Bialas:1998ci,Bialas:1999ad}. The nature and entropy of SDT spacetimes has been investigated analytically in great detail in \cite{Ambjorn:1996ny,Gabrielli:1997zy,Gionti:1998jy,Ambjorn:1998ec}.

\section{Scaling}
\label{secscaling}

In the elongated phase the components of (\ref{numcomp}) are supposed to be average-size four-spheres with small `caps' taken out and glued along the caps' boundaries. Such a configuration can be mapped to a tree graph, in which vertices correspond to components and links to `gluings'. A typical example is a branched polymer graph with average coordination number of the vertices not very different from 2. Indeed, in the elongated phase the SDT spacetimes have characteristics of a statistical ensemble of branched polymers \cite{Ambjorn:1995dj,AmbjornDJ1997,Bialas:1995xq}.
One such characteristic is the value of the entropy exponent $\gm\approx 1/2$, another is the scaling behavior of $n(r)$ \cite{Ambjorn:1995dj}.

In \cite{deBakker:1994zf}
we investigated scaling of the probability of geodesic distance $r$ between two simplices, $p(r)=n(r)/N_4$. Suppose that a length $r_*$ can be chosen to depend on $\kp_2$ and $N_4$, and $\kp_2$ to depend on $N_4$, such that the following limiting procedure makes sense:
\bea
&&
\frac{r_* n(r)}{N_4}\equiv \rh_*(r/r_*,\kp_2,N_4) \to \tilde\rh_*(x,\ta),
\quad x_r\equiv r/r_* \to x, \quad N_4\to\infty,
\label{defrh}
\\&&
\int_0^\infty dx \tilde\rh_*(x,\tau) = \lim_{N_4\to\infty} \sum_r \Delta x\,\rh_*(x_r,\kp_2,N_4) =
\sum_r p(r) = 1,
\quad
\Delta x = \frac{1}{r_*}.
\label{rhnorm}
\eea
The label $\tau$ distinguishes different shapes of $\tilde\rh$ as a function of $x$ resulting from different ways of limit taking.
The scale $r_*$ was taken to be $\rmm$, the value of $r$ where $n(r)$ is maximal, and different sequences $\kp_2(N_4)$ were envisioned that produce scaling functions
$\rh_{\rm m}(x,\kp_2,N_4)$ of different shapes, e.g.\ $N_{4,j} = 8000\times 2^j$, $j=0,1,\ldots$, $\ta = \kp_2(N_{4,0}$). Another possibility for $r_*$ is the average value $r_* = \rav = \sum_r p(r)\, r$.
Assuming $\langle x\rangle_m$ is finite in the limit $N_4\to\infty$,
the two are proportional for large $N_4$: using $r_*=\rmm$ we have
$\rav/\rmm\to\int dx\, \tilde\rh_{\rm m}(x,\tau)\, x = \langle x\rangle_m$.
For simplicity the labels $\kp_2$, $N_4$ of $\rh$ and $\tau$ of $\tilde\rh$ will be dropped in the following.

\FIGURE{\includegraphics[width=7cm]{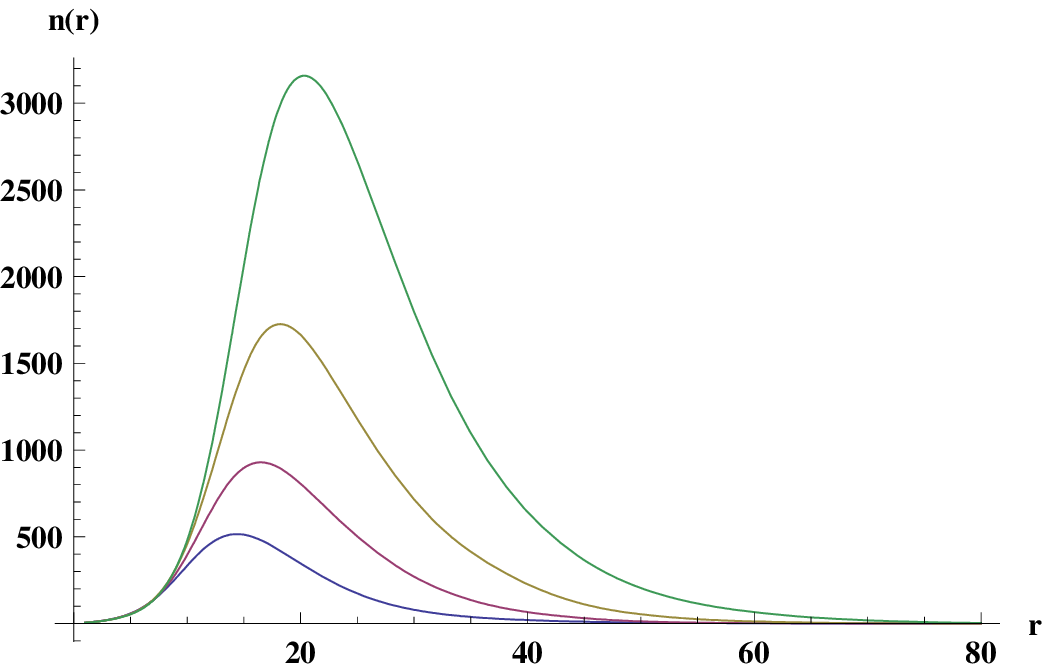}
\includegraphics[width=7cm]{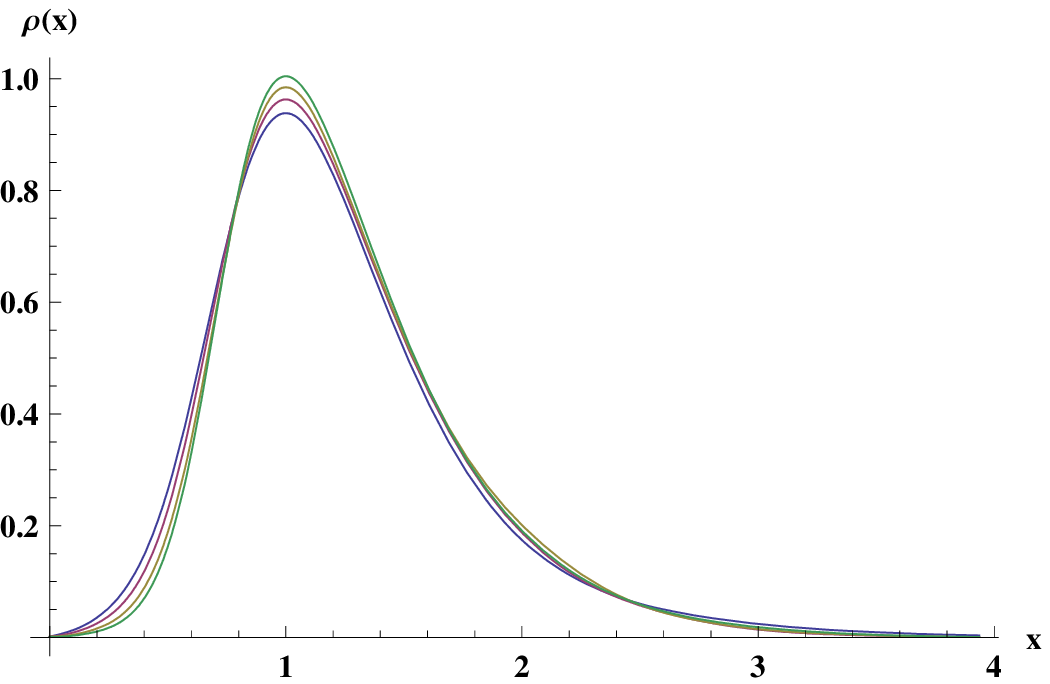}
\caption{Left: Crumpled phase $n(r)$ data at $(N_4,\,\kp_2)=(8000, 1.17)$ (blue), (16000, 1.21) (red), (32000, 1.23) (brown) and (64000, 1.26) (green). Right: same data scaled:
$\rmm n(\rmm x)/N_4 =\rh_{\rm m}(x)$.
}
\label{fignp17213060}
}

Figure \ref{fignp17213060} shows an example in the crumpled phase.
The scaled data match within statistical errors and for clarity interpolated data is shown without error bars.
Data were chosen from an available set to give approximate matching by eye. Also quantitatively they corresponded to smallest differences in the norm
$\int dx\, |\rh(x)-\rh'(x)|$, and these differences converge towards zero (for $\rh_{\rm m}$ they turned out slightly smaller than for $\rh_{\rm av}$).
The DOB-fit parameters of the sequence are
\be
\begin{array}{cccc}
(N_4\,\kp_2)&s&r_0&c\\
(8000,\,1.17)&-2.23&11.7&0.127\\
(16000,\,1.21)&-2.25&9.90&0.112\\
(32000,\,1.23)&-2.25&8.26&0.0831\\
(64000,\,1.26)&-2.59&7.76&0.0727
\end{array}
\label{tDOBfitscal}
\ee

The dependence of $\rav$ or $\rmm$ on $N_4$ can be fitted with a power law, e.g.\
$\rmm \propto N_4^{1/\Ds}$, with a scaling dimension $\Ds\simeq 6.2$, but it can be fitted slightly better  by a logarithm,
$\rmm= a+b\ln N_4$, which corresponds to $\Ds\to\infty$;
see figure \ref{figscalrm}. An infinite scaling dimension was also considered to be most likely in \cite{Catterall:1994pg,Ambjorn:1995dj}, based on a comparison of data at fixed $\kp_2$ deeper in the crumpled phase. The ratio $\rmm/\rav\simeq 0.81$, 0.82, 0.80, 0.80, is practically constant compared to the change in $\rav$ itself, which suggests that $<x>_m$ is finite indeed in the limit $N_4\to\infty$.

\FIGURE{\includegraphics[width=7cm]{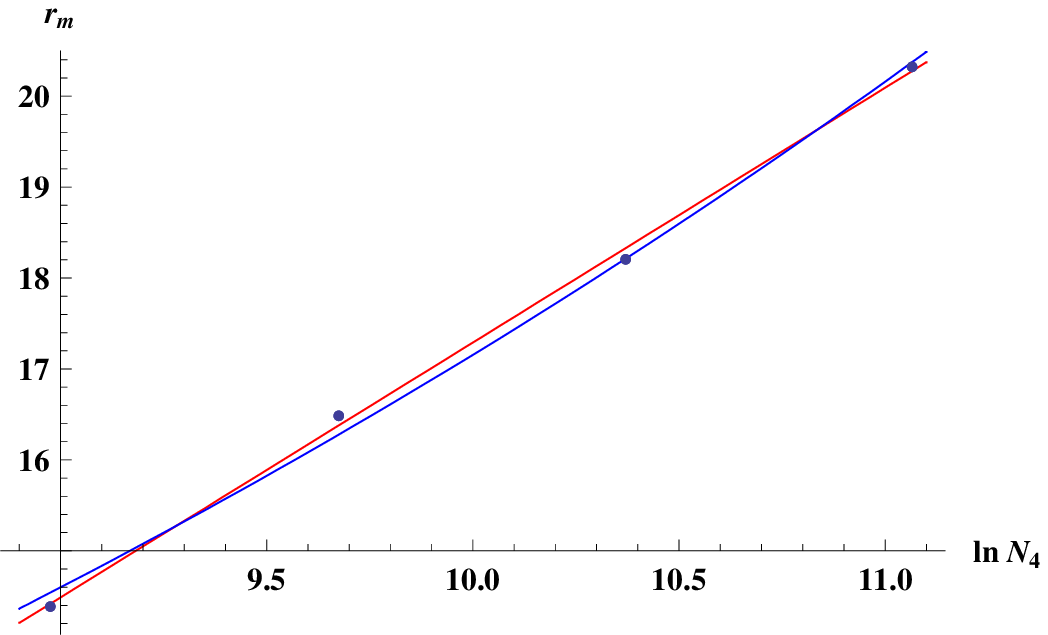}
\includegraphics[width=7cm]{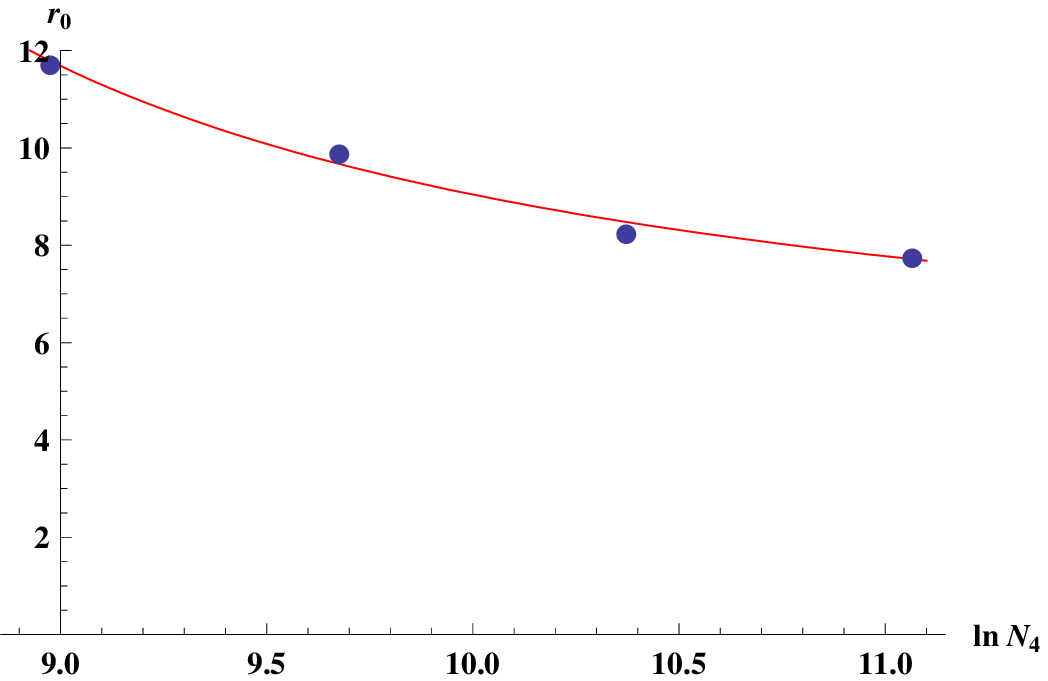}
\caption{Left: $\rmm$ versus $\ln(N_4)$, logarithmic fit (red) $\rmm=-10.74 + 2.803 \ln N_4$, and power fit (blue) $\rmm = \exp[1.228+(1/6.195) \ln N_4]$,
for the scaling sequence in figure \ref{fignp17213060}.
Right: Curvature radii (blue) and fit (red) via
$r_0= \rmm x_0$ with $\rmm$ taken from the logarithmic fit in the left plot and
$x_0 = 1.49/(-7.16 + \ln N_4)$.
}
\label{figscalrm}
}

The evidence for scaling presented in \cite{deBakker:1994zf} was compatible with the idea that sequences $\kp_2(N_4)$ could be found for which the curvature radius $r_0$ grows with $\rmm$, and that $\rh$ would have a semiclassical form for all $x$, with $\Ds\to 4$.
The subsequent discovery of the first order nature of the transition between the crumpled and elongated phase made such an outcome unlikely, and instead of growing one expects the $r_0$ to approach a constant in lattice units. The curvature radii
of the DOB-fits in table \ref{tDOBfitscal} are decreasing, but they can be fitted well by a function that approaches a finite limit\footnote{For the B-fit with $c$ fixed by the result in table \ref{tBfit} the $r_0$ are already almost constant (cf.\ table \ref{tBfitscal}). The A-fit deteriorates for the two smaller values of $N_4$.}
($\approx 4$) as $N_4\to\infty$ (right plot in figure \ref{figscalrm}).

The near constancy of $r_0$
can be used to argue that negative curvature is the reason for the logarithmic dependence of $\rav$ and $\rmm$ on $N_4$.
Since the mother universe has a macroscopic fraction of the total volume
its average local curvature will be close to the total average -- negative with radius $r_0$ -- and it will dominate the shape and properties of the distribution $n(r)$ in regions with substantial probability. One such property is the position $\rmm$ of the maximum of $n(r)$.  Consider the integral $p=\int_{1/\rmm}^1 dx\,\rh_{\rm m}(x) = (1/N_4)\int_1^{\rmm} dr\, n(r)$, which is almost independent of $N_4$ in the scaling sequence: $p\simeq 0.34247$, 0.34248, 0.34249, 0.34240. We write $p$ as a fraction $f$ of the integral obtained by replacing $n(r)$ by its constant-curvature fit {\em in the whole region $r<\rmm$}:
\be
 p =\frac{f}{N_4} \int_1^{\rmm}dr\, c \left[ r_0 \sinh\frac{r-s}{r_0}\right]^3\simeq
 \frac{f c r_0^4}{24 N_4}\, e^{3(\rmm-s)/r_0}\left[1 + \mO(e^{-2(\rmm-s)/r_0})\right].
\ee
Then
\be
\rmm \simeq \frac{r_0}{3}\,\ln N_4 + \frac{r_0}{3}\ln\frac{24 p}{f c r_0^4} + s.
\ee
Assuming that $f$ is depends only modestly on $N_4$, less than linear, the leading dependence is given by the explicit $\ln N_4$.
Its coefficient $r_0/3$ is for the scaling sequence $3.9$, 3.3, 2.8, 2.6, not far\footnote{For the B-fit the numbers are $3.0$, 3.0, 2.9, 2.9.}  from the fitted coefficient 2.8 of figure \ref{figscalrm}.

Unfortunately, similar matching data in the elongated phase is inaccessible to us now. In \cite{Ambjorn:1995dj}, $\rav/N_4^{1/2}$ was found to become independent of $\kp_2$ deep in the elongated phase and scaling was observed with $r_* = N_4^{1/2}$, with a scaling function $\rh(x)$ corresponding to generic branched polymers (no need for a parameter $\tau$).
The scaling function can be characterized by a
so-called Hausdorff dimension $\Df$
that is identified from the small $x$ behavior
\be
\rh(x) \propto x^{\Df-1}, \quad x\to 0.
\label{rhxsmall}
\ee
General arguments based on a scaling assumption \cite{Ambjorn:1995dj} relate the large distance behavior to the scaling dimension of $r_* \propto N^{1/\Ds}$,
\be
\rh(x) \propto x^\al
\exp\left(-c_1 x^\frac{\Ds}{\Ds-1}\right), \quad x \to \infty,
\label{rhBPAJ}
\ee
and it is assumed that $\Df=\Ds$. For branched polymers the natural definition of distance is the number of links between two vertices, and an analog $p_{\rm BP}(r)$ -- the probability of distance $r$ between two vertices -- scales in the generic case with $\Ds=\Df=2$ and $\al=1$ \cite{AmbjornDJ1997,Bialas:1995xq}. With a suitable choice of $r_*$ we may write,
\be
\rh(x) = x \, e^{-x^2/2}, \qquad \mbox{generic branched-polymers}.
\label{rhBP}
\ee
Evidence was presented in \cite{Ambjorn:1995dj} that this form also applies to the $n(r)$ of SDT sufficiently deep in the elongated phase and good fits of (\ref{rhBPAJ}) to $n(r)$ were obtained for all $r\gg 1$, with $\al=1$ and $\Ds=2$. The fits used only one parameter, $c_1$, with $r_* = N_4^{1/2}$, since the normalization of $\rh$ is fixed by
$\int dx\, \rh(x) = 1$. Rewriting this as $n(r) = (N_4/r_1^2)\, r \exp(-r^2/2 r_1^2)$,  the parameter $r_1 = r_*$, which is also the position of the maximum of this analytic form for $n(r)$, and it would be fixed to $r_1=\rmm$ if the fit were perfect.
\FIGURE{\includegraphics[width=7cm]{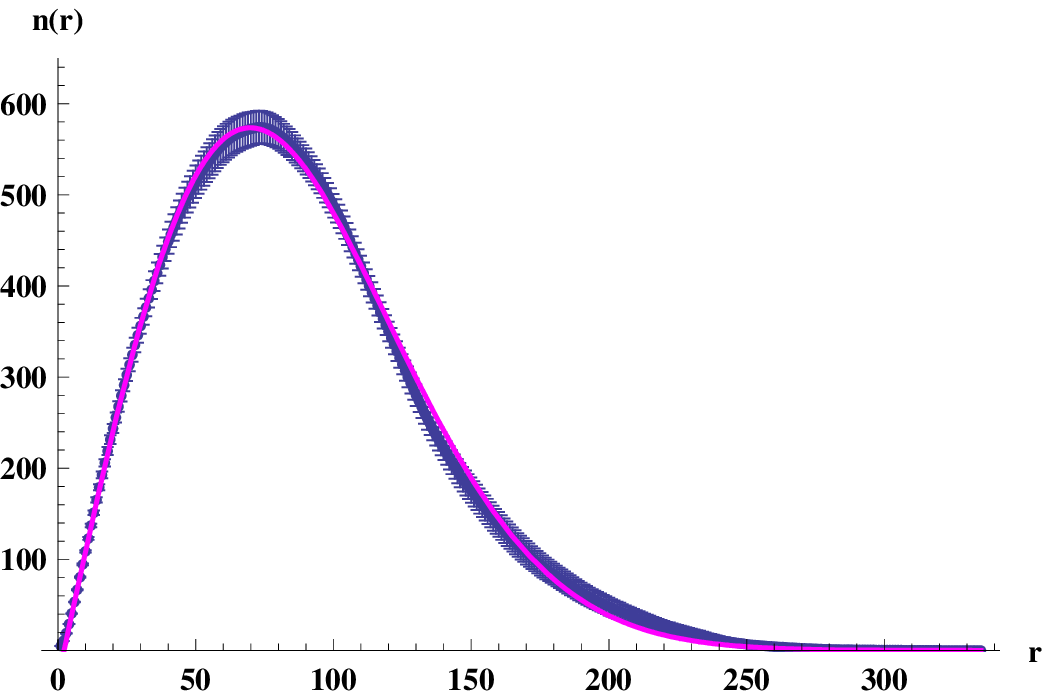}
\includegraphics[width=7cm]{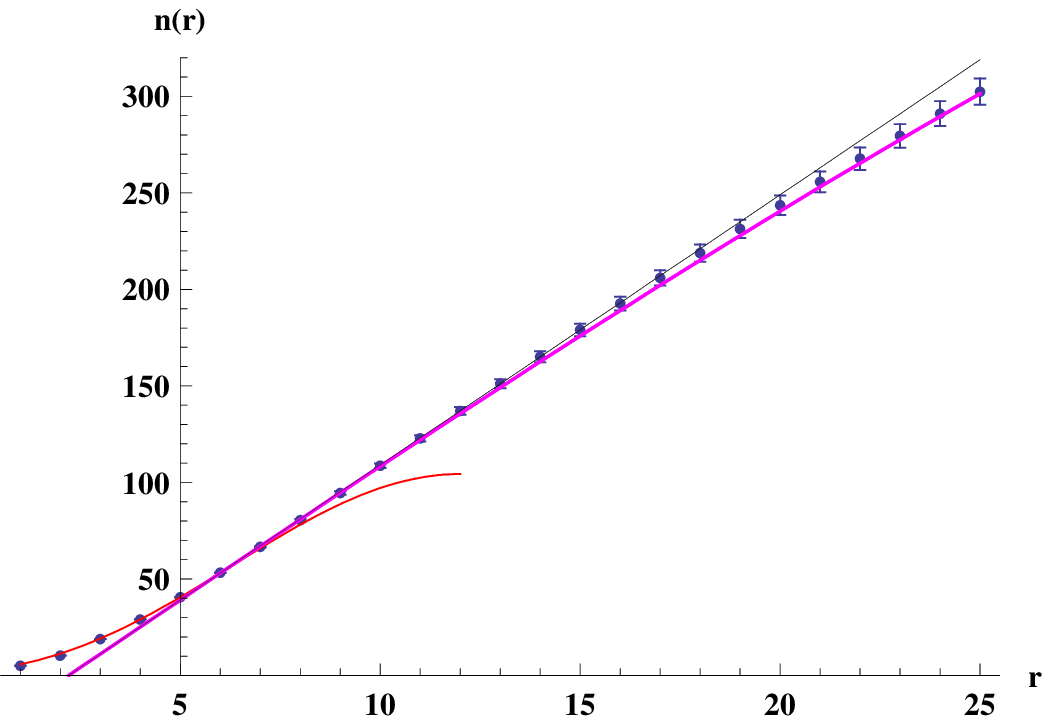}
\caption{
Left: Volume-distance correlator $n(r)$ (blue) in the elongated phase ( $(N_4,\kp_2) = (64000,1.29)$) and fit (\ref{fitBPJS}) (magenta). Right: close-up showing also the asymptote $(N_4 /r_1^2)\,(r-s_1)$
(dark grey) and the $S^4$ DOB-fit $c (r_0 \sin[(r-s_0)/r_0])^3$ (red).}
\label{figpDR}
}
We find that such a `zero parameter fit' is poor to the data at $(N_4,\kp_2) = (64000,1.29)$ (the same values as used in figure 7 in \cite{Ambjorn:1995dj}), and even keeping $r_1$ free does not give very good fits. However, with a small shift in the fitting function the linear behavior (\ref{rhxsmall}) can be made compatible with the data, i.e.\
\be
n(r) = \frac{N_4}{r_1}\, x\, e^{-x^2/2},
\quad x=\frac{r-s_1}{r_1}, \quad r_*=r_1.
\label{fitBPJS}
\ee
In this way we obtained good fits in the region $r\geq 7$ to the data at $\kp_2 = 1.29$ and 1.3. The first case is shown figure \ref{figpDR}, for which $s_1 \simeq 2.2$, $r_1 \simeq 68$ ($\rmm \simeq 73$). The slope at $r=s_1$ and the
DOB-fit are also shown in the right plot, which exhibits a smooth transition
from curved to linear behavior.
Closer to the transition, $\kp_2\leq 1.285$, the form $\rh(x) = x\exp(-x^2/2)$
is not able to fit the data well anymore even with the shifted variable $x=(r-s_1)/r_1$.
Presumably, the crumpled phase is too near for these $\kp_2$ and the two-state nature of the first order transition makes itself felt by contaminating the statistics through crumpled-like configurations.

In the elongated phase the components in (\ref{numcomp}) are small average-size baby universes which corresponds a vertices of a branched-polymer graph. The relation between the SDT lattice-geodesic distance and the branched-polymer distance is somewhat fuzzy, but the same limiting scaling function appears to emerge.
In the crumpled phase there is in addition a large mother-universe component;
how this reflects on the $\rh$ observable is unclear.
As mentioned in section \ref{secsmooth}, there are many possibilities of finite-volume hyperbolic spacetimes and an analytic scaling form may not exist or be very complicated.
In the following we compare qualitative features of $\rh$ in both phases.

\FIGURE{\includegraphics[width=6.5cm]{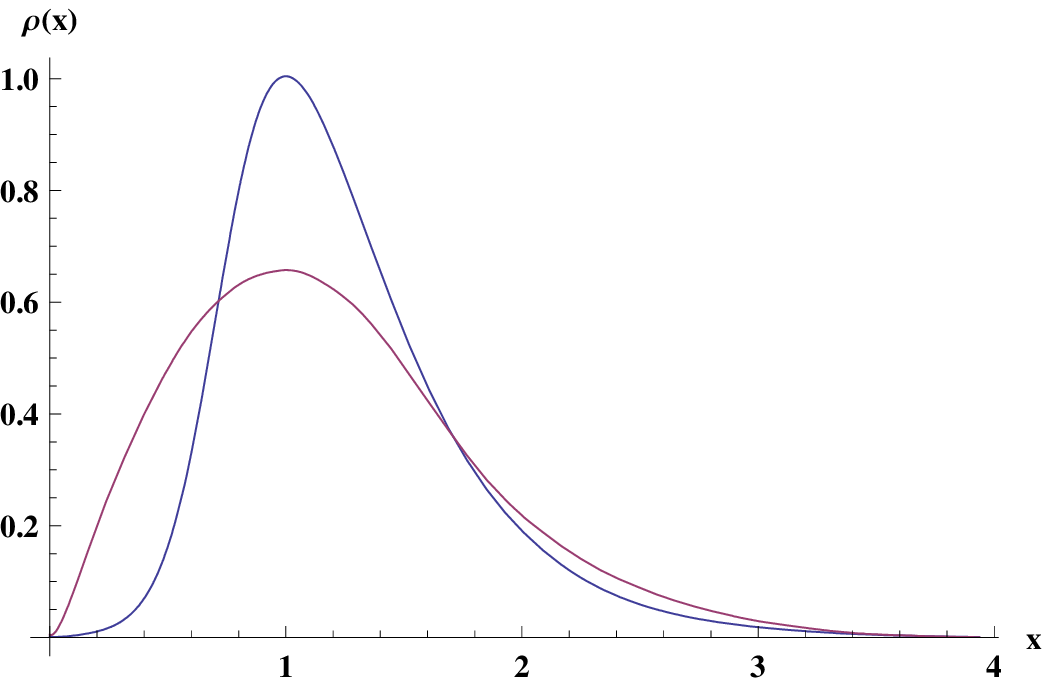}
\caption{Scaling function $\rh_{\rm m}(x)$ for $\kp_2=1.26$ in the crumpled phase (blue) and 1.29 in the elongated phase (red), $N_4=64$ k.
}
\label{figrhm260290}
}

Figure \ref{figrhm260290} shows scaling functions at $\kp_2=1.29$ and $\kp_2=1.26$ for $N_4=64000$, using $r_*=\rmm$ to match the maxima at $x=1$.
In the elongated phase the `one-component region'
$r<r_0$ corresponds to\footnote{DOB-fit $r_0$ values, unless otherwise mentioned.}
$x<x_0=r_0/\rmm=0.128$.
The inflexion point in $x<1$, the point of maximum slope,
is not much larger, $x_{\rm ms}^-= 0.144$. Presumably $x_0$ and $x_{\rm ms}^-$ vanish like $1/\rmm\propto 1/N_4^{1/2}$ as $N_4\to\infty$ \cite{Ambjorn:1995dj} and there is no reason to doubt that the pure generic branched-polymer form
$\tilde\rh_{\rm m}(x)=x \exp(-x^2/2)$ emerges in the limit.

In the crumpled phase the scaled curvature radius is larger, $x_0=0.38$. For the other three members of the scaling sequence in figure \ref{fignp17213060} the scaled radii are $x_0 = 0.82$, 0.60, 0.45 (increasing $N_4$).
Since we expect the $r_0$ to approach a finite limit as $N_4\to\infty$, the $x_0$ will vanish like $1/\rmm$, i.e.\ only logarithmically  $\propto 1/\ln N_4$. They can indeed be approximated by
the function $x_0=b/(c + \ln N_4)$, a fit gives $(b,\,c)=(1.47,\,-7.16)$.
This gives also a fit to the $r_0$ of the scaling sequence using only two parameters
via $r_0=\rmm x_0$, which is undistinguishable from the right plot of figure \ref{figscalrm}.
The inflexion points are much larger than in the elongated phase, for the scaling sequence\footnote{The error bars correspond to the envelope of the Jackknife errors of the $n(r)$ data.}  $x_{\rm ms}^- \simeq 0.6628(10)$, 0.6776(3), 0.6966(2) and  0.7025(1). They seem to approach a finite limit and can be fitted well by the form $a+ b/(c+\ln N_4)$.
However, the three parameters are too correlated to get a meaningful result if we want to extrapolate beyond $N_4=64000$. Assuming $a=1$ gives a good fit with $(b,\,c)=(-4.96,\,5.77)$.
There is also an inflexion point $x_{\rm ms}^+$ in the region $x>1$; its scaling sequence 1.342(21), 1.334(25), 1.308(49), 1.333(9), will be discussed below.

\FIGURE{\includegraphics[width=7cm]{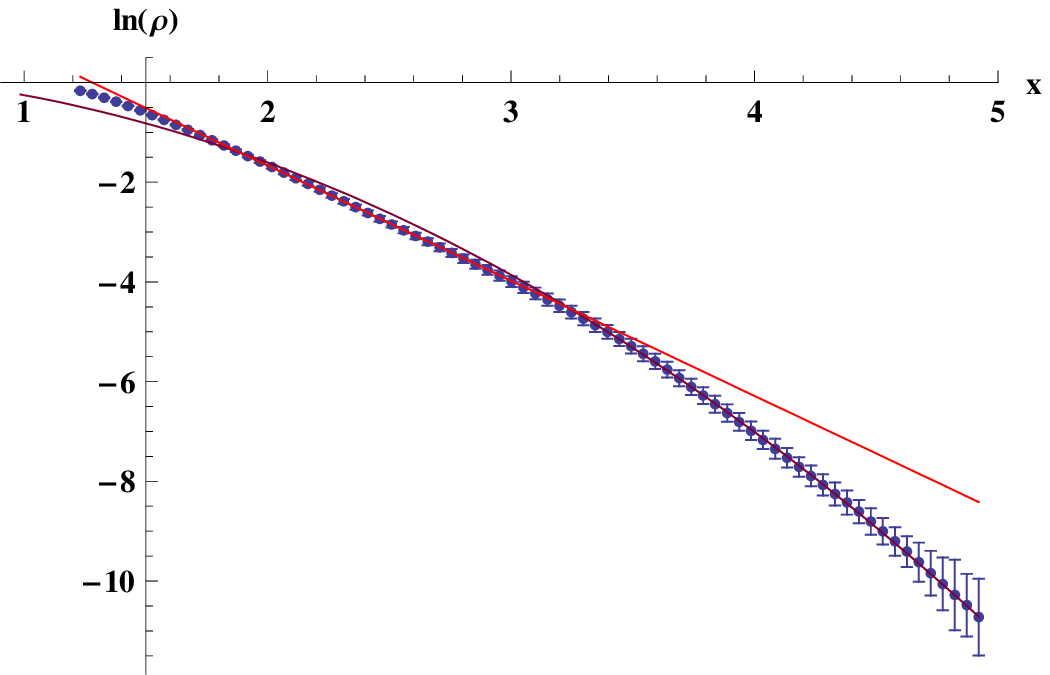}
\includegraphics[width=7cm]{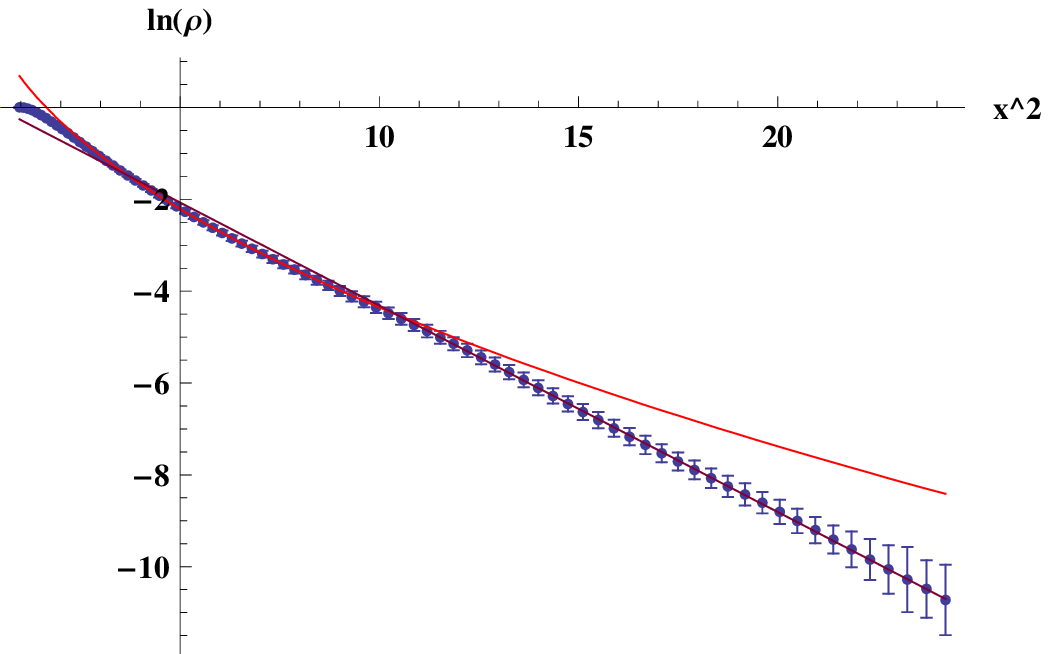}
\caption{Left: $\ln\rh_{\rm m}(x)$ as a function of $x$ for $\kp_2=1.26$ with exponential fit in $1.8 < x < 3.0$ (red) and Gaussian fit in $3.5 < x < 4.9$ (blue-brown). Right: same as a function of $x^2$.
}
\label{figrhm260}
}
Turning to the long-distance region $x\gg 1$, the form (\ref{rhBPAJ}) with $\Ds\to\infty$ implies exponential behavior \cite{Ambjorn:1995dj}. The left plot in figure \ref{figrhm260} shows $\ln\rh_{\rm m}$, which can be fitted well by a function linear in $x$ in the interval
$1.8<x < 3.0$. For larger $x$ there appears to be a turnover to a different behavior. The right plot
shows $\ln\rh_{\rm m}$ versus $x^2$, which suggests Gaussian behavior at large $x$ and a good fit $\ln\rh = b_0 - x^2/2 x_1^2$ can be obtained in $x^2> 12$, or $x>3.5$. For comparison both fits are shown in the left and right plots. The turn-over from exponential to Gaussian behavior is so `abrupt' that a linear + quadratic fit to $\ln\rh$ in the whole region $1.8 < x <  5$ does not look convincing.
We saw similar behavior of $\ln\rh$ for the other members of the scaling sequence. The fitted values of $x_1$ are $1.29$, 1.00, 0.94, 1.05, with errors that are hard to quantify as they are dominated by systematics related to the fitting domain.

\FIGURE{\includegraphics[width=7cm]{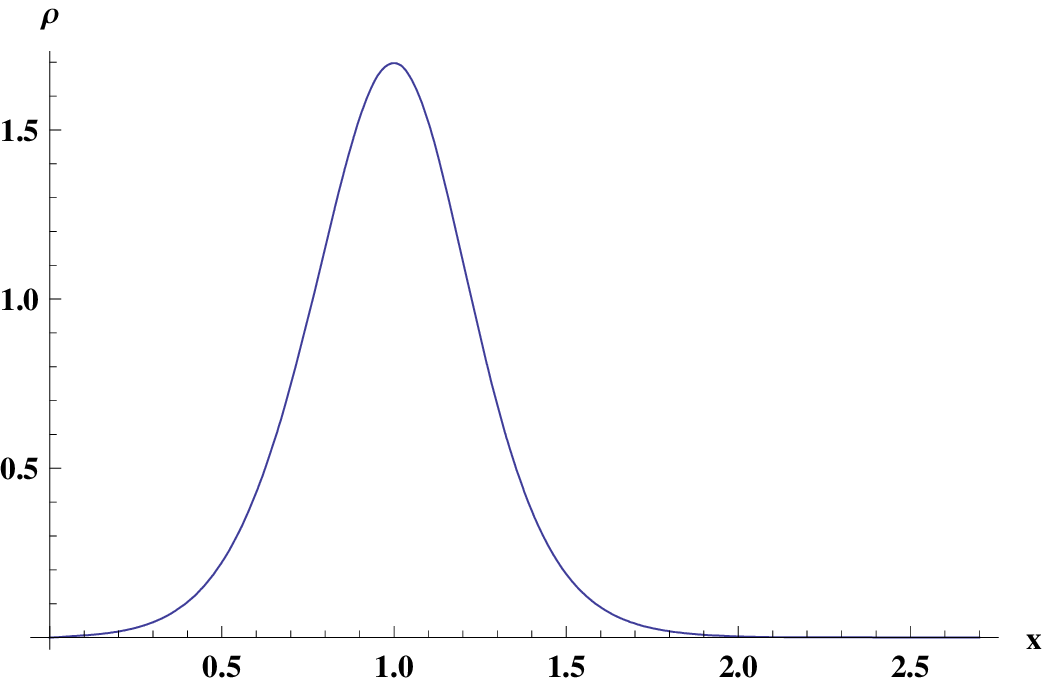}
\includegraphics[width=7cm]{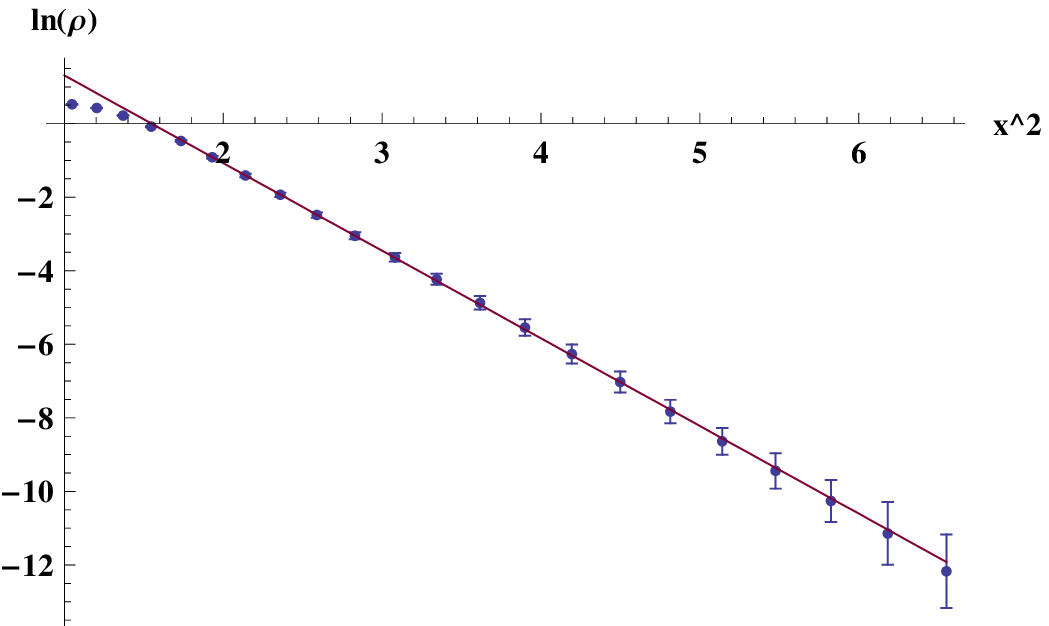}
\caption{Left: $\rh_{\rm m}(x)$ for $(\kp_2,\,N_4)=(1,\,16000)$. Right:
$\ln \rh_{\rm m}$ (blue) with fit $b_0 - x^2/2x_1^2$ in $2.1 < x^2 < 6.6$ (blue-brown), as a function of $x^2$.
}
\label{figrhm00}
}
Accepting tentatively the Gaussian behavior of the tail of the distribution, its interpretation might be that even in the crumpled phase, the baby universes are sufficiently many and small in size to cause branched polymer-like features at the largest distance scale. This idea can be tested by going deeper in the crumpled phase, at smaller curvature radii. The best example in our data set involves somewhat unfortunately a smaller number of simplices,
$(\kp_2,\,N_4)=(1,\, 16000)$. Since $\kpc(16000)\approx 1.22$, $\kp_2=1$ is here much deeper in the crumpled phase than the data we have shown thus far. Its DOB-fit parameters are $s=-1.98$, $c=0.143$ ($\lm=0.360$), $r_0=6.03$, and the maximum of $n(r)$ is at $\rmm=13.7$.
Despite the relatively small volume the putative number of components (\ref{numcomp}) is fairly large\footnote{The B-fit gives even $\numc\simeq 126$.},
$\numc\simeq 63$.
Its scaling function $\rh_{\rm m}$ is shown in figure \ref{figrhm00};
$x_0=0.441$, $x_{\rm ms}^-=0.7846(6)$ and $x_{\rm ms}^+=1.210(2)$.
The right plot shows Gaussian behavior, $\ln\rh_{\rm m}$ is linear in $x^2$ over eleven e-folds in the tail region $x^2>2.1$ ($x> 1.5$).
We found no convincing\footnote{A linear function of $x$ fits the data well only in the rather small region $2.14 <x^2 <3.34$ ($1.5 < x < 1.8$).}
indication of exponential behavior in $x>1$.

Consider again the right plot in figure \ref{fignp17213060}. The scaling violation in $x<1$ shows a systematic steepening of the scaling function with crossings near $x=0.8$ and an increasing maximum, with a probability $\int_{1/\rmm}^1 dx\, \rh_{\rm m}(x)$ that stays nearly constant as $N_4$ in creases, as noted above.
The systematics in the region $x>1$ is somewhat less clear but the curves for the largest two $N_4$ cross again near $x=1.6$, indicating a narrowing of the distribution. If this trend continues the scaling function might start looking like the one in the left plot of figure \ref{figrhm00}. The true limit $N_4\to\infty$ might even be a Dirac delta function
(note the difference in hight of the maxima in figures \ref{fignp17213060} and \ref{figrhm00}).

\FIGURE{\includegraphics[width=7cm]{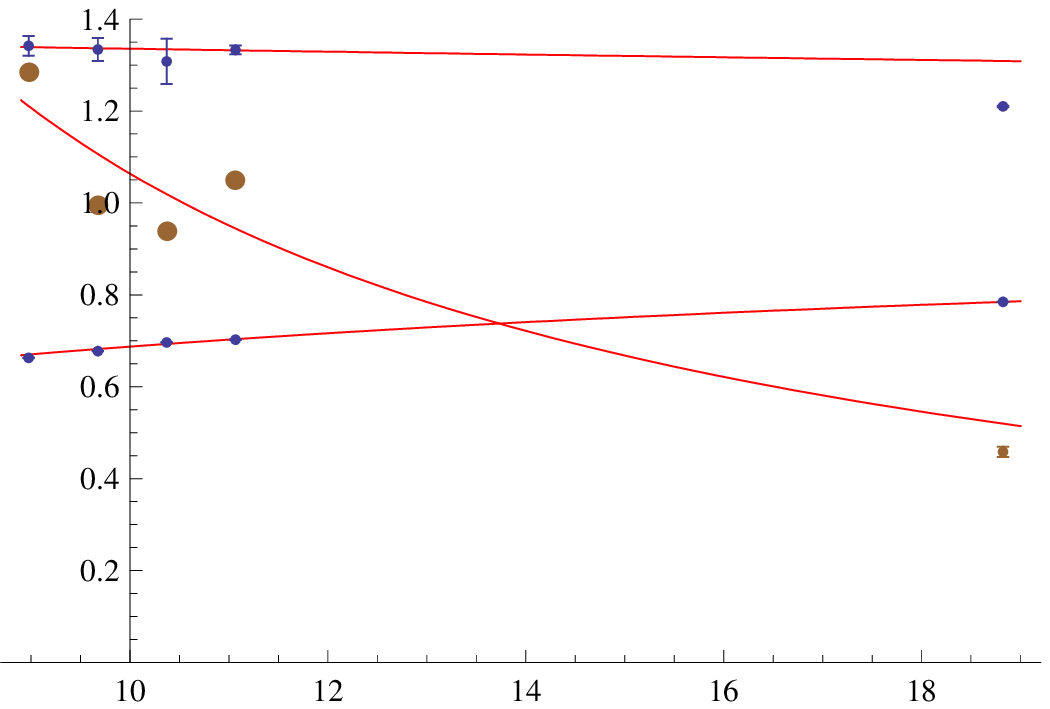}
\caption{Fits to the scaling sequence of $x_{\rm ms}^+$ (blue, upper), $x_{\rm ms}^-$ (blue, lower), $x_1$ (brown, middle), and extrapolations to the region around the shifted data of $(\kp_2,\,N_4)=(1,\,16000)$.
}
\label{figxmsG}
}

To investigate this quantitatively we extrapolated the characteristics
$x_{\rm ms}^{\pm}$ and $x_1$ to larger volumes to see whether they approach those of the scaling function in figure \ref{figrhm00}: $\rhta$.
The $x_{\rm ms}^-$ of the scaling sequence have the smallest errors. They can be fitted by
$x_{\rm ms}^- =1- 6.11/(9.54 +\ln N_4)$, and extrapolating to larger $N_4$ this matches the
$x_{\rm ms}^-$ value of $\rhta$ at
$\ln N_4=18.8$ ($N_4\approx 1.5\times 10^8$).
The fit to the $x_0$ data does not share such a matching property, but $x_0$ characterizes scaling-{\em violation}.
In $x>1$, fits to the $x_{\rm ms}^+$ and $x_1$ data
do approach those of $\rhta$ shifted to $\ln N_4=18.8$ but they miss
by quite a lot, see figure\footnote{Because of the substantial systematic uncertainty in the Gaussian fits leading to the $x_1$ data points, these have not been given error bars but large dots, and the fit to their scaling sequence is of the least-squares type.} \ref{figxmsG}.
Hence, the three extrapolations in this figure do not appear to support the idea that a unique scaling function emerges in the crumpled phase at very large $N_4$.

However, the reason might be finite size effects. The first order nature of the phase transition was seen only in the data at $N_4 \geq 32000$ and $N_4=16000$ may be too small to give a good indication of the limiting scaling function, no matter how deep in the crumpled phase.
The finite-size effects appear to be larger in the region $x>1$ than in $x<1$, since in the former we could match $x_{\rm ms}^-$ at `a reasonable' $\ln N_4= 18.8$, whereas matching $x_{\rm ms}^+$ would require `an unreasonable' $\ln N_4\approx 71$. Furthermore, we found no evidence of an exponential region in the tail of $\rhta$, instead it is Gaussian.
For
$\rh_{\rm m}^{(1.26,64000)}$ the tail is mostly exponential\footnote{The associated probabilities are small: $\int_{1.5}^\infty dx\, \rhta(x)\simeq 0.024$; for $(\kp_2,\,N_4)=(1.26,\, 64000)$ the probability in the exponential region $1.5<x<3.5$ is $\simeq 0.127$, in the Gaussian tail $x>3.5$ it is $\simeq 0.0014$.}.

In practice, the crumpled-phase scaling function is not unique because of the logarithmic slowness of its change with $N_4$, and the shape-distinguishing parameter $\tau$ in (\ref{defrh}) is needed indeed.

\section{Constraint effective action}
\label{seceffact}

We wish to compare $\Rc$ with the average Regge curvature
$\langle\RR\rangle = \langle\bar\RR\rangle$ (the volume-averaged Regge curvature $\bar\RR$ was defined in (\ref{RRegge2})). An overview of the latter in a relatively large region of $\kp_2$ values can be found in \cite{Agishtein:1992xx,Bruegmann:1992jk,deBakker:1995yb}, but here we need a closeup near $\kpc$. It can be obtained by integrating the susceptibility  $\langle(\bar\RR-\langle\bar\RR\rangle)^2\rangle \propto (\partial/\partial\kp_2)\langle\bar\RR\rangle$, for which suitable data at $N_4=64000$ is shown in \cite{deBakker:1996zx}. We shall do this by fitting a model for the {\em constraint effective potential} of $\bar\RR$ to a set of data values.

The distribution of $\bar\RR$ is described by the constraint effective action $\Gmu$,
\be
e^{-\Gmu(\Ru)} = \sum_{\mathcal{T}(N_4)} e^{\kp_2 N_2}\,\dl(\ellt^2 \bar \RR,\ellt^2 \Ru),
\quad
Z(\kp_2,N_4) = \ellt^2\int_{-\infty}^\infty d\Ru\, e^{-\Gmu(\Ru)},
\label{effactbare}
\ee
where the delta function stands for applying a Dirac function after interpolating a histogram of $\bar \RR$.
Properties of the constraint effective action in scalar field theory and its relation with the usual effective potential are discussed in \cite{O'Raifeartaigh:1986hi}; see e.g.\ \cite{Dimitrovic:1991qg} for a numerical study and \cite{Fodor:1994sj} for a study of a composite gauge-invariant operator ($\vr^\dagger\vr$) in electroweak theory. Rewriting (\ref{RRegge2}) in the form
\be
N_2 = \frac{V}{4\pi v_2}\left(\bar\RR + 20\theta\frac{v_2}{v_4}\right),
\quad V=N_4 v_4,
\ee
we see that $e^{\kp_2 N_2}$ can be taken out of the SDT sum, such that we can write
\bea
\Gmu &=& \Su+\Sgu,
\quad \Su=\kpu V(-\Ru + 2\Lmu),
\quad
\kpu\equiv\frac{\kp_2}{4\pi v_2}=\frac{1}{16\pi G_0},
\;\;
\Lmu \equiv -10\theta \frac{v_2}{v_4},\\
\Sgu &=& -\ln \sum_{\mathcal{T}(N_4)} \dl(\ellt^2 \bar \RR,\ellt^2 \Ru).
\eea
Here, $S$ is the Einstein-Hilbert action specialized to constant curvature $\Ru$. Note that $\Sgu$, the constraint selfenergy function, does not depend on $\kp$.
It follows from these definitions that
\bea
\Rav &\equiv&\langle\bar\RR\rangle=
\frac{1}{V}\,\frac{\partial}{\partial \kpu}\, \ln Z + 2\Lm_0,
\label{ravz}\\
\ch&\equiv& V(\langle \bar\RR^2\rangle - \langle\bar\RR\rangle^2) = \frac{1}{V}\,\frac{\partial^2}{\partial \kpu^2}\,
\ln Z.
\label{chiz}
\eea

One expects $\Sgu/V$ to depend only moderately on the volume. For large $V$ the integral in (\ref{effactbare}) can then be done in a saddle-point approximation. For a first order phase transition $\Sgu$ is supposed to have two minima, not necessarily at equal depth. Although $\Sgu$ is independent of $\kp$, the minima in $\Gmu$ do depend on $\kp$;
at a certain value $\kps$ they are at equal depth, i.e.
\be
\kps(-R + 2\Lm_0) + \frac{\Sgu(R)}{V} = \gm + \frac{1}{2 s_\pm}\,(R-R_\pm)^2 + \mO((R-R_\pm)^3),
\quad R\to R_\pm,
\ee
where we assume $+$ and $-$ to correspond to the elongated and crumpled phase, respectively. We then have
\be
\Gm(R) \approx V\left[\kp(-R + 2\Lm_0) + \gm + \frac{1}{2 s_\pm}\,(R-R_\pm)^2\right],
\quad R\approx R_\pm,
\quad \kp\equiv \kpu - \kps = \frac{1}{16\pi G}.
\label{Gaussmod}
\ee
Here $\kp$ can be seen to define a renormalized large-scale $G$.
Near $\kp=0$ we have to keep both saddle points,
\bea
Z&\approx& e^{-V(2\kp\Lm_0 + \gm)} z,\quad z=z_+ + z_-,
\label{Zz}
\\
z_\pm &=& \ellt^2\int dR\, e^{V[\kp R - (R-R_\pm)^2/2s_\pm]}
= \left(\frac{2\pi s_\pm \ellt^4}{V}\right)^{1/2} e^{V(\kp R_\pm + \kp^2 s_\pm/2)}.
\label{zGauss}
\\
p(R)&=&p_+(R) + p_-(R),
\quad
p_{\pm}(R) = \frac{1}{z}\,e^{V[\kp R -(R-R_{\pm})^2/2s_{\pm}]},
\eea
where $p_\pm(R)$ are the probabilities of $R$ in the elongated (+) and crumpled (--) phase.
Using the notation
\be
R_\pm = \Rs \pm \Rd,\quad
s_\pm = \sss \pm \sd,\quad
\dl = \quart\, \ln\frac{s_+}{s_-},\quad
w = \kp\Rd + \kp^2 \sd/2,
\ee
the average curvature and susceptibility following from (\ref{ravz}), (\ref{chiz}) and (\ref{Zz}) can be expressed as
\bea
\Rav&=& \Rs + \kp\sss + (\Rd + \kp \sd)\tanh(w V + \dl),
\label{ravGauss}\\
\ch&=&\chp+\chb,\\
\chp&=& V(\Rd + \kp \sd)^2 \left[1-\tanh^2(w V + \dl)\right],\\
\chb&=& \sss + \sd \tanh(w V+\dl)].
\label{chbGauss}
\eea
The susceptibility $\ch$ splits naturally into a peak component $\chp$ that is proportional to the explicit $V$ and a background $\chb$ that does not have this $V$ dependence; the other parameters are expected to become volume-independent for large $V$. The value $\kp_{\rm c}$ where $\chp$ is maximal is close to zero,
\be
\kp_{\rm c} = -\frac{\dl}{\Rd V} +\left(\frac{3}{2} \sd -\half \dl^2\sd\right)\frac{1}{\Rd^3 V^2} +
\mO(V^{-3}).
\label{kp0Gauss}
\ee
With $\ch$ given, $\Rav$ in (\ref{ravGauss}) is its first integral with respect to $\kp$, in which $\Rs$ plays the role of integration constant.

It turns out that the Gaussian model (\ref{Gaussmod}) is not quite good enough to fit the susceptibility data outside the peak region. An extension, in which the generating functions $\ln z_\pm$ are obtained from a Legendre transformation of an effective potential that includes also quartic terms, is able to give a good fit with only two new parameters $t_\pm$. At $\kpu=\kps$ the potential near the minima is
\be
f_\pm(R) = \frac{1}{2s_\pm}\,(R-R_\pm)^2 + \frac{t_\pm^2}{27 s_\pm^3}\, (R-R_\pm)^4.
\label{quarticmod1}
\ee
Solving the `semi-classical equation' $(\partial/\partial R) [\kp R-f_\pm(R)] =0$ gives $R=R^{\rm soln}_\pm(\kp)$, and
\be
\ln z_\pm = V w_\pm(\kp),\quad w_\pm(\kp) =
\kp R^{\rm soln}_\pm(\kp) - f_\pm(R^{\rm soln}_\pm(\kp)).
\label{quarticmod2}
\ee
For $t_\pm=0$ this reduces to the Gaussian model without the root prefactor in (\ref{zGauss}), such that $\dl=0$ in (\ref{ravGauss}) -- (\ref{kp0Gauss}). Details are in appendix \ref{appquartic}. The variance $w''_\pm(\kp)$ falls like $\kp^{-2/3}$ for $\kp t_\pm \gg 1$, unlike the Gaussian model for which it is constant, $w''_\pm(\kp)=s_\pm$. The series expansion in $\kp$,
\be
w_\pm (\kp) = \kp R_\pm +\kp^2 s_\pm \left(\half\, -\frac{1}{27}\,\kp^2 t_\pm^2 + \frac{8}{729}\, \kp^4 t_\pm^4 +
\mO(\kp^6 t_\pm^6)\right),
\ee
shows the behavior at small $\kp$. The difference of $w''_\pm$ with that of Gaussian model is  only a function of $\kp t_\pm$.

\section{Phase transition and renormalized Regge curvature}
\label{secPT}

In this section we renormalize the average Regge-curvature $\langle \bar\RR \rangle$ (\ref{RRegge2}) and compare it with the continuum curvature $\Rc$ of figure \ref{figpRcosc} and with the continuum curvature that follows from the Gauss-Bonnet formula (\ref{chiGB}). This Gauss-Bonnet curvature,
$\RGB=\pm (384\pi^2/V)^{1/2}$,
is that of a smooth spacetime of constant curvature and volume $V=N_4 v_4$.

The Regge curvature can be compared with `the plaquette' in $SU(N)$ lattice gauge theory. Classically, the Wilson loop of the discretized gauge field around an elementary square $(x,\mu\nu)$, $\Tr\, U_{\mu\nu x}$, is related to the field strength $F_{\mu\nu}(x)$ appearing in the continuum action, by
$4\sum_{\mu\nu}\Tr(1- U_{\mu\nu x})= \ell^{4} F_{\mu\nu}(x)F_{\mu\nu}(x) + \mO(\ell^6)$. In the quantum theory, the numbers change: for an expectation value in the ground state at zero or finite temperature,
$4\sum_{\mu\nu}\Tr\left(1- \langle U_{\mu\nu x}\rangle\right)= \zg +
Z_{F^2}\, \ell^{4} \langle F_{\mu\nu}(x)F_{\mu\nu}(x)\rangle + \mO(\ell^6)$, where $\zg$ and $Z_{F^2}$ are dimensionless renormalization constants which depend on the gauge coupling and the temperature. In the standard model there are in addition to the gluon condensate $\langle F^2\rangle$ also condensates bilinear in the quark fields, and the Higgs condensate.
We assume a similar property for the Regge curvature,
\be
\Rav \equiv \langle \bar\RR \rangle = \zg\, \ellt^{-2} + Z_R R_{\rm ren} + \mO(\ellt^2),
\label{Rren}
\ee
where $R_{\rm ren}$ is a renormalized `Regge condensate'. The $\mO(\ellt^2)$ is somewhat inappropriate, since a continuum limit is not available at this stage.
In QCD renormalization `constants' like $\zg$ and $Z_{F^2}$ are computed in perturbation theory, which serves to define the non-perturbative condensates (see for example \cite{Rakow:2005yn} and references therein; a similar example in electroweak theory is in \cite{Fodor:1994sj}). Here we shall turn this around and consider $R_{\rm ren}$ as given, and see what this implies for $Z_R$, after having defined $\zg$. Our candidates for $R_{\rm ren}$ are $\Rc$ and $\RGB$. First we need to obtain $\Rav$ from the susceptibility data at the phase transition.

Indications of a first-order nature of the transition were presented in \cite{Bialas:1996wu,deBakker:1996zx}.
A two-state signal was found in the average number of nodes $\langle N_0\rangle$, for $N_4=32000$ and 64000. Furthermore, the exponents $\Delta$ and $\Gm$ characterizing the growth rate and width of the peak of its susceptibility
\be
\ch_{N_0}
=  N_4 \left(\left\langle \frac{N_0^2}{N_4^2}\right\rangle - \left\langle \frac{N_0}{N_4}\right\rangle^2\right),
\label{defchnode}
\ee
$\ch_{N_0} \propto N_4^\Delta$, $\dl\kp_2\propto N_4^{-\Gm}$, were estimated
to be $\Delta = 0.81(4)$ and
$\Gm = 1.24(18)$ \cite{deBakker:1996zx}. For a first-order transition these exponents are 1.

The number of nodes at fixed $N_4$ is equivalent to the volume-averaged Regge curvature (\ref{RRegge2}) because of the relation $N_0 = N_2/2 - N_4 + \chi_{\rm E}$, with
$\ch_{\rm E}=2$ for spherical topology. Hence, the node susceptibility is proportional to the variance of the volume-averaged Regge curvature, i.e.\ the susceptibility defined in (\ref{chiz}),
\be
\ch_{N_0}= c_{\ch\ch}\, \ch,
\quad c_{\ch\ch} = \quart\,\frac{v_4}{(4\pi v_2)^2} \simeq 1.97\times 10^{-4}.
\ee

\FIGURE{\includegraphics[width=7cm]{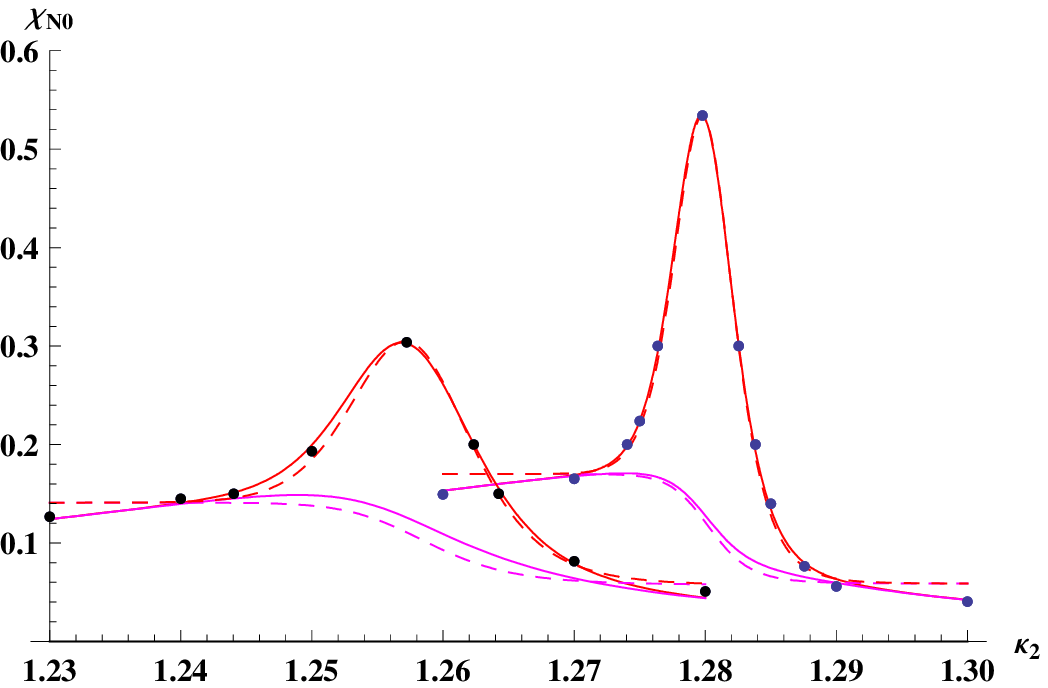}
\includegraphics[width=7cm]{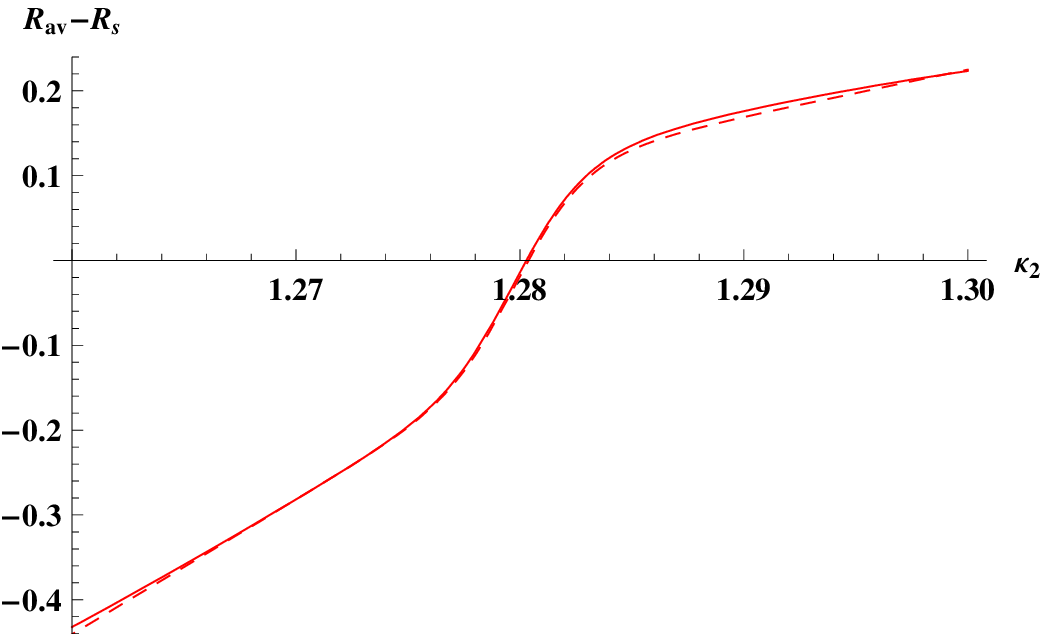}
\caption{Left: Fit of the Gaussian (dashed) and quartic (fully drawn) model-$\ch$ to $\ch_{N_0}$ data (blue) from \cite{deBakker:1996zx}; red: $c_{\ch\ch}\ch$, magenta: $c_{\ch\ch}\chb$,  low peak: $N_4=32$ k, high peak: $N_4=64$ k.
Right: corresponding subtracted Regge curvature $\Rav-\Rs$ for $N_4=64$ k.
}
\label{figch}
}

The left plot in figure \ref{figch} shows a least-squares fit of the quartic model (\ref{quarticmod1},\,\ref{quarticmod2}) to $\ch_{N_0}$ data at $N_4=32000$ and 64000, taken from figure 6 in \cite{deBakker:1996zx}, with the constraint that the $t$-parameters are the same for the two volumes. For comparison, also shown are Gaussian-model fits (dashed), for which $\ch$ becomes constant away from the transition.
This quartic-model fit to the combined data looks good enough. The constraint on the $t$-parameters reduces correlations between the $s$- and $t$-parameters in the fit.
Without the constraint, the resulting fits look `perfect', but the parameters $s_+$ and $t_+$ differ wildly between the two volumes.
The fitted parameters depend moderately on the volume:
\be
\begin{array}{cccccccccc}
\ellt^2 \Rs&\ellt^2\Rd  &s_+&\ellt^{-2} t_+&s_-&\ellt^{-2} t_-&\kp_2^*&\kpc&\quad&N_4\\
&0.113&288&   0&717&   0&1.25905&1.25712&&32000\\
&0.107&375&4308&798&1574&1.25898&1.25683&&32000\\
&0.117&299&   0&864 &  0&1.28038&1.27976&&64000\\
7.06&0.116&366&4308&887&1574&1.28029&1.27971&&64000
\label{chfit}
\end{array}
\ee
The value of $\Rs$ was obtained from figure 2.3 in \cite{deBakker:1995yb}, which shows that $\langle N_2/N_4\rangle \simeq 2.41$ at $\kp_2 = 1.3$ and $N_4=8000$, \ldots, 32000. Assuming the same value for 64000 this gives  $\Rav\simeq 7.28\, \ellt^{-2}$ with (\ref{RRegge2}), and since the quartic fit has given the combination $[\Rav-\Rs]_{\kp_2=1.3}$, we get
$\Rs\simeq 7.06\,\ellt^{-2}$. The deviations $\pm \Rd$ from this central value are smaller by almost two orders of magnitude. The dimensionless distance $2\sqrt{V}\,\Rd=\sqrt{V}(R_+-R_-)$ between the minima of the critical effective potential,
$Vf_\pm(R) = \half \left(R\sqrt{V}-R_\pm\sqrt{V}\right)^2/s_\pm^2 + \cdots$,
becomes larger than the largest width, $\sqrt{s_-}\approx 30$, only for $N_4 \gtrsim 8000$.

The background curves $\chb$ of the two models are also shown in figure \ref{figch}. Especially for the smaller volume the Gaussian-model background is substantially smaller at the transition than the quartic one. On the other hand, the fits to the total $\ch$ appear equally good in the transition region. Hence, systematic errors are larger in $\chb$ than in $\ch$. The background-subtracted peak of the susceptibility,  $\ch-\chb$, is equal to $\Rd^2 V$ at $\kp=0$, i.e.\ $\kp_2=\kp_2^*$, which is close but not equal to the pseudo-critical point $\kpc$, the position of the maximum where $\partial \ch/\partial \kp_2=0$, which is listed also in (\ref{chfit}). After subtracting the background, the `improved' exponents $\Delta$ come out larger than one
\be
\frac{\left[\ch-\ch_{\rm b}\right]_{64000}^{\rm peak}}{\left[\ch-\ch_{\rm b}\right]_{32000}^{\rm peak}}\simeq 2.11,\;2.29,
\qquad \Delta\simeq 1.08,\; 1.19,
\label{Deltaimproved}
\ee
respectively for the Gaussian and quartic model. Similar ratios at $\kp_2^*$ follow from the ratios of $\Rd^2 V$: $\Delta\simeq 1.11$ and $1.24$.

The right plot in figure \ref{figch} shows the Regge curvatures corresponding to the left plot for $N_4=64000$, shifted by $\Rs$, i.e.\ $\Rav-\Rs$ which vanishes at $\kp=0$.
For the discussion of the renormalization to $\Rc$ it is useful and interesting to make a fit also to the $\Rc$ data. The Gaussian model can be used for this purpose. The form (\ref{ravGauss}) for $\Rav$ is by itself not able to fit the $\Rc$ data well because its slope away from the transition (where $\tanh\to\pm 1$) is tightly coupled to the width of the transition. However, with an additional overall scale factor $1/Z'$, i.e.\
\be
\Rc = (1/Z')\left[\Rs' + \sss'\kp'
+ (\Rd' + \sd'\kp')\tanh[V (\Rd'\kp' + \sd'\kp^{\prime 2}/2]\right],
\quad \kp'=(\kp_2-\kp_2^{\prime *})/(4\pi v_2),
\label{Rcfit}
\ee
a good fit is obtained, as shown in the left plot of figure \ref{figRRren}.  For comparison we have also shown $\Rav$. Away from the transition the fit-$\Rc$ becomes linear in $\kp_2$ and the same holds for Gaussian model-$\Rav$. These linear forms,
\be
R_{\rm c,lin}^{\pm} = (1/Z')[\Rs'\pm\Rd' + (\sss'\pm\sd')\kp],\quad
R_{\rm av,lin}^{\pm} =  \Rs \pm \Rd + (\sss\pm\sd) \kp,
\label{Rlinear}
\ee
extrapolated to $\kp_2=1.28$, are also shown in the plot. In the large volume limit the width of the transition region shrinks to zero and the Gaussian model should become exact (assuming the parameters stabilize in the limit). At finite volume we can then replace the curves towards the transition by the above linear forms.

The renormalization constants $\zg$ and $Z_R$ can be specified as follows. We {\em define} $\zg=\Rs$. Then $\Rav$ vanishes at $\kp=0$, or $\kp_2=\kp_2^*$ which is almost equal to $\kpc$.
Given $\Rc$, the factor $Z_R$ is given by the ratio $(\Rav-\Rs)/\Rc$, which is shown in the right plot of figure \ref{figRRren}. Towards the transition the ratio is not meaningful for a first order transition, but we can replace it by the ratio of the linear extrapolations (\ref{Rlinear}), which is also shown in the plot. The values extrapolated from the elongated(crumpled) side are $Z_R=\pm\Rd/((\Rs'\pm\Rd')/Z')\simeq 0.20$ (0.14).

\FIGURE{\includegraphics[width=7cm]{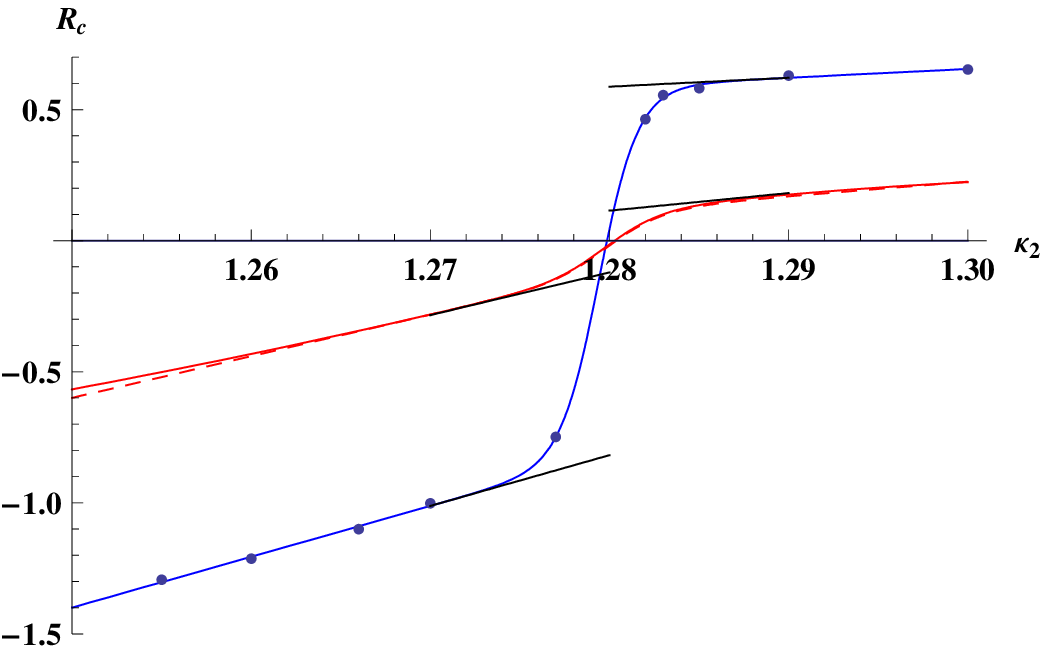}
\includegraphics[width=7cm]{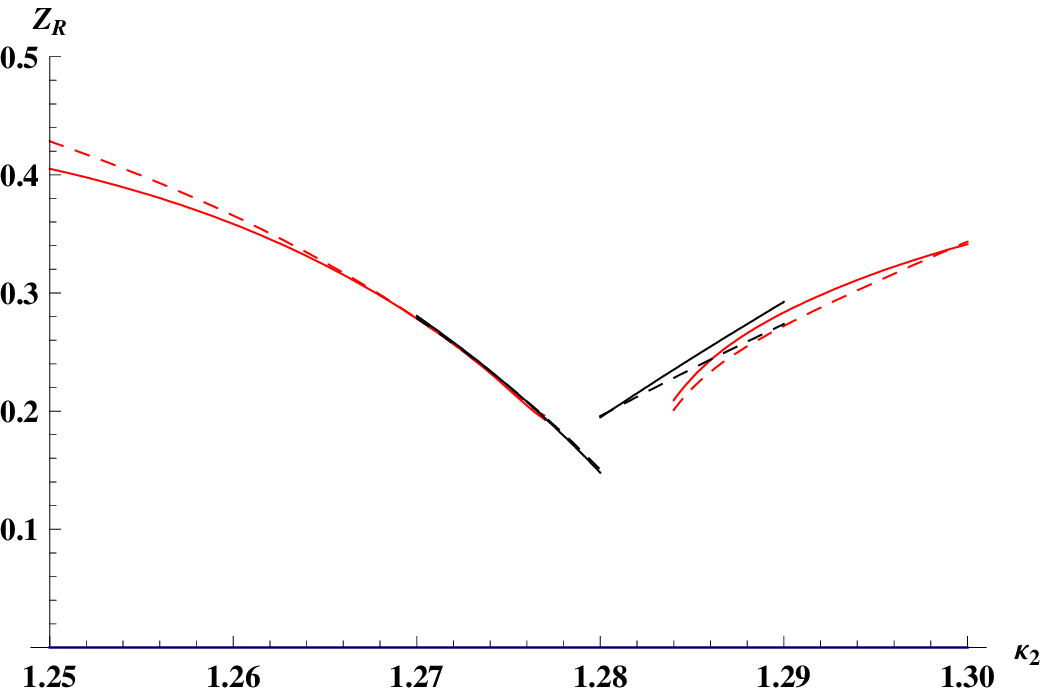}
\caption{Left: Fit (blue) to the $\Rc$ data (blue dots) and linear extrapolations to $\kpc$ (black). Also shown is $\Rav-\Rs$ from the right plot in figure \ref{figRRren} and its linear extrapolations (black). Right:
$\kp_2$ dependence of $Z_R =(\Rav-\Rs)/\Rc$ (same color scheme as in figure \ref{figch}). Also shown are the ratios of the linear forms (\ref{Rlinear}) (black).}
\label{figRRren}
}

The Gauss-Bonnet curvature is given by $\RGB = \pm 12/L^2$, $V=(8\pi^2/3) L^4$, which gives
$\ellt^2 \RGB\simeq 0.23$,  0.16  ($L/\ellt\simeq7.295$, 8.675), respectively for $N_4=32000$, 64000. For the quartic-model the extrapolated $\Rav-\Rs=\pm \Rd$ then give at the transition
$Z_R = \pm \Rd/\RGB \simeq 0.47$ and 0.73, respectively for $N_4=32$ k and 64 k; for the Gaussian-model these number are 0.50 and 0.73.
The value 0.73  is much closer to 1 than the 0.20 (0.14) found above for $\Rc$ case. In other words, after a multiplicative renormalization, which approaches 1 going from 32 k to 64 k, the subtracted Regge curvature $\Rav-\Rs$ is that of a smooth spacetime of volume $V=N_4 v_4$ and curvature $\RGB$.

\section{Conclusions}
\label{secdisc}

The continuum curvature $\Rc$ is obtained from the volume-distance correlation-function by making a small-distance, constant-curvature, approximation which smoothes out `distortions' caused by the lattice, and subsequently making a change of scale from the lattice- to the continuum-distance. The resulting $\Rc$ is quite sensitive to changes in the way the numerical data are fitted, but the qualitative behavior of $\Rc$ near the phase transition is robust. In this sense a qualitatively consistent picture has emerged in which curvature radii are in continuum units not much larger than the lattice spacing\footnote{Recall $\ell=\sqrt{10}\,\ellt\simeq 3.2\, \ellt$}.

The positive curvature in the elongated phase leads naturally to the picture of an average spacetime consisting solely of baby universes glued together into a branched-polymer structure, whereas the negative curvature in the crumpled phase accommodates easily a mother-universe. In such hyperbolic spacetimes of large volume, conflicts with the Gauss-Bonnet theorem have to be avoided somehow by singularities. This gives a new look on the occurrence of the `singular' vertices and links observed in this phase.
Remarkably, these `singular structures' seem innocuous to observables on the dual lattice and the scaling of the volume-distance correlator.

The precise nature of scaling in the crumpled phase is hard to establish, because the slow logarithmic change of scale requires exponentially large volumes. Surprising is the branched-polymer character of its scaling function at the largest distances\footnote{The associated probability is small.}.

The phase transition data of \cite{deBakker:1996zx} could be well described by models of the constraint effective-action, which also gave analytic expressions for the background of the susceptibility peak. The model with Gaussian potentials in the two phases appeared sufficient within the peak region and more robust than its extension to quartic potentials, which was needed for a good description outside the peak. Subtracting the background from the peak led to critical exponents closer (and even larger) than the 1 of a first order transition\footnote{Since we refrained from estimating errors all numbers should be viewed with caution.}.
The renormalized coupling $1/G$ in the effective action changes sign at the transition. Of course, this $G$ need not be the same as the renormalized Newton constant $G_{\rm N}$ characterizing the strength of the gravitational interaction.
A simple adaptation also gave a good fit to $\Rc$. Its rapid passing through zero as a function of $1/G_0$ is caused by the mixed contribution of both phases, away from the transition it is slowly varying.

The quartic-model fit led to a detailed description of the average Regge curvature $\Rav$ through the transition, up to a constant $\Rs$ which could be found from \cite{deBakker:1995yb}.
Subtracting this constant, a renormalized Regge curvature $R_{\rm ren}=(\Rav-\Rs)/Z_R$ emerges that passes through zero at the transition. This was compared with $\Rc$, and also with the Gauss-Bonnet curvature $\RGB$, the
continuum curvature that relates volume and Euler index by the Gauss-Bonnet theorem. The multiplicative renormalization constant $Z_R$ in this comparison came out rather small compared to 1 for $\Rc$ at the transition, but much closer to 1 for $\RGB$,
and it increased with volume.

The average Regge curvature is quite different from $\Rc$ in its sensitivity to quantum fluctuations. It can be expressed as a volume average of the curvature at a triangle and in this sense it is associated to a large distance-scale,
whereas $\Rc$ is derived from an averaged quantity at a scale not much larger than lattice spacing.
The Gauss-Bonnet curvature is associated with a length scale $L$ derived from the volume $V=N_4 v_4$: $L\equiv [V/(8\pi^2/3)]^{1/4}$, $\RGB=\pm 12/L^2$. It is in a sense a `target' curvature that one would like to get out of a detailed understanding of the renormalization of the curvature scalar in the bare Einstein-Hilbert action.

The fact that the subtracted
$\Rav-\Rs$ is not very different from $\RGB$ for the largest simulation example analyzed here ($N_4=64$ k) is intriguing and results at larger volumes are needed to establish that it is not a mere coincidence. If $Z_R$ would keep approaching the target 1, then the phase transition would not be of first order\footnote{Assuming that the jump $2\Rd$ of $\Rav$ at the transition stabilizes in lattice units, as expected for a 1st-order transition, a doubling of the volume would bring $Z_R$ within a few percent of 1, a further doubling to about $\sqrt{2}$, etc.; in the contrary case $\Rd \propto 1/L^{2}$ for large $L$.}.

In any case,
a better understanding is needed of the physics of the branched-polymer structures, which are also present in the spatial slices of the spacetimes in the `De Sitter phase' of CDT \cite{Ambjorn:2005qt}.

\section*{Acknowledgement}
The numerical simulations on which this work is based were performed by Bas de Bakker around 1996 (the plots in figure \ref{figRV} were also made by him), with support from FOM/NWO.

\appendix
\section{Simple lattice models in flat spacetimes}
\label{appmodels}
\FIGURE{\includegraphics[width=7cm]{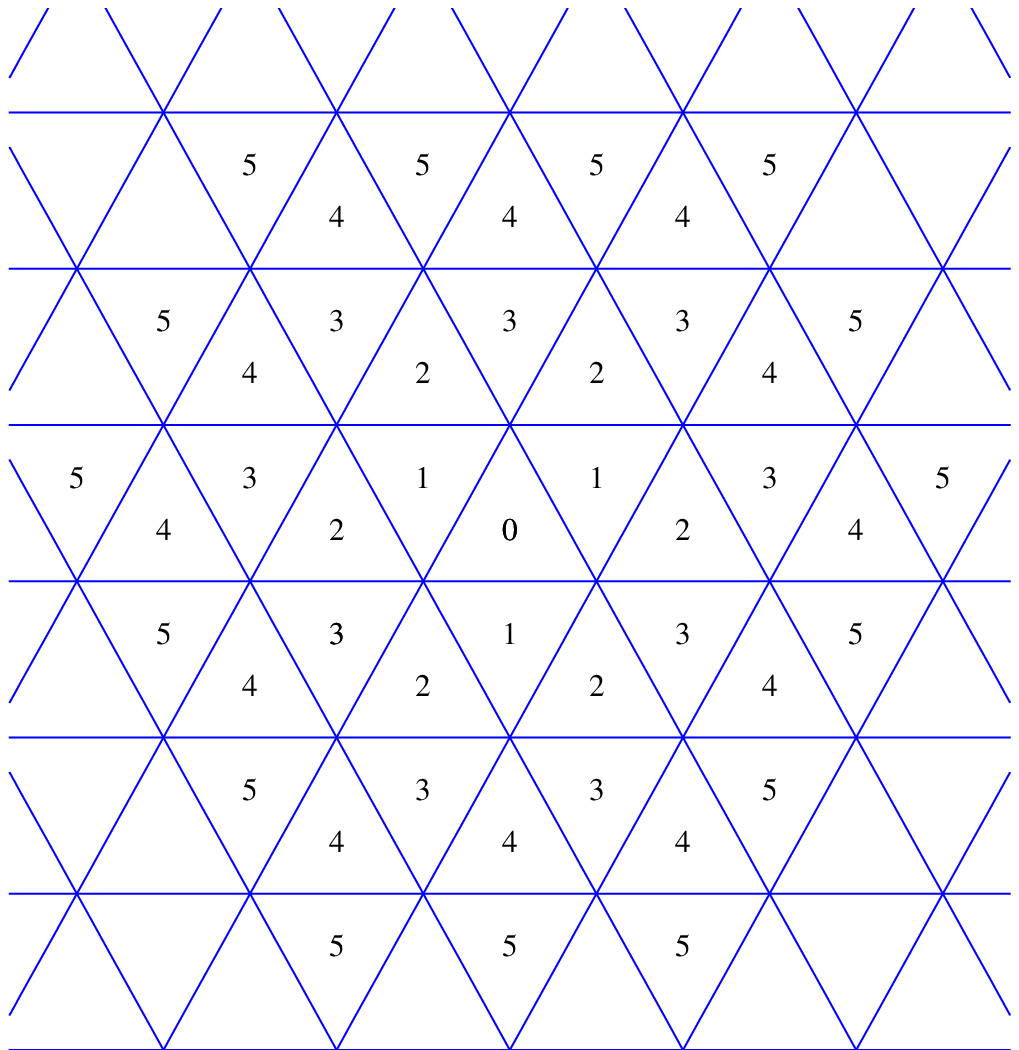}
\caption{Equilateral triangular lattice with lattice-geodesic distances indicated up to $r=5$.}
\label{figtriangle}
}

In this appendix we study the volume-distance relation for a 2D triangular lattice and for hyper-cubic lattices in $D=2$, 3 and 4 four dimensions, and also the retrieval of the metric scale factor $a(r)$ of the line element $ds^2 = dr^2 + a(r)^2 d\Om_{D-1}^2$. In flat spacetime, $a(r) =r$ and this is to be the target result in the limit of zero lattice spacing, the continuum limit. Ensembles of these lattice configurations are supposed to contain only one member, so there are no effects due to averaging.

Figure \ref{figtriangle} shows a triangular lattice in which the lattice geodesic-distance from an arbitrary origin is given by the numbers in the centers of the triangles. In these units, the link length $\ell=\sqrt{3}$. The centers at even $r$ form a hexagonal shape; at odd $r$ the shape is somewhat different, but this becomes negligible as $r$ increases. However, the difference between the lattice geodesic-distance $r$ and the continuum distance $\dcon$ depends on direction and does not diminish as $r\to\infty$.  For example, along a link direction, $\dcon = \frac{\sqrt{3}}{2}\, r$ (even $r$), whereas in a direction perpendicular to a link direction (a direction along a dual-lattice link), $\dcon = \frac{3}{4}\, r$ for $r=4$, 8, 12, \ldots and $\dcon = \frac{3}{4}\, r + \quart$ for $r=1$, 5, 9, \ldots.
Inspection shows that $n(r)$ is given by $n(r) = 3\, r$, $r=1$, 2, \dots. With $N(0) = 1$ it follows that \be
N(r)=1+\sum_{r'=1}^r n(r') = 1+\frac{3}{2}r+ \frac{3}{2}\, r^2, \quad r=0, \,1, \ldots.
\ee
As polynomials in $r$, $n(r)$ and $N(r)$ are naturally extended from integers to real numbers.

To avoid confusion we indicate a continuum object by a subscript c, for example $\rcon$, $V_{\rm c}(\rcon) = \pi \rcon^2$.  Lattice objects could be given the subscript $\ell$, but for notational convenience we drop this, so $\rlat=r$.
Consider the volume (area) $v_2 N(r)$.
Making explicit the lattice spacing $\ell$ and the spacing of the dual lattice $\ellt = \ell/\sqrt{3}$, the dimension-full lattice geodesic-distance $r$ is an integer multiple of $\tilde\ell$ and the volume of a triangle is $v_2= \frac{\sqrt{3}}{4}\,\ell^2 = \frac{3}{4}\,\sqrt{3}\,\tilde\ell^2$. Then
\bea
v_2 N(r/\tilde\ell) &=& \frac{3}{4}\,\sqrt{3}\, \left(\frac{3}{2}\,r^2  + \frac{3}{2}\, r\ellt +\ellt^2\right) 
\nonumber\\
&\to& \frac{9}{8}\,\sqrt{3}\, r^2
\label{N2incon}
\eea
in the limit $\ellt\to 0$, $r$ fixed.
It has the same behavior, $\propto r^2$, as $V_{\rm c} \propto \rcon^2$.
Equating $V_{\rm c}(\rcon)= \lim_{\ellt\to 0} v_2 N(r/\ellt)$,
it follows that $\rcon = \lm r$ with $\pi \lm^2 = \frac{3}{2}\, v_2/\ellt^2$, $\lm=(9\sqrt{3}/8\pi)^{1/2}\simeq 0.79$. As might be expected, this factor $\lm$ lies between the direction coefficients $\frac{3}{4}=0.75$ and $\frac{\sqrt{3}}{2}\simeq 0.87$ found above.

At finite $\ellt\ll r$, $v_2 N(r/\ellt)\approx \Vcon(\lm r)$.
This approximation can be improved by including a shift of order of the lattice spacing,
chosen to cancel the leading $\mathcal{O}(\ellt)$ contribution. This shift turns out to be $1/2$ in lattice units:
\be
\ellt^2 N((r/\ellt)-1/2) =  \frac{3}{2} \,r^2 +\frac{5}{8} \ellt^2.
\label{Ntrmh}
\ee
In fact, not only the leading lattice artifact linear in $\ellt$ is canceled, but also the coefficient of the $\ellt^2$ term has been reduced by the shift. Furthermore, for a slowly varying function $F(r)$, $F(r)-F(r-\ellt) = \ellt (d/dr)F(r-\half\ellt) + \mathcal{O}(\ellt^3)$, and using this for $F(r)=N(r/\ellt)$ we have
\bea
n(r/\ellt) &\equiv& N(r/\ellt)-N((r/\ellt) -1) = \ellt \frac{d}{dr}\, N((r/\ellt)-1/2) + \mathcal{O}(\ellt^3)
\nonumber\\
& =&
N'((r/\ellt)-1/2) + \mathcal{O}(\ellt^3).
\label{Npvsn}
\eea
For the simple triangular case there are no $\mathcal{O}(\ellt^3)$ corrections: $\ellt N'((r/\ellt)-1/2) =3r=\ellt\, n(r/\ellt)$.

Using $\rcon=\lm r$ and $V_{\rm c}(\rcon) \approx v_2 N((r/\ellt)-1/2)$,
the metric scale-factor $a_{\rm c}(\rcon)$ can be retrieved from $n$ or $N'$ through
\bea
2\pi a_{\rm c}(\rcon)
=
\frac{d}{d\rcon}\, V_{\rm c}(\rcon)
&\approx& \frac{d}{\lm d r}\, v_2 N ((r/\ellt)-1/2)
= \frac{v_2}{\lm\ellt}\,N'((r/\ellt)-1/2) = \frac{v_2}{\lm\ellt}\,n(r)
\label{aVtriangle}\\
&=& 2\pi \rcon.
\eea
In this 2D example
there are no finite $\ellt$ corrections,
$a_{\rm c}(\rcon) =\rcon$.

The constant $\lm$ can be absorbed by changing units. Suppose the function $V_{\rm c}$, which in $D$ spacetime dimensions has engineering dimension $D$, is described in terms of a length scale
$r_{0{\rm c}}$, e.g.\ a curvature radius. We can then introduce objects without the subscript c, $V$ and $a$, that `look' exactly like their continuum version $\Vcon$ and $a_{\rm c}$,
\bea
V_{\rm c}(\rcon; r_{0{\rm c}}) &=& \lm^{D} V_{\rm c}(r;r_{0})
\equiv\lm^{D} V(r,r_0),
\quad r=\rcon/\lm, \quad r_0 \equiv r_{0{\rm c}}/\lm,
\\
a_{\rm c}(\rcon; r_{0{\rm c}})&\equiv& \lm \,a(r;r_0).
\label{avsacon}
\eea
In short, $V_{\rm c}(\rcon) = \lm^{D} V(r)$, $a_{\rm c}(\rcon) = \lm\, a(r)$, and we have to keep in mind that distances derived from $r$ are larger by a factor $1/\lm$ than those corresponding to $\rcon$.
In terms of $a$, the generalization of (\ref{aVtriangle}) to a lattice in $D$ dimensions
is tentatively
\bea
D C_D a(r)^{D-1} &=&  D c_{D}\,\lm^{-(D-1)} a_{\rm c}(\rcon)^{D-1} = \lm^{-(D-1)} V'_{\rm c}(\rcon)
\\
&\approx& \frac{v_D}{\ellt\lm^D}\, N'((r/\ellt)-1/2),
\label{adeftent}\\
C_D &=& \pi,\, 4\pi/3,\, \pi^2/2, \quad D=2,\,3,\,4,
\eea
where $C_D$ is the volume of the unit ball in $D$ dimensions.
The factor $\lm$ is determined by the coefficient $\al$ that characterizes the small-distance behavior of $N$ {\em after} taking the continuum limit (cf.\ the discussion after (\ref{N2incon})),
\bea
\lim_{\ellt\to 0}\,\ellt^D N((r/\ellt)-1/2) &=& \al\, r^D + \mathcal{O}(r^{D+2}),\\
v_D N((r/\ellt)-1/2) &=& V_{\rm c}(\rcon) = C_D \lm^D r^D + \mathcal{O}(r^{D+2},\ellt^2),\\
\lm &=& \left(\frac{ v_D\, \al}{\ellt^D C_D}\right)^{1/D}.
\label{gmdef}
\eea
The order of the corrections $\mO(r^{D+2})$ and $\mO(\ellt^2)$ applies to the lattices in this appendix.

The tentative equations (\ref{adeftent})--(\ref{gmdef}) above can now be turned around to {\em define} a metric scale factor $a(r)$ from the lattice $N(r)$ at finite lattice spacing. For reasons to become clear below and in the main part of this paper, we shall allow for a shift of order of the lattice spacing in the relation between $\rcon$ and $r$,
\be
\rcon=\lm (r-s),\quad s=\mathcal{O}(\ellt).
\label{rrc}
\ee
Then
\bea
a_{\rm c}(\rcon)&=&\lm\, a(r-s),
\\
a(r-s) &=& \left[\frac{\veff}{D c_{D}}\, N'((r/\ellt)-1/2)\right]^{1/(D-1)},
\quad \mbox{or}\quad
\left[\frac{\veff}{D c_{D}}\, n(r) \right]^{1/(D-1)},
\label{aVdef}\\
\veff &\equiv& \frac{v_D}{\ellt\lm^D}.
\label{veffgm}
\eea
is a concrete realization of (\ref{aNappr}).
Note that
\be
\left[a(r-s)\right]_{\ellt\to 0} = r  + \mathcal{O}(r^3),
\ee
in accordance with the proper-time interpretation of $r$ in the continuum. The version with  $n(r)$ in (\ref{aVdef}) will have somewhat larger lattice artifacts at non-zero $\ellt$.

Let us now see how this works out for a flat hyper-cubic lattice.
In two dimensions, the centers of the plaquettes at $\dlat=r$ form a square rotated by $45^\circ$.
With points on the dual lattice labeled by integers $(x_1,\ldots,x_D)$, the lattice geodesic distance is $\dlat(x,y) = (|x_1-y_1| + \cdots +|x_D-y_D|)\ellt$ ($\ellt=\ell$). As for the triangular lattice, $\dlat(x,y)$ differs from the continuum distance through the interior of the lattice,
$\dcon(x,y) = \sqrt{\sum_i(x_i-y_i)^2}\,\ellt$, even at arbitrarily large distances and there can be many lattice geodesic paths with the same $r=\dlat(0,y)$.
Along a dual link direction, $\dcon = r$, whereas in a direction along a link of the original lattice, $\dcon = r/\sqrt{2}$. The breaking of rotational symmetry of the set of centers at distance $r$
is larger here than for the triangular lattice.
For $n$ and $N$ we find, reverting for simplicity to lattice units $\ellt=1$,
\be
\begin{array}{llcr}
n(r)& N(r)&\lm&D\\
4r&1+2r+2 r^2=\half + 2(r+\half)^2&0.80&2\\
2 + 4 r^2&1+ \frac{8}{3} r + 2 r^2 + \frac{4}{3} r^3= \frac{5}{3}(r+\half) + \frac{4}{3}(r+\half)^3&0.68&3\\
\frac{16}{3}r+\frac{8}{3}r^3& 1 + \frac{8}{3}r+\frac{10}{3}r^2 +\frac{4}{3}r^3+\frac{2}{3}r^4=
\frac{3}{8} + \frac{7}{3}(r+\half)^2 + \frac{2}{3}(r+\half)^4 & 0.61&4\\
\end{array}
\label{ncubic}
\ee
where we used $v_D = \ell^D = \ellt^D$ for the cubic lattices to calculate the $\lm$.
Similar to the triangular case, $n$ and $N$ are naturally extended as polynomials from integers to real numbers. The same result is obtained by using interpolation of sufficiently high order. The powers of $r$ in $n(r)$ differ by 2, so the lattice artifacts start at order $\ellt^2$. In contrast, $N(r)$ contains all powers up to $D$. But for $N(r-1/2)$ the powers of $r$ differ by 2 again, compatible with (\ref{Npvsn}). In $D=3$ and 4 the coefficients of the non-leading powers of $r$ in $N'(r-1/2)$ are slightly smaller than those in $n(r)$, so the lattice artifacts in $N'(r-1/2)$ are somewhat smaller than in $n(r)$.

The formulas (\ref{adeftent})--(\ref{gmdef}) result in a scale factor $a(r)$ with lattice artifacts in the small $r$ region. These artifacts cause a non-zero scalar curvature $R(r)$ (by (\ref{RRW})) that rapidly vanishes as $r\to\infty$ (the continuum limit in lattice units).
In four dimensions, choosing the version with $n(r)$ in (\ref{aVdef}) with shift $s=0$,
we have  $a(r) = (2 r + r^3)^{1/3}$, with $2\pi^2/\veff=8/3$. The scalar curvature has the large-$r$ expansion\footnote{In the version using $N'(r-1/2)$,
$a(r)= [(7/4) r +r^3]^{1/3}$, $R(r) = (49/8) r^{-6} + \mathcal{O}(r^{-8})$.}
$R(r)=8 r^{-6}+\mathcal{O}(r^{-8})$. It becomes accurate for $r\gtrsim 3$, where $R(r)< 0.01$, whereas $R(1)\simeq 0.74$.
In three dimensions, $a(r)=(1/2+r^2)^{1/2}$, $R(r)=-4 a^{\prime\prime}/a-2 a^{\prime 2}/a^2+2/a^2 = -r^{-4}+r^{-6} + \mathcal{O}(r^{-8})$, and $R(1)\simeq -0.44$ but already smaller than 0.004 for $r\gtrsim 4$.
It is no surprise that $R(r)$ is of order of the discretization scale for $r$ of order 1, but it is comforting to find it to be already less than percent of that for $r\gtrsim 4$.
In two dimensions we have simply $a(r)=r$, as for the triangular lattice.

In 4D simplicial gravity we do not have the luxury of determining $\lm$ and $\veff$ from the behavior of $n(r)$ or $N'(r-1/2)$ at arbitrarily large $r$, i.e.\ the continuum limit, because the distances at which a  description in terms of a metric scale factor $a(r)$ might apply is not very much larger than the lattice spacing. Typically, we find in the main body of this paper that the maximum $r$ has to be less than about  12. An important quantity is the effective volume $\veff$ in the relation
$a(r-s) = [n(r)\veff/2\pi^2]^{1/3}$ ($s$ is the optional shift introduced in (\ref{veffgm})).
We can estimate it from intermediate distances\footnote{Recall that $\rmm$ is the position of the maximum in $n(r)$.}
$1\ll r\ll \rmm$ by fitting a constant-curvature model for $V'(r)$ to $n(r)$ or $N'(r-1/2)$, e.g.\
\be
n(r) \approx c \left(r_0 \sin\frac{r-s}{r_0}\right)^3 ,
\quad
c=\frac{2\pi^2}{\veff},
\quad
r=\rfitmin,\,\rfitmin+1,\cdots \rfitmax,
\label{veffdet1}
\ee
or
\be
[N'(r-1/2)]^{1/3} \approx
c^{1/3}r_0 \sin\frac{r-s}{r_0},
\label{veffdet2}
\ee
and its $\sinh$ analogues.
The sine and hyperbolic sine cases can be combined into an explicit function of $y=\pm r_0^2$, with $y>0$($<0$) corresponding to positive(negative) curvature.

\FIGURE{\includegraphics[width=7cm]{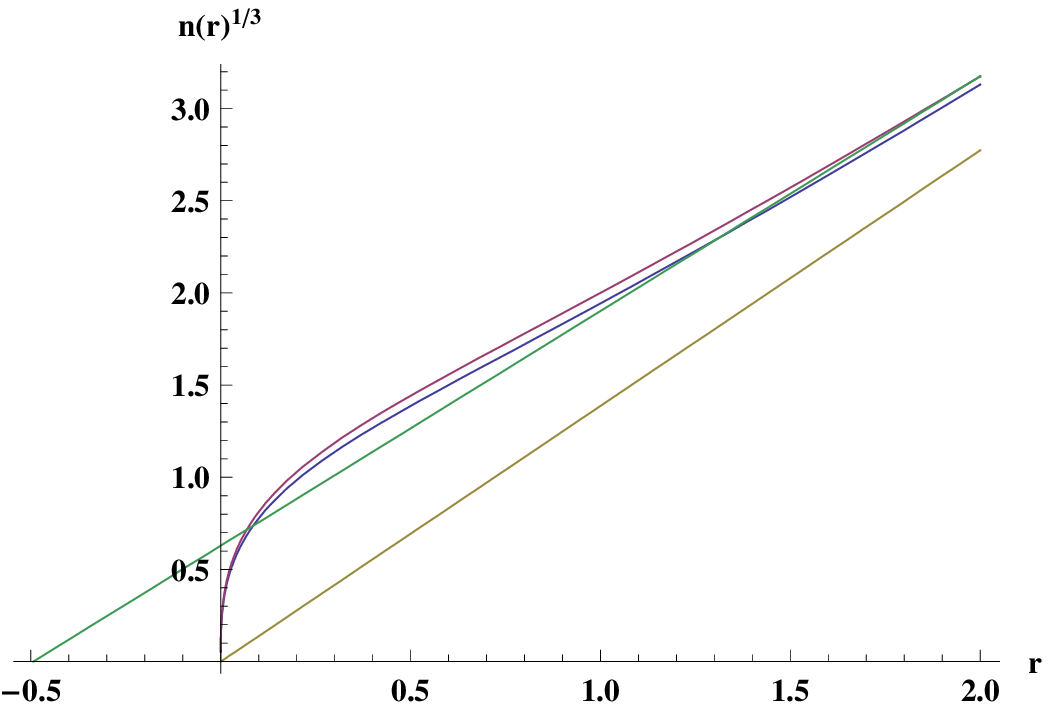}
\caption{$N'(r-1/2)^{1/3}$ (blue), $n(r)^{1/3}$ (red), their continuum limit $(8/3)^{1/3} r$ (brown), and the linear inter/extrapolation $f(r)$ (green), for the 4D cubic lattice.}
\label{fcubicnpthrd}
}

It is instructive to test this on the cubic lattice case in four dimensions.
We shall do the fitting to
$n(r)^{1/3}$ as in the main text.
Figure (\ref{fcubicnpthrd}) shows $N'(r-1/2)^{1/3}$, $n(r)^{1/3}$,  their continuum limit $(8/3)^{1/3} r$, and a function $f(r)$ that is the linear interpolation of
$n(r)$, $r=2$, 3, \dots, and its linear extrapolation into the region $r<2$, where it has a zero at $s_0\simeq-0.49$, $f(s_0)=0$.
Its slope at the zero point, $f'(s_0)\simeq 1.27$, smaller than
$(8/3)^{1/3}\simeq 1.39$.
Note that the curve for $N'(r-1/2)^{1/3}$ lies slightly below that of $n(r)$, closer to the continuum limit.
Figure \ref{figcubic2fits} show the result of two least-squares fits, one with $s=0$ to the $n(r)^{1/3}$ data at $r=6$, 7, \ldots, 10, and one with $s=s_0$ to the data at $r=1$, 2, \ldots, respectively the A-fit and B-fit.
\FIGURE{\includegraphics[width=7cm]{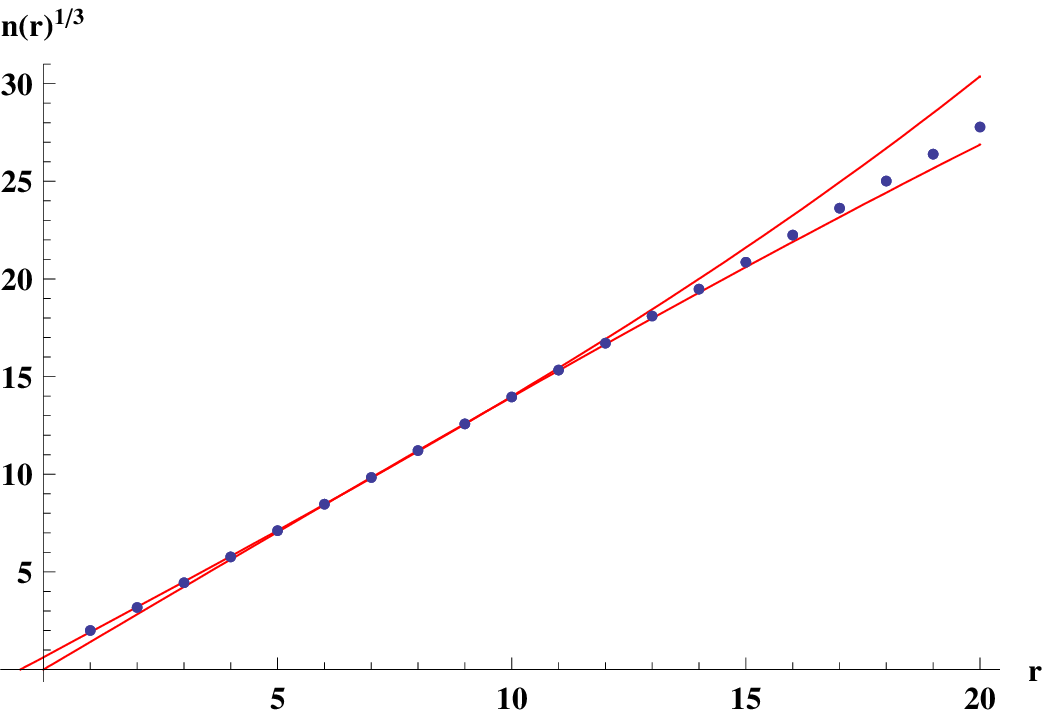}
\includegraphics[width=7cm]{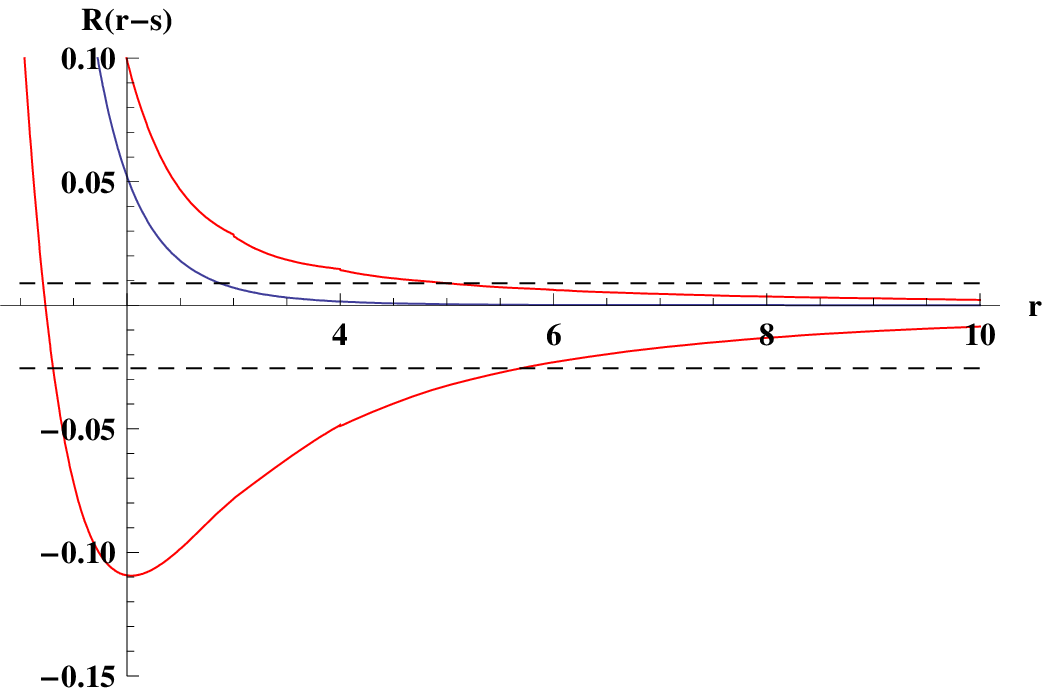}
\caption{Left: two fits (red) to the $n^{1/3}$ cubic lattice data (blue dots); lower curve: A-fit ($s=0$),
$c= 2.82$ ($\lm=0.61$), $r0=36.7$ ($R=+12/r_0^2=0.0089$), fitted data at $r=7$, 8, \ldots, 10; upper curve: B-fit ($s=s_0$), $c=2.11$ ($\lm=0.57$), $r_0=21.7$ ($R=-12/r_0^2=-0.025$), fitted data at $r=1$, 2, \ldots, 10.
Right: curvature $R(r)$ from the A-fit (upper red), from the exact $c=8/3$ ($s=0$, middle blue curve), and $R(r-s)$ from the B-fit (lower red). The straight dashed lines represent $\pm 12/r0^2$ of the A- and B-fit. }
\label{figcubic2fits}
}
The A-fit gives $y>0$, positive curvature with a $c$ larger than the exact value $8/3\simeq 2.67$ of the continuum limit, the B-fit $y<0$, negative curvature with $c$ smaller than $8/3$. The B-fit shows nice agreement to the smaller-$r$ data all the way down to $r=1$, but the values for $r_0$ and $c$ are actually less accurate than for the A-fit. Of course, the curvature radii increase to infinity and $\lm$ to the value in table \ref{ncubic} ($c\to 8/3$), as $\rfitmax\to\infty$.

Having determined $\veff$ we can then construct the scale factor $a(r)$ by (\ref{aVdef}) and compute the metric curvature $R(r-s)$ from (\ref{RRW}). The latter is shown in the right plot of figure \ref{figcubic2fits}. This curvature is not constant but becomes reasonably close to zero in the fitting region.
The plot also shows the more accurate $R(r)$ obtained from the exact value $c=8/3$, for which
$a(r)= (2r +r^3)^{1/3}$, $R(r) = 8 r^{-6} + \mathcal{O}(r^{-8})$.
The metric curvature $R(r)$ is sensitive to the magnitude of $a(r)$ through the term $6/a^2$ in (\ref{RRW}), an error in $c$ leads to an error in $R(r)$, with new terms in its asymptotic expansion starting already at order $r^{-2}$. In fact, the $-6a''/a$ and $-6a^{\prime 2}/a^2 + 6/a^2$ contributions are separately rather large but of opposite sign and they nearly cancel for the exact value of $c$.

In this cubic case the curvature $R(r)$ resulting from the A- and B-fit becomes accurate at the percent level of $\ell^{-2}$ at continuum distances $\rcon\gtrsim 3.1$ and 5.6, respectively. The accuracy of the B-fit lessens when we determine $s_0$ by linear extrapolation from $=1,\, 2$ instead of $r=2,\,3$ above. It leads to a 42\% larger $|s_0|$ and a slope $f'(s_0)$ deviating further from the continuum limit. Fortunately, in the SDT case this difference is only about 15\%.

The DOB-fit plays tricks here: $\Rosc(r+1/2)$ has an apparent minimum at the left boundary of the $r$-values for which (\ref{osceq}) has a solution, which is $r=3$. For $r=2$ there is no solution\footnote{This happens also in SDT cases, but then there are true minima in $r>3$.} and the region $r<3$ cannot be reached by interpolation, which leads one to reject the apparent minimum. As might be expected, the principle of minimum sensitivity then leads the DOB-fit to `slide' to a stationary point which lies at infinity, $r_{\rm stat}=\infty$, where the fit becomes exact. For $r+1/2=8$, the resulting $\Rc$ is already more accurate than the one from the B-fit with fitting domain $\{7,8,9,10\}$.

\section{Results of A- and B-fits}
\label{appAB}

\FIGURE{\includegraphics[width=7cm]{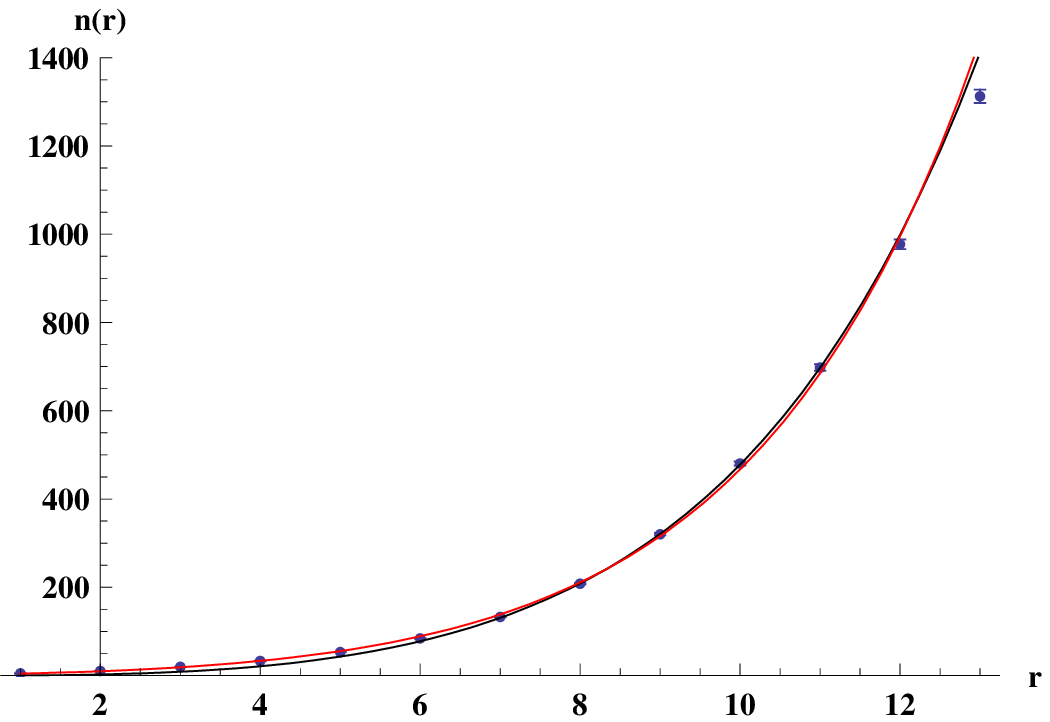}
\includegraphics[width=7cm]{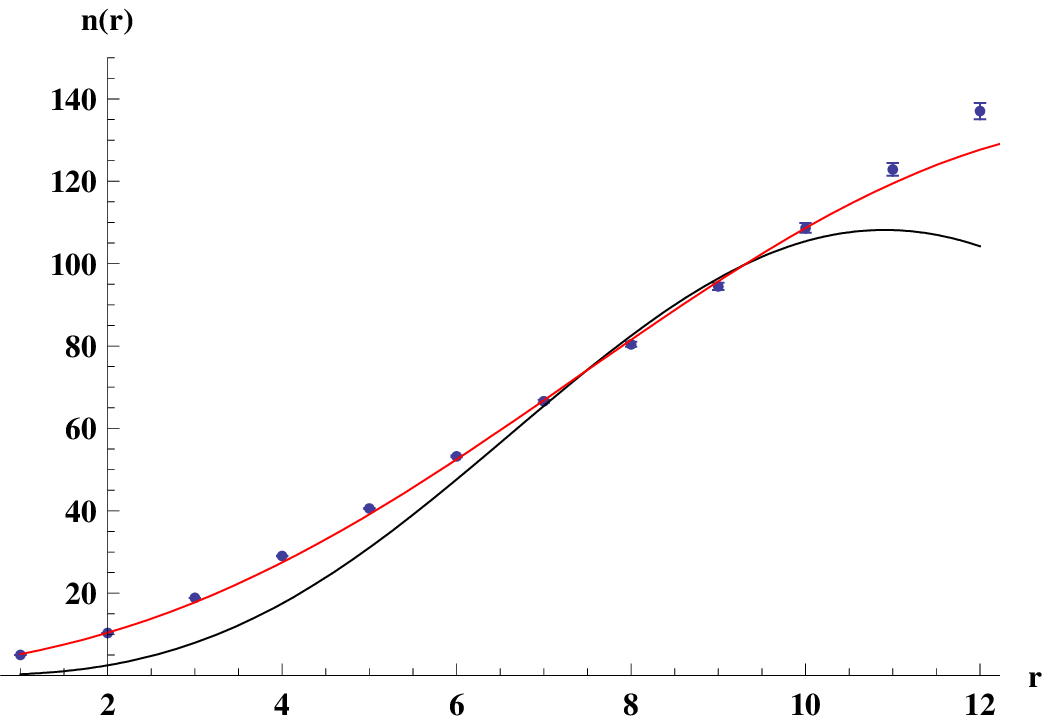}
\caption{Left: results of fits to the $n(r)$ data in the crumpled phase for $\kp_2=1.26$;
black curve: A-fit in $7\leq r\leq 12$, red curve: B-fit in $1\leq r\leq 12$. Right: similar for the elongated phase for $\kp_2=1.29$; A-fit $7\leq r\leq 10$, B-fit $1\leq r\leq 10$. Blue: numerical data with jackknife error bars.}
\label{fignAB260290}
}

We start with the A-fits.
The left plot in figure \ref{fignAB260290} shows the result of an A-fit of $c[\sinh(r/r_0)]^3$ to the $n(r)$ data in the crumpled phase at $\kp_2=1.255$, 1.260, 1.266, 1.270, for $N_4 = 64000$. To avoid cluttering only the curve for $\kp_2=1.26$ is shown in black.
The right plot shows a result with $\sinh\to\sin$ in the elongated phase (black curve,  $\kp_2=1.29$). In this case the fit was done simultaneously to the
data at $\kp_2=1.282$, 1.283, 1.285, 1.290, 1.300.
The parameters of the fits are $c=2\pi^2/\veff$ (independent of $\kp_2$ within a phase) and the $r_0$ depending on $\kp_2$; their values are in the following table:
\be
\begin{array}{llcllc}
\mbox{crumpled}&\mbox{phase    }&&
\mbox{elongated}&\mbox{phase}&\\
\mbox{c=0.305}&(\lm=0.436)&&
\mbox{c=0.323}&(\lm=0.442)&\\
\kp_2&r_0&&\kp_2&r_0\\
1.255&9.54&&1.282&8.28\\
1.260&10.4&&1.283&7.43\\
1.266&11.9&&1.285&7.21\\
1.270&13.5&&1.290&6.94\\
1.277&40.7&&1.300&6.85
\end{array}
\qquad\qquad \mbox{A-fit, $s=0$}.
\label{tAfit}
\ee
The lattice-continuum conversion factor $\lm$ corresponding to $c$ is also listed in (\ref{tAfit}).
The deviations in figure \ref{fignAB260290} at $r\leq 6$ are larger in the elongated phase than in the crumpled phase (but note that the vertical scale is enlarged by an order of magnitude).
\FIGURE{\includegraphics[width=12cm]{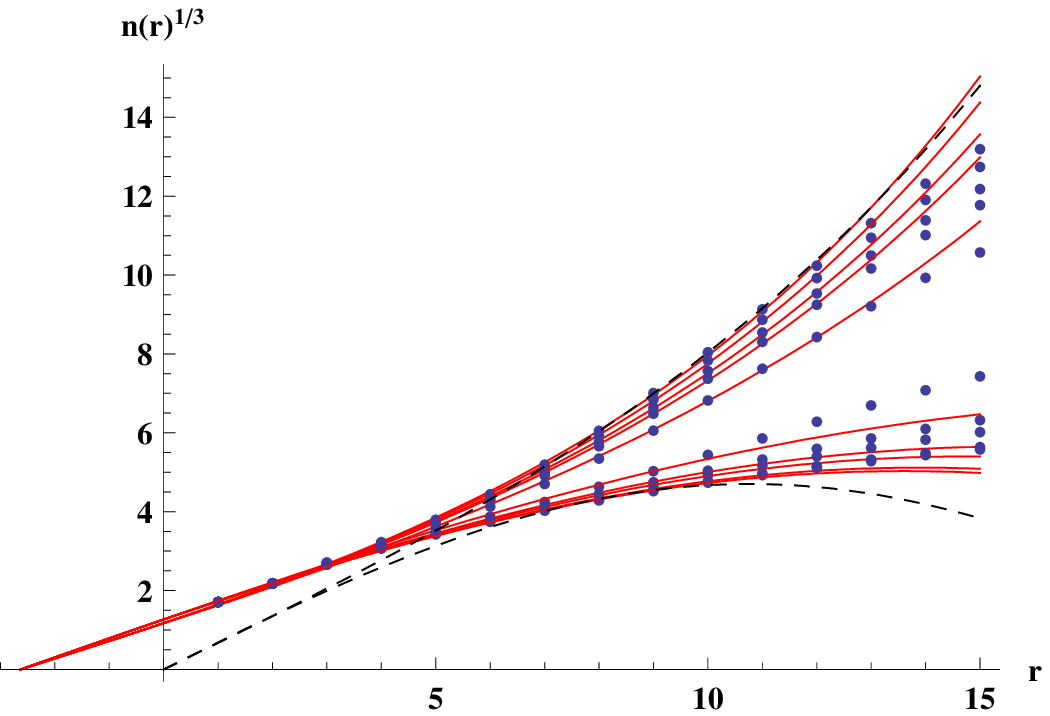}
\caption{B-Fit to the data in the crumpled phase (upper set of red curves, fitting domain $1\leq r\leq 12$) and elongated phase (lower set of red curves, fitting domain $1\leq r\leq 10$), with fixed $s_0=-2.63$. The black dashed curves represent the A-fit for $\kp_2=1.255$ and 1.3 .}
\label{figpnthrdfit}
}

In this respect, the B-fits that include the shift $s$ do visually much better.
Figure \ref{figpnthrdfit} shows results of a B-fit with fixed $s=-2.63$, which is the average of the two $s_0$ of figure \ref{figpnthrd260290}. The same $\kp_2$ and $N_4$ were chosen as in figure \ref{fignAB260290}. In the crumpled phase, $c^{1/3}r_0\sinh[(r-s)/r_0]$ was fitted with the least-squares method\footnote{Once $s$ is fixed by the condition $n(s)=0$, similar results can be obtained with chi-squared fits to $n(r)$ data in the region $r\geq 7$.} to $n(r)^{1/3}$; in the elongated phase $\sinh\to\sin$.
For comparison we have also shown the result of the A-fit in figure \ref{figpnthrdfit} for $\kp_2=1.255$ and 1.300 (dashed curves), and on the $n(r)$ plot in figure \ref{fignAB260290} the result of the B-fit (red curves).

The B-fit gives clearly a better description of the data. (However, for the 4D cubic test case in appendix \ref{appmodels} the A-fit is more accurate than the B-fit in estimating $\veff$ and in reproducing the zero curvature.) It is comforting to note in passing that two-parameter (`devil's advocate') fits of the form $\al r + \bt r^3$ to $n(r)^{1/3}$ at individual $\kp_2$ give larger least-squares deviations than two-parameter ($c$ and $r_0$) B-fits. Note also that the B-fits reasonably approximate $n(r)$ over a larger domain than the osculation fits in figure \ref{figpthrd260290osc}.
The B-fit analog of the A-fit table in (\ref{tAfit}) is
\be
\begin{array}{llcllc}
\mbox{crumpled}&\mbox{phase    }&&
\mbox{elongated}&\mbox{phase}&\\
\mbox{c=0.0853}&(\lm=0.317)&&
\mbox{c=0.115}&(\lm=0.341)&\\
\kp_2&r_0&&\kp_2&r_0\\
1.255&8.31&&1.282&13.9\\
1.260&8.65&&1.283&11.6\\
1.266&9.13&&1.285&11.1\\
1.270&9.55&&1.290&10.5\\
1.277&11.2&&1.300&10.3
\end{array}
\qquad\qquad \mbox{B-fit, $s=-2.63$}.
\label{tBfit}
\ee
The $c$ parameters are about a factor of three smaller than for the A-fit, implying also smaller scaling factors $\lm$. Using the above values of $s$ and $c$ in a B-fit with one fit parameter $r_0$ to the scaling-sequence data discussed in section \ref{secscaling} leads to
\be
\begin{array}{ccccc}
(N_4,\,\kp_2)&(8000,\,1.17)&(16000,\,1.21)&(32000,\,1.23)&(64000,\,1.26)\\
r_0&8.87&8.96&8.67&8.65
\end{array}
\label{tBfitscal}
\ee
where we have listed again the value for $(N_4,\,\kp_2)=(64000,\,1.26)$.
Figure \ref{figpRcAB} shows the $\lm$-scaled curvatures $\pm 12/(\lm r_0^2)$ for the A and B fit at $N_4=64000$ as a function of $\kp_2$, and we have also included the result of the DOB-fit in section \ref{secLocMet}.
Qualitatively, the switching from negative to positive at the phase transition is present in all fits, and in the B- and DOB-fits the switch is very close to $\kpc=1.280(1)$ determined by the peak in the node susceptibility in \cite{deBakker:1996zx}. Quantitatively, the systematic differences are evidently large.

\FIGURE{\includegraphics[width=12cm]{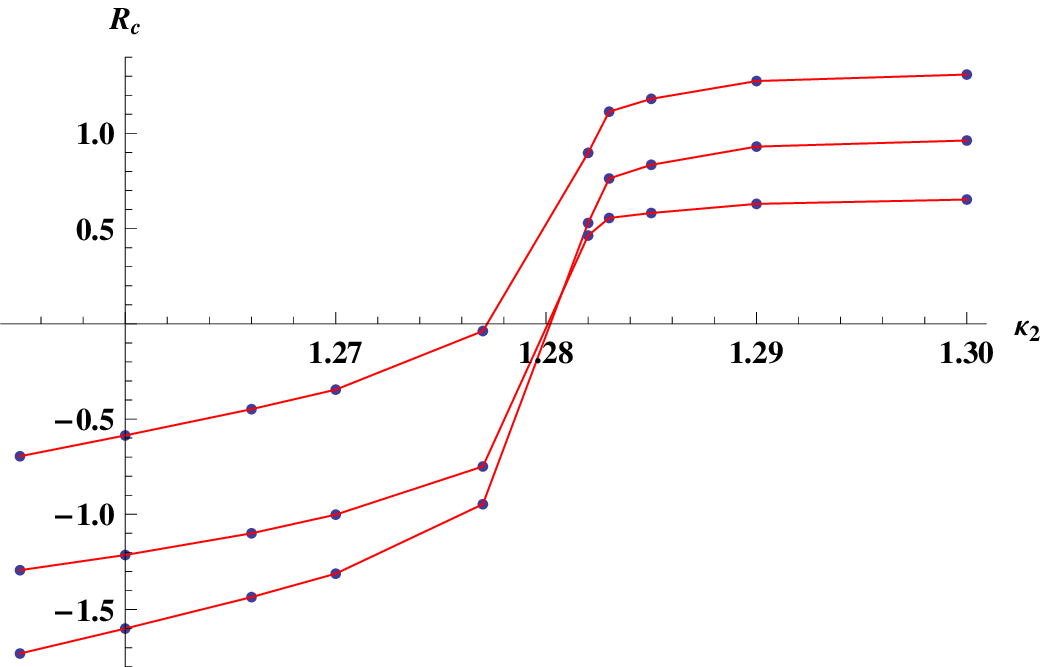}
\caption{Curvatures $R_{\rm c}=\pm 12/(\lm r_0)^2$ with linear interpolation. From top to bottom in $\kp_2 > 1.28$: A-fit, B-fit, DOB-fit.}
\label{figpRcAB}
}

\FIGURE{\includegraphics[width=7cm]{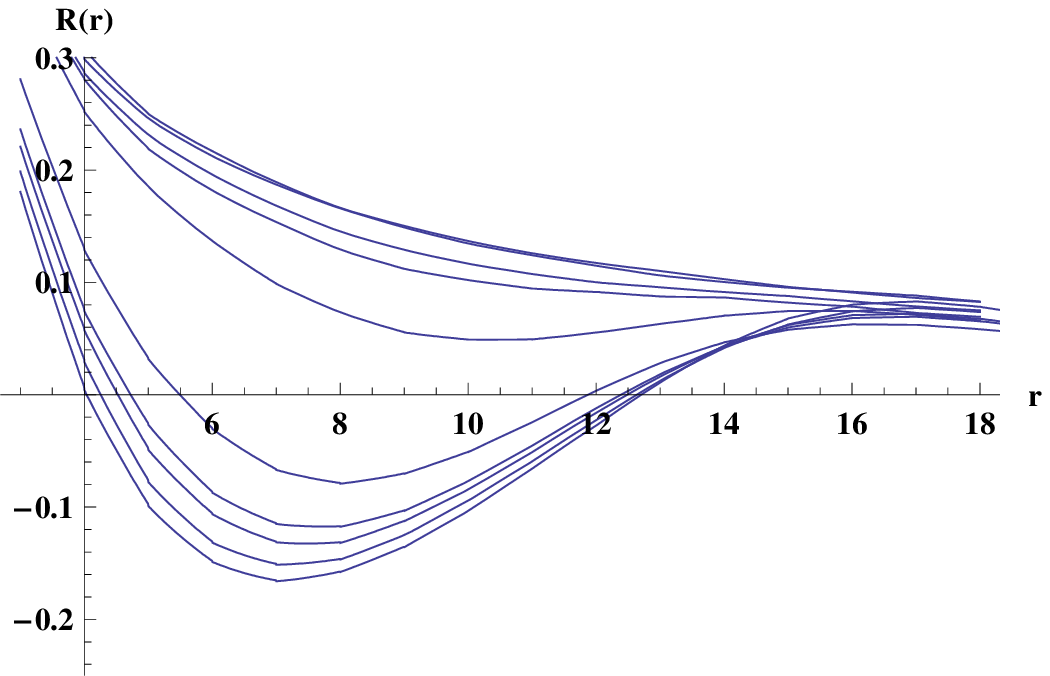}
\includegraphics[width=7cm]{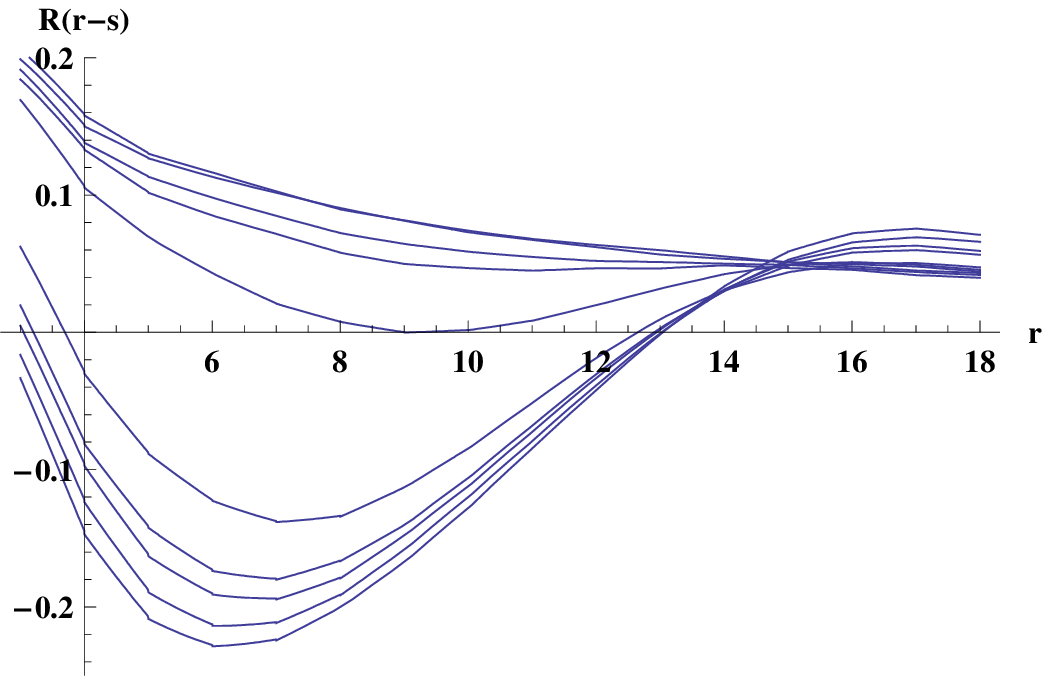}
\caption{Left: curvatures $R(r)$ from the A-fit for the $\kp_2$ values of figure \ref{figpRcAB}. Right: $R(r-s)$ from the B-fit. A cubic polynomial was used for interpolating $a(r)$.}
\label{figRRW}
}
Given the parameter $c=2\pi^2/\veff$, the metric scale factor $a(r-s)=[n(r)/c]^{1/3}$ can be obtained again by interpolation. The curvature $R(r)$ following from (\ref{RRW}) is shown in figure \ref{figRRW}.
As for the DOB-fits, this $R(r)$ is not constant, although it has shallow regions both phases. It is evidently an observable that is very sensitive to small deviations of $a(r)$ from the constant-curvature form. In particular the nice looking match of the B-fit to the data in figure \ref{figpnthrdfit} contrasts with the non-constancy of the curves in the right plot of figure \ref{figRRW}. The plots are similar to those in figure \ref{figcubic2fits} in appendix \ref{appmodels},  in which lattice artifacts produced by similar A- and B-fits are shown for the case of a 4D hypercubic lattice in flat space. Surely, there are discretization artifacts in figures \ref{figpRcAB} and \ref{figRRW}, but there is a qualitative difference with figure \ref{figcubic2fits}: in the flat cubic case, the A-fit produced a fake positive curvature and the B-fit a fake negative curvature, whereas here the A-fit in figure \ref{figRRW} shows positive as well as negative curvature, and similarly for the B-fit (right plot), with an upwardly shifted R(A) compared to R(B).
This supports the notion that the curvature plots reflect a genuine aspect of quantum spacetimes. A non-perturbative regularization in the continuum might may well show similar behavior. Note the small continuum distance $\rcon$ at which the matching of continuum spacetimes to the SDT results is done: $r=8.5$ in the matching region of the A-fit corresponds to $\rcon=\lm r\simeq 4.5$ and 3.8 for the crumpled and elongated phase; for the B-fit these are respectively $\rcon=\lm (8.5-s)\simeq 3.5$ and 3.8. These $\rcon$ are just a little larger than the lattice spacing $\ell=\sqrt{10}\, \ellt \simeq 3.2$.

The results in this appendix do not change much upon relaxing some chosen constraints on the parameter freedom. Instead of using one value of $s$ it can be set to the zero $s_0$ found from the linear extrapolation of $n(r)^{1/3}$ at each $\kp_2$ individually as in figure \ref{figpnthrd260290}, and also $c$ can be fitted individually. This gives very similar results, with values of $c$ differing only by a few percent within a phase. Of course, the individual fits improve somewhat, although hardly visible. However, for $N_4$ smaller than 32000 this improvement becomes important.

\section{Quartic effective potential model}
\label{appquartic}
The relevant solution of the semiclassical equation can be expressed as
\be
R_\pm^{\rm soln}(\kp) = R_\pm + \frac{3 s_\pm}{2 t_\pm} \left(\frac{1}{y_\pm}-y_\pm\right),\quad
y_\pm=\left(-\kp t_\pm + \sqrt{1+\kp^2 t_\pm^2}\right)^{1/3}.
\ee
Then $\Rav$, $\ch$ and $\chb$ follow from $\ln z_\pm=V w_\pm$,
\bea
\Rav &=& r_+ w^{\prime}_+ + r_- w^{\prime}_-, \quad
r_\pm = z_\pm/z,
\\
\ch &=& \chp+\chb,
\\
\chp&=& V\left[ r_+(1-r_+) w^{\prime\, 2}_+ + r_-(1-r_-) w^{\prime\, 2}_-
-2 r_+ r_- w^{\prime}_+ w^{\prime}_-\right],
\\
\chb &=& r_+ w^{\prime\prime}_+ + r_- w^{\prime\prime}_-,
\eea
where the prime denotes differentiation with respect to $\kp$. Writing $R_\pm = \Rs\pm\Rd$, it turns out that $\chb$ depends only on $s_\pm$ and $t_\pm$ (apart from $\kp$), $\chp$ depends also on $\Rd$ but not on $\Rs$, and $\Rav$ in addition on $\Rs$. In the Gaussian model (\ref{zGauss}) $r_-=1-r_+$.

\bibliography{lit}
\end{document}

%% file: newcomm.tex
\newcommand{\intx}{\int d^4 x\,}

\newcommand{\al}{\alpha}
\newcommand{\bt}{\beta}
\newcommand{\gm}{\gamma}
\newcommand{\dl}{\delta}

\newcommand{\et}{\eta}
\newcommand{\kp}{\kappa}
\newcommand{\lm}{\lambda}
\newcommand{\rh}{\rho}

\newcommand{\ta}{\tau}

\newcommand{\vr}{\varphi}
\newcommand{\ch}{\chi}

\newcommand{\Om}{\Omega}

\newcommand{\Gm}{\Gamma}
\newcommand{\Lm}{\Lambda}
\newcommand{\Sg}{\Sigma}

\newcommand{\half}{\frac{1}{2}}
\newcommand{\quart}{\frac{1}{4}}

\newcommand{\Tr}{\mbox{Tr}}

\newcommand{\eela}[1]{\label{#1}\end{equation}}
\newcommand{\eeala}[1]{\label{#1}\end{eqnarray}}
\newcommand{\be}{\begin{equation}}
\newcommand{\ee}{\end{equation}}
\newcommand{\bea}{\begin{eqnarray}}
\newcommand{\eea}{\end{eqnarray}}

%% file: background.bbl
\providecommand{\href}[2]{#2}\begingroup\raggedright\begin{thebibliography}{10}

\bibitem{Ambjorn:1998xu}
J.~Ambjorn and R.~Loll, {\it {Nonperturbative Lorentzian quantum gravity,
  causality and topology change}},  {\em Nucl.Phys.} {\bf B536} (1998)
  407--434, [\href{http://xxx.lanl.gov/abs/hep-th/9805108}{{\tt
  hep-th/9805108}}].

\bibitem{Ambjorn:2001cv}
J.~Ambjorn, J.~Jurkiewicz, and R.~Loll, {\it {Dynamically triangulating
  Lorentzian quantum gravity}},  {\em Nucl. Phys.} {\bf B610} (2001) 347--382,
  [\href{http://xxx.lanl.gov/abs/hep-th/0105267}{{\tt hep-th/0105267}}].

\bibitem{Ambjorn:2005qt}
J.~Ambjorn, J.~Jurkiewicz, and R.~Loll, {\it {Reconstructing the universe}},
  {\em Phys. Rev.} {\bf D72} (2005) 064014,
  [\href{http://xxx.lanl.gov/abs/hep-th/0505154}{{\tt hep-th/0505154}}].

\bibitem{Ambjorn:2008wc}
J.~Ambjorn, A.~Gorlich, J.~Jurkiewicz, and R.~Loll, {\it {The Nonperturbative
  Quantum de Sitter Universe}},  {\em Phys. Rev.} {\bf D78} (2008) 063544,
  [\href{http://xxx.lanl.gov/abs/0807.4481}{{\tt arXiv:0807.4481}}].

\bibitem{Ambjorn:2010fv}
J.~Ambjorn, A.~Gorlich, J.~Jurkiewicz, and R.~Loll, {\it {Geometry of the
  quantum universe}},  {\em Phys.Lett.} {\bf B690} (2010) 420--426,
  [\href{http://xxx.lanl.gov/abs/1001.4581}{{\tt arXiv:1001.4581}}].

\bibitem{Ambjorn:2012ij}
J.~Ambjorn, S.~Jordan, J.~Jurkiewicz, and R.~Loll, {\it {Second- and
  First-Order Phase Transitions in CDT}},  {\em Phys.Rev.} {\bf D85} (2012)
  124044, [\href{http://xxx.lanl.gov/abs/1205.1229}{{\tt arXiv:1205.1229}}].

\bibitem{Ambjorn:2012jv}
J.~Ambjorn, A.~Goerlich, J.~Jurkiewicz, and R.~Loll, {\it {Nonperturbative
  Quantum Gravity}},  \href{http://xxx.lanl.gov/abs/1203.3591}{{\tt
  arXiv:1203.3591}}.

\bibitem{Agishtein:1991cv}
M.~E. Agishtein and A.~A. Migdal, {\it {Simulations of four-dimensional
  simplicial quantum gravity}},  {\em Mod. Phys. Lett.} {\bf A7} (1992)
  1039--1062.

\bibitem{Ambjorn:1991pq}
J.~Ambjorn and J.~Jurkiewicz, {\it {Four-dimensional simplicial quantum
  gravity}},  {\em Phys. Lett.} {\bf B278} (1992) 42--50.

\bibitem{Thorleifsson:1998jr}
G.~Thorleifsson, {\it {Lattice gravity and random surfaces}},  {\em
  Nucl.Phys.Proc.Suppl.} {\bf 73} (1999) 133--145,
  [\href{http://xxx.lanl.gov/abs/hep-lat/9809131}{{\tt hep-lat/9809131}}].

\bibitem{Krzywicki:1999ai}
A.~Krzywicki, {\it {Random manifolds and quantum gravity}},  {\em
  Nucl.Phys.Proc.Suppl.} {\bf 83} (2000) 126--130,
  [\href{http://xxx.lanl.gov/abs/hep-lat/9907012}{{\tt hep-lat/9907012}}].

\bibitem{Ambjorn:1995dj}
J.~Ambjorn and J.~Jurkiewicz, {\it {Scaling in four-dimensional quantum
  gravity}},  {\em Nucl. Phys.} {\bf B451} (1995) 643--676,
  [\href{http://xxx.lanl.gov/abs/hep-th/9503006}{{\tt hep-th/9503006}}].

\bibitem{Ambjorn:1996ny}
J.~Ambjorn, M.~Carfora, and A.~Marzuoli, {\it {The Geometry of Dynamical
  Triangulations}},  {\em Lecture Notes in Physics Monographs} {\bf 50} (1997),
  (Springer) [\href{http://xxx.lanl.gov/abs/hep-th/9612069}{{\tt
  hep-th/9612069}}].

\bibitem{Gabrielli:1997zy}
D.~Gabrielli, {\it {Polymeric phase of simplicial quantum gravity}},  {\em
  Phys.Lett.} {\bf B421} (1998) 79--85,
  [\href{http://xxx.lanl.gov/abs/hep-lat/9710055}{{\tt hep-lat/9710055}}].

\bibitem{Gionti:1998jy}
S.~Gionti, Gabriele, {\it {Simplicial quantum gravity in the elongated phase}},
   {\em J.Math.Phys.} {\bf 39} (1998) 6593--6602.

\bibitem{AmbjornDJ1997}
J.~Ambjorn, B.~Durhuus, and T.~Jonsson, {\em Quantum {G}eometry}.
\newblock Cambridge University Press, Cambridge, {UK}, 1997.

\bibitem{Hotta:1995ud}
T.~Hotta, T.~Izubuchi, and J.~Nishimura, {\it {Singular Vertices in the Strong
  Coupling Phase of Four-Dimensional Simplicial Gravity}},  {\em Nucl. Phys.
  Proc. Suppl.} {\bf 47} (1996) 609--612,
  [\href{http://xxx.lanl.gov/abs/hep-lat/9511023}{{\tt hep-lat/9511023}}].

\bibitem{Hotta:1995ca}
T.~Hotta, T.~Izubuchi, and J.~Nishimura, {\it {Singular vertices in the strong
  coupling phase of four-dimensional simplicial gravity}},  {\em Prog. Theor.
  Phys.} {\bf 94} (1995) 263--270,
  [\href{http://xxx.lanl.gov/abs/hep-lat/9709073}{{\tt hep-lat/9709073}}].

\bibitem{Catterall:1995ig}
S.~Catterall, G.~Thorleifsson, J.~B. Kogut, and R.~Renken, {\it {Singular
  Vertices and the Triangulation Space of the D- sphere}},  {\em Nucl. Phys.}
  {\bf B468} (1996) 263--276,
  [\href{http://xxx.lanl.gov/abs/hep-lat/9512012}{{\tt hep-lat/9512012}}].

\bibitem{Catterall:1997xj}
S.~Catterall, R.~Renken, and J.~B. Kogut, {\it {Singular structure in 4-D
  simplicial gravity}},  {\em Phys.Lett.} {\bf B416} (1998) 274--280,
  [\href{http://xxx.lanl.gov/abs/hep-lat/9709007}{{\tt hep-lat/9709007}}].

\bibitem{Bialas:1996wu}
P.~Bialas, Z.~Burda, A.~Krzywicki, and B.~Petersson, {\it {Focusing on the
  fixed point of 4d simplicial gravity}},  {\em Nucl. Phys.} {\bf B472} (1996)
  293--308, [\href{http://xxx.lanl.gov/abs/hep-lat/9601024}{{\tt
  hep-lat/9601024}}].

\bibitem{deBakker:1996zx}
B.~V. de~Bakker, {\it {Further evidence that the transition of 4D dynamical
  triangulation is 1st order}},  {\em Phys. Lett.} {\bf B389} (1996) 238--242,
  [\href{http://xxx.lanl.gov/abs/hep-lat/9603024}{{\tt hep-lat/9603024}}].

\bibitem{Jurkiewicz:1996yd}
J.~Jurkiewicz and A.~Krzywicki, {\it {Branched polymers with loops}},  {\em
  Phys. Lett.} {\bf B392} (1997) 291--297,
  [\href{http://xxx.lanl.gov/abs/hep-th/9610052}{{\tt hep-th/9610052}}].

\bibitem{Antoniadis:1996pb}
I.~Antoniadis, P.~O. Mazur, and E.~Mottola, {\it {Criticality and scaling in
  4-D quantum gravity}},  {\em Phys.Lett.} {\bf B394} (1997) 49--56,
  [\href{http://xxx.lanl.gov/abs/hep-th/9611145}{{\tt hep-th/9611145}}].

\bibitem{Antoniadis:1992xu}
I.~Antoniadis, P.~O. Mazur, and E.~Mottola, {\it {Conformal symmetry and
  central charges in four-dimensions}},  {\em Nucl.Phys.} {\bf B388} (1992)
  627--647, [\href{http://xxx.lanl.gov/abs/hep-th/9205015}{{\tt
  hep-th/9205015}}].

\bibitem{Bilke:1997sc}
S.~Bilke, Z.~Burda, A.~Krzywicki, B.~Petersson, J.~Tabaczek, and
  G.~Thorleifsson, {\it {4d simplicial quantum gravity interacting with gauge
  matter fields}},  {\em Phys. Lett.} {\bf B418} (1998) 266--272,
  [\href{http://xxx.lanl.gov/abs/hep-lat/9710077}{{\tt hep-lat/9710077}}].

\bibitem{Bilke:1998vj}
S.~Bilke, Z.~Burda, A.~Krzywicki, B.~Petersson, J.~Tabaczek, and
  G.~Thorleifsson, {\it {4d simplicial quantum gravity: Matter fields and the
  corresponding effective action}},  {\em Phys. Lett.} {\bf B432} (1998)
  279--286, [\href{http://xxx.lanl.gov/abs/hep-lat/9804011}{{\tt
  hep-lat/9804011}}].

\bibitem{Horata:2000eg}
S.~Horata, H.~S. Egawa, N.~Tsuda, and T.~Yukawa, {\it {Phase structure of
  four-dimensional simplicial quantum gravity with a U(1) gauge field}},  {\em
  Prog. Theor. Phys.} {\bf 106} (2001) 1037--1050,
  [\href{http://xxx.lanl.gov/abs/hep-lat/0004021}{{\tt hep-lat/0004021}}].

\bibitem{Horata:2003hm}
S.~Horata, T.~Yukawa, and H.~S. Egawa, {\it {Matter dependence of the string
  susceptibility exponent in four-dimensional simplicial quantum gravity}},
  {\em Prog. Theor. Phys.} {\bf 108} (2002) 1171--1176.

\bibitem{Bruegmann:1992jk}
B.~Bruegmann and E.~Marinari, {\it {4-d simplicial quantum gravity with a
  nontrivial measure}},  {\em Phys.Rev.Lett.} {\bf 70} (1993) 1908--1911,
  [\href{http://xxx.lanl.gov/abs/hep-lat/9210002}{{\tt hep-lat/9210002}}].

\bibitem{Ambjorn:1999ix}
J.~Ambjorn, K.~Anagnostopoulos, and J.~Jurkiewicz, {\it {Abelian gauge fields
  coupled to simplicial quantum gravity}},  {\em JHEP} {\bf 9908} (1999) 016,
  [\href{http://xxx.lanl.gov/abs/hep-lat/9907027}{{\tt hep-lat/9907027}}].

\bibitem{Laiho:2011ya}
J.~Laiho and D.~Coumbe, {\it {Evidence for Asymptotic Safety from Lattice
  Quantum Gravity}},  {\em Phys.Rev.Lett.} {\bf 107} (2011) 161301,
  [\href{http://xxx.lanl.gov/abs/1104.5505}{{\tt arXiv:1104.5505}}].

\bibitem{Laiho:2011zz}
J.~Laiho and D.~Coumbe, {\it {Asymptotic safety and lattice quantum gravity}},
  {\em PoS} {\bf LATTICE2011} (2011) 005.

\bibitem{Nambu:1961tp}
Y.~Nambu and G.~Jona-Lasinio, {\it {Dynamical Model of Elementary Particles
  Based on an Analogy with Superconductivity. 1.}},  {\em Phys.Rev.} {\bf 122}
  (1961) 345--358.

\bibitem{Buballa:2003qv}
M.~Buballa, {\it {NJL model analysis of quark matter at large density}},  {\em
  Phys.Rept.} {\bf 407} (2005) 205--376,
  [\href{http://xxx.lanl.gov/abs/hep-ph/0402234}{{\tt hep-ph/0402234}}].

\bibitem{Elitzur:1979uv}
S.~Elitzur, R.~Pearson, and J.~Shigemitsu, {\it {The Phase Structure of
  Discrete Abelian Spin and Gauge Systems}},  {\em Phys.Rev.} {\bf D19} (1979)
  3698.

\bibitem{Horn:1979fy}
D.~Horn, M.~Weinstein, and S.~Yankielowicz, {\it {Hamiltonian approach to Z(N)
  lattice gauge theories}},  {\em Phys.Rev.} {\bf D19} (1979) 3715.

\bibitem{Ukawa:1979yv}
A.~Ukawa, P.~Windey, and A.~H. Guth, {\it {Dual Variables for Lattice Gauge
  Theories and the Phase Structure of Z(N) Systems}},  {\em Phys.Rev.} {\bf
  D21} (1980) 1013.

\bibitem{Yankielowicz:1981ug}
S.~Yankielowicz, {\it {Phase in gauge theories}},  {\em Structural Elements in
  Particle Physics and Statistical Mechanics,} {\bf Plenum Press, New York}
  (1983) 115.

\bibitem{Alessandrini:1982ju}
V.~Alessandrini, {\it {Dynamical generation of a Coulomb phase in the mean
  field approach to Z(N) lattice gauge theories}},  {\em Nucl.Phys.} {\bf B215}
  (1983) 337.

\bibitem{deBakker:1994zf}
B.~V. de~Bakker and J.~Smit, {\it {Curvature and scaling in 4-d dynamical
  triangulation}},  {\em Nucl. Phys.} {\bf B439} (1995) 239--258,
  [\href{http://xxx.lanl.gov/abs/hep-lat/9407014}{{\tt hep-lat/9407014}}].

\bibitem{Bialas:1996eh}
P.~Bialas, Z.~Burda, B.~Petersson, and J.~Tabaczek, {\it {Appearance of mother
  universe and singular vertices in random geometries}},  {\em Nucl.Phys.} {\bf
  B495} (1997) 463--476, [\href{http://xxx.lanl.gov/abs/hep-lat/9608030}{{\tt
  hep-lat/9608030}}].

\bibitem{Egawa:1996fu}
H.~S. Egawa, T.~Hotta, T.~Izubuchi, N.~Tsuda, and T.~Yukawa, {\it {Scaling
  behavior in 4D simplicial quantum gravity}},  {\em Prog. Theor. Phys.} {\bf
  97} (1997) 539--552, [\href{http://xxx.lanl.gov/abs/hep-lat/9611028}{{\tt
  hep-lat/9611028}}].

\bibitem{deBakker:1995yb}
B.~V. de~Bakker, {\it {Simplicial quantum gravity}},
  \href{http://xxx.lanl.gov/abs/hep-lat/9508006}{{\tt hep-lat/9508006}}.

\bibitem{Bruegmann:1995qz}
B.~Bruegmann and E.~Marinari, {\it {Monte Carlo simulations of 4-D simplicial
  quantum gravity}},  {\em J.Math.Phys.} {\bf 36} (1995) 6340--6352,
  [\href{http://xxx.lanl.gov/abs/hep-lat/9504004}{{\tt hep-lat/9504004}}].

\bibitem{Ambjorn:1998ec}
J.~Ambjorn, M.~Carfora, D.~Gabrielli, and A.~Marzuoli, {\it {Crumpled
  triangulations and critical points in 4-D simplicial quantum gravity}},  {\em
  Nucl.Phys.} {\bf B542} (1999) 349--394,
  [\href{http://xxx.lanl.gov/abs/hep-lat/9806035}{{\tt hep-lat/9806035}}].

\bibitem{Ratcliffe2006}
J.~G. Ratcliffe, {\em Foundations of {H}yperbolic {M}anifolds (2nd ed.)}.
\newblock Springer, New {Y}ork, {USA}, 2006.

\bibitem{RatcliffeVolumeSpectrum}
J.~G. Ratcliffe, Steven, and S.~T. Tschantz, {\it The volume spectrum of
  hyperbolic 4-manifolds},  {\em Experiment. Math} {\bf 9} (2000) 101--125.

\bibitem{Ratcliffe:1998rb}
J.~Ratcliffe and S.~Tschantz, {\it {Gravitational instantons of constant
  curvature}},  {\em Class.Quant.Grav.} {\bf 15} (1998) 2613--2627.

\bibitem{Anderson:2003js}
M.~Anderson, S.~Carlip, J.~Ratcliffe, S.~Surya, and S.~Tschantz, {\it {Peaks in
  the Hartle-Hawking wave function from sums over topologies}},  {\em
  Class.Quant.Grav.} {\bf 21} (2004) 729--742,
  [\href{http://xxx.lanl.gov/abs/gr-qc/0310002}{{\tt gr-qc/0310002}}].

\bibitem{Carlip:2001wq}
S.~Carlip, {\it {Quantum Gravity: A Progress Report}},  {\em Rept.Prog.Phys.}
  {\bf 64} (2001) 885, [\href{http://xxx.lanl.gov/abs/gr-qc/0108040}{{\tt
  gr-qc/0108040}}].

\bibitem{Bialas:1998ci}
P.~Bialas, Z.~Burda, and D.~Johnston, {\it {Phase diagram of the mean field
  model of simplicial gravity}},  {\em Nucl.Phys.} {\bf B542} (1999) 413--424,
  [\href{http://xxx.lanl.gov/abs/gr-qc/9808011}{{\tt gr-qc/9808011}}].

\bibitem{Bialas:1999ad}
P.~Bialas, L.~Bogacz, Z.~Burda, and D.~Johnston, {\it {Finite size scaling of
  the balls in boxes model}},  {\em Nucl.Phys.} {\bf B575} (2000) 599--612,
  [\href{http://xxx.lanl.gov/abs/hep-lat/9910047}{{\tt hep-lat/9910047}}].

\bibitem{Bialas:1995xq}
P.~Bialas, {\it {Long range correlations in branched polymers}},  {\em
  Phys.Lett.} {\bf B373} (1996) 289--295,
  [\href{http://xxx.lanl.gov/abs/hep-lat/9511024}{{\tt hep-lat/9511024}}].

\bibitem{Catterall:1994pg}
S.~Catterall, J.~B. Kogut, and R.~Renken, {\it {Phase structure of
  four-dimensional simplicial quantum gravity}},  {\em Phys.Lett.} {\bf B328}
  (1994) 277--283, [\href{http://xxx.lanl.gov/abs/hep-lat/9401026}{{\tt
  hep-lat/9401026}}].

\bibitem{Agishtein:1992xx}
M.~E. Agishtein and A.~A. Migdal, {\it {Critical behavior of dynamically
  triangulated quantum gravity in four-dimensions}},  {\em Nucl. Phys.} {\bf
  B385} (1992) 395--412, [\href{http://xxx.lanl.gov/abs/hep-lat/9204004}{{\tt
  hep-lat/9204004}}].

\bibitem{O'Raifeartaigh:1986hi}
L.~O'Raifeartaigh, A.~Wipf, and H.~Yoneyama, {\it {The Constraint Effective
  Potential}},  {\em Nucl.Phys.} {\bf B271} (1986) 653.

\bibitem{Dimitrovic:1991qg}
I.~Dimitrovic, J.~Nager, K.~Jansen, and T.~Neuhaus, {\it {Shape of the
  constraint effective potential: A Monte Carlo study}},  {\em Phys.Lett.} {\bf
  B268} (1991) 408--414.

\bibitem{Fodor:1994sj}
Z.~Fodor, J.~Hein, K.~Jansen, A.~Jaster, and I.~Montvay, {\it {Simulating the
  electroweak phase transition in the SU(2) Higgs model}},  {\em Nucl.Phys.}
  {\bf B439} (1995) 147--186,
  [\href{http://xxx.lanl.gov/abs/hep-lat/9409017}{{\tt hep-lat/9409017}}].

\bibitem{Rakow:2005yn}
P.~E. Rakow, {\it {Stochastic perturbation theory and the gluon condensate}},
  {\em PoS} {\bf LAT2005} (2006) 284,
  [\href{http://xxx.lanl.gov/abs/hep-lat/0510046}{{\tt hep-lat/0510046}}].

\end{thebibliography}\endgroup
